\documentclass[11pt,a4paper]{article}
\usepackage{jheppubedit}
\usepackage{enumitem}
\usepackage{amsmath}
\usepackage{amsfonts}
\usepackage{graphicx}
\usepackage{caption}
\usepackage{colortbl}
\usepackage{epstopdf}
\pdfoutput=1

\title{String Theory in Polar Coordinates and the Vanishing of the One-Loop Rindler Entropy}

\author[a]{Thomas G. Mertens,}
\author[b]{Henri Verschelde}
\author[c,d,e]{and Valentin I. Zakharov}
\affiliation[a]{Joseph Henry Laboratories, Princeton University, Princeton, NJ 08544, USA,} 
\affiliation[b]{Ghent University, Department of Physics and Astronomy\\
Krijgslaan, 281-S9, 9000 Gent, Belgium}
\affiliation[c]{ITEP, B. Cheremushkinskaya 25, Moscow, 117218 Russia,}
\affiliation[d]{Moscow Inst Phys \& Technol, Dolgoprudny, Moscow Region, 141700 Russia,}
\affiliation[e]{School of Biomedicine, Far Eastern Federal University, Sukhanova str 8, 
Vladivostok 690950 Russia}

\emailAdd{tmertens@princeton.edu}
\emailAdd{henri.verschelde@ugent.be}
\emailAdd{vzakharov@itep.ru}
\abstract{We analyze the string spectrum of flat space in polar coordinates, following the small curvature limit of the $SL(2,\mathbb{R})/U(1)$ cigar CFT. We first analyze the partition function of the cigar itself, making some clarifications of the structure of the spectrum that have escaped attention up to this point. The superstring spectrum (type 0 and type II) is shown to exhibit an involution symmetry, that survives the small curvature limit. We classify all marginal states in polar coordinates for type II superstrings, with emphasis on their links and their superconformal structure. This classification is confirmed by an explicit large $\tau_2$ analysis of the partition function. Next we compare three approaches towards the type II genus one entropy in Rindler space: using a sum-over-fields strategy, using a Melvin model approach as in \cite{He:2014gva} and finally using a saddle point method on the cigar partition function. In each case we highlight possible obstructions and motivate that the correct procedures yield a vanishing result: $S=0$. We finally discuss how the QFT UV divergences of the fields in the spectrum disappear when computing the free energy and entropy using Euclidean techniques.}

\begin{document}

\maketitle

\section{Introduction}
Perturbative string theory in curved backgrounds has many applications (e.g. holography, dualities etc.), but at the same time, a thorough understanding is still lacking. The main reason is of course that the worldsheet theory is not free anymore, and solving for the string spectrum directly proves to be nearly impossible. The only curved backgrounds where significant progress could be made are group and coset manifolds: their additional underlying structure allows an almost complete solution of the theory in many backgrounds (see e.g. \cite{Gepner:1986wi}\cite{Goddard:1984vk}\cite{Gawedzki:1988hq} for some of the seminal works). Surprisingly though, one of the most simple backgrounds, flat space described in polar coordinates, was never analyzed in detail in the past. \\

\noindent One can distinguish two main reasons for being interested in the string spectrum in polar coordinates.
Firstly, it is apparent that one should understand this basic situation in our endeavor to understand string theory in its full generality. \\
Secondly, and this is our main motivation, flat space in polar coordinates provides the thermal manifold for Rindler space. Hence the polar coordinate description is expected to contain some relevant information concerning string theory near black hole horizons; this is a story that we analyzed extensively in \cite{Mertens:2013pza}\cite{Mertens:2013zya}\cite{Mertens:2014nca}\cite{Mertens:2014cia}\cite{Mertens:2014dia}\cite{Mertens:2014saa}\cite{Mertens:2015hia}\cite{Mertens:2015adr} where we studied the link between the near-horizon random walking long string and the singly wound string in polar coordinates, which turns out to be precisely massless in this case. We argued there that this singly wound state (the thermal scalar) is the most important contribution to thermodynamical quantities of the near-horizon string gas, in a very similar way as happens for near-Hagedorn thermodynamics in flat space. \\

\noindent Recent firewall paradoxes related to black hole horizons, caused a revival of the study of string theory near black hole horizons \cite{Giveon:2012kp}\cite{Giveon:2013ica}\cite{Giveon:2014hfa}\cite{Giveon:2013hsa}\cite{Giveon:2015cma}\cite{Giribet:2015kca}\cite{Ben-Israel:2015mda}\cite{Ben-Israel:2015etg}\cite{Silverstein:2014yza}\cite{Dodelson:2015toa}\cite{Dodelson:2015uoa}\cite{Martinec:2014gka}\cite{Martinec:2015pfa}. \\
Also, a more general understanding of Rindler space in quantum gravity and holography is an active research area at the moment (see e.g. \cite{Czech:2012be}\cite{Parikh:2012kg}\cite{Halyo:2014rra}\cite{Halyo:2015ffa}\cite{Halyo:2015oja}\cite{Halyo:2016ofb}\cite{Afshar:2015wjm} for some recent work). \\

\noindent One way of obtaining flat space in a polar coordinate description, is to follow a certain parametric limit in a curved coset model and find the resulting string spectrum. Of course, some information is readily known because polar coordinates simply correspond to a different description of flat space. For instance, the type II partition function vanishes due to spacetime supersymmetry and this is independent of the coordinate system used. \\

\noindent The individual states and their description are completely different when comparing Cartesian to polar coordinates. For instance, in polar coordinates one chooses a priviledged origin and translational invariance for each mode is broken. However, coordinate invariant quantities such as the partition function will experience an emergent translational invariance as a result of summing over modes. \\

\noindent For the thermodynamical application, one is also interested in conical spaces to which we will also briefly turn further on. In string theory, one is restricted to studying cones with opening angles $2\pi/N$ with $N\in\mathbb{N}$, the $\mathbb{C}/\mathbb{Z}_N$ orbifold models. Such conical spaces have been largely studied in the past \cite{Dabholkar:1994ai}\cite{Lowe:1994ah}, but the way the $N=1$ limit is realized, has not. One of the motivations in this regard is to find out whether other marginal states (such as the singly wound state) exist and whether they are of relevance to thermodynamics.\\

\noindent Basically, we want to ask: ``Are there other marginal states on the thermal manifold of Rindler space (and finite mass black holes)? If so, when do we expect these to appear and in what sense are they relevant for thermodynamics?" \\
One reason to anticipate the appearance of other marginal states is that the type II partition function in flat space is proportional to 
\begin{equation}
\vartheta_3^4 - \vartheta_4^4-\vartheta_2^4
\end{equation}
and hence vanishes by Jacobi's obscure identity. This is so regardless on whether one expands this partition function in Cartesian or polar coordinates. From a polar coordinate perspective, this partition function computes the sum of the Rindler vacuum energy and the Rindler free energy of the system. We know that there exists at least one marginal state, the thermal scalar state. Hence the vanishing of the partition function requires there to be other marginal states that contribute with opposite sign to the partition function. Our goal is therefore to further elucidate this fact. \\

\noindent In \cite{Mertens:2013zya}\cite{Mertens:2014saa} we obtained partial results on the spectrum in polar coordinates,  only the parts that were directly relevant to our goal there. Here however, we are more ambitious on this front and set out to fully construct the superstring spectrum in this space. \\

\noindent From the many different group and coset models studied in the literature, there are two main candidates that yield flat space in polar coordinates (and its orbifolds) under a suitable parametric limit. The first is the Melvin background \cite{Tseytlin:1994ei}\cite{Russo:1995tj}\cite{Russo:1995ik}, in which the $R\to0$ limit would correspond to flat space $\mathbb{C}/\mathbb{Z}_N$ orbifolds \cite{Takayanagi:2001jj}\cite{He:2014gva}.\footnote{We will make this more precise further on.} The second is the 2d cigar $SL(2,\mathbb{R})_k/U(1)$ background \cite{Witten:1991yr}\cite{Dijkgraaf:1991ba} for which the small curvature (large $k$) limit would yield flat space \cite{Giveon:2012kp}\cite{Giveon:2013ica}\cite{Giveon:2014hfa}.\footnote{The $SL(2,\mathbb{R})_k/U(1)$ model continues to be a valuable guide towards understanding quantum black holes, see e.g. \cite{Giveon:2015cma}\cite{Giribet:2015kca}\cite{Ben-Israel:2015mda}\cite{Ben-Israel:2015etg} for some recent studies.} In this paper, we mainly focus on the second avenue (in line with our previous work). Near the end, we will briefly look at the Melvin background as well. \\

\noindent When computing the string spectrum, one usually imposes the on-shell condition on physical states as
\begin{equation}
L_0 \left|\psi\right\rangle = \bar{L}_0 \left|\psi\right\rangle = 0.
\end{equation}
In this paper, we will only impose 
\begin{align}
L_0 - \bar{L}_0 \in \mathbb{Z},
\end{align}
which corresponds to off-shell string states consistent with modular invariance. In string path integrals, such string states are tachyonic, marginal or massive whenever $L_0 - \bar{L}_0 $ is negative, zero or positive respectively. It is this criterion that we will utilize. \\
The reader interested in determining the on-shell states can easily select the required subset of the states we will display here. \\

\noindent We will also not discuss the additional CFT required to make the total central charge vanish, but we will assume that it is unitary on its own and hence cannot make a state more tachyonic. The only thing we will need of this internal CFT is the zero-mode weight of the additional worldsheet fermions in the Ramond sectors: we will add $3/8$ to the conformal weights whenever a Ramond sector is discussed. \\

\noindent This work started by the computation in \cite{He:2014gva} where it is suggested that the one loop entropy for type II superstrings in Rindler space actually vanishes. This then would again require the contribution from the thermal scalar state to be compensated by other states in the thermal spectrum. \\

\noindent The main goals of this paper can be summarized as:
\begin{itemize}
\item Obtain the type II string spectrum in polar coordinates.
\item Explain the vanishing of the type II partition function in terms of the polar coordinate description.
\item Provide arguments to show that the one loop entropy vanishes in type II string theory in Rindler space.
\end{itemize}

\begin{figure}[h]
\centering
\includegraphics[width=0.85\textwidth]{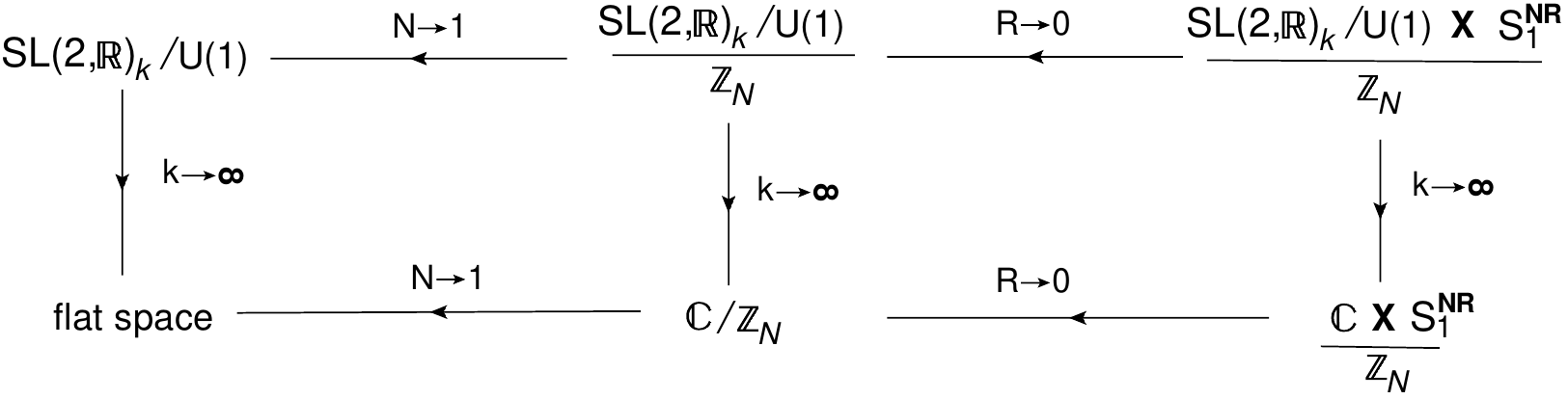}
\caption{Scheme of the different conformal models and their link to flat space. Top row (from left to right): cigar model, cigar orbifold model, Melvin regularized cigar model. Bottom row (from left to right): flat space, flat space orbifold model, Melvin model. To get from the Melvin models to the other models as $R\to0$, an additional flat dimension emerges that is not depicted here.}
\label{scheme}
\end{figure}

\noindent Throughout this paper, we will continually switch between the different ways of thinking about the polar description of flat space, either as the large $k$ limit of a cigar model, or as the $N\to1$ limit of the flat cones. In the later sections of this paper, we will also discuss the conical orbifolds of the cigar, and the Melvin model and its cigar generalization at the very end. A scheme of these models is shown in figure \ref{scheme}. \\

\noindent Recently, a new study was conducted in the cigar model \cite{Giveon:2016dxe}. The authors consider a generalization of the FZZ duality in terms of an isomorphism between what we call the thermal scalar state (the $w=1$ state) and the discrete dilaton mode (the $w=0$ state). They argue these to have different target-space interpretations and propose a duality between them, depending on the excitation level of the mode. \\
On a technical level, this isomorphism is the same as the involution symmetry we will uncover for type II superstrings. \\
Whereas they approach the problem from the CFT perspective, we approach the problem from the path integral (and character decomposition) perspective. \\

\noindent This paper is organized as follows. \\
The first half focuses on obtaining the type II (and type 0) string spectrum in polar coordinates where we aim to be very careful in recovering \emph{all} marginal states in the spectrum. We will do this by taking the large $k$ limit of the cigar model. Along the way, we will uncover several subtleties missed in the existing literature and we elucidate the superconformal structure of the worldsheet theory. \\
In section \ref{motivations} we provide some additional motivation for a closer study of the partition functions. This is to set the stage for the later discussions. Sections \ref{boso} and \ref{supe} analyze the spectral content of the known partition functions on the cigar. At the same time, this resolves a puzzle about negative $r$ quantum numbers in the partition function due to the construction of a map into the normal states. This map, when restricting to states with $r=0$, will boil down to an involution symmetry of the spectrum that we discuss in detail. In section \ref{classification} we provide a full classification of all discrete marginal states on the cigar geometry for superstrings. We focus on the lowest weight states in each of the four type II sectors and clarify the underlying $\mathcal{N}=2$ structure. Although this decomposition was done quite elaborately before in the literature \cite{Hanany:2002ev}\cite{Israel:2004ir}\cite{Eguchi:2004yi}\cite{Sugawara:2012ag}, the main difference with previous treatments is that we focus on the primary states themselves and we attempt to be as thorough as possible to classify \emph{all} marginal states. Also, our interest is in \emph{generic} states that are present for any value of $k$, whereas previous research has mainly focused on integer-valued $k$ \cite{Israel:2004ir} or rational $k$ \cite{Eguchi:2004yi}\cite{Sugawara:2012ag}. Of course, $k$ can get quantized by a suitable choice of additional compact group manifold (such as $SU(2)$), but we will not focus on this situation.\footnote{This situation arises for instance in the near-horizon limit of $k\in\mathbb{N}$ near-extremal NS5 branes; the level $k$ is in that case automatically an integer.} \\
After all this preliminary work, we finally arrive at flat space in polar coordinates in section \ref{largek} where we fully classify all marginal states in polar coordinates. The continuous sector of states is treated in section \ref{continuous} where we immediately discuss the flat limit. \\
In the second half of this work, we make contact with thermodynamics. In the black hole interpretation, the angular periodicity in polar coordinates is interpreted as the (inverse) Hawking temperature of the black hole. So firstly, section \ref{cones} analyzes how the different marginal states on the plane evolve as the temperature is varied (by introducing a conical singularity). In section \ref{directlim} we analyze the large $\tau_2$ limit of the cigar partition functions directly, without first going through the character decomposition. This is interesting in that it does not require any involved mathematical machinery but instead gives a hands-on approach to the most dominant states. Then, in section \ref{entropy}, we take a closer look at the black hole entropy one would compute from three complementary perspectives. We conclude that in a proper treatment, the one-loop entropy vanishes, in agreement with the recent analysis of \cite{He:2014gva}. Finally section \ref{UV} is a somewhat standalone section aimed at better understanding how the UV divergences in the Lorentzian QFT spectrum get compensated within string theory to obtain a finite free energy and entropy. We end with a summary in section \ref{concl} and the appendices contain much of the technical details.

\section{Motivation: where are the negatively wound states?}
\label{motivations}
In this first section, we will mention an additional motivation for analyzing the partition functions on both the cigar and the (flat) orbifold models more closely. Namely, from this perspective, it is not that clear how precisely the negatively wound companion of each thermal state is encoded in the partition function. Here we try to formulate this question more clearly and then we provide a first clue to its answer. The sections after this will then provide an in-depth analysis of the partition functions themselves.

\subsection{The field theory of the primaries on the cigar CFT}
\begin{figure}[h]
\centering
\includegraphics[width=0.45\textwidth]{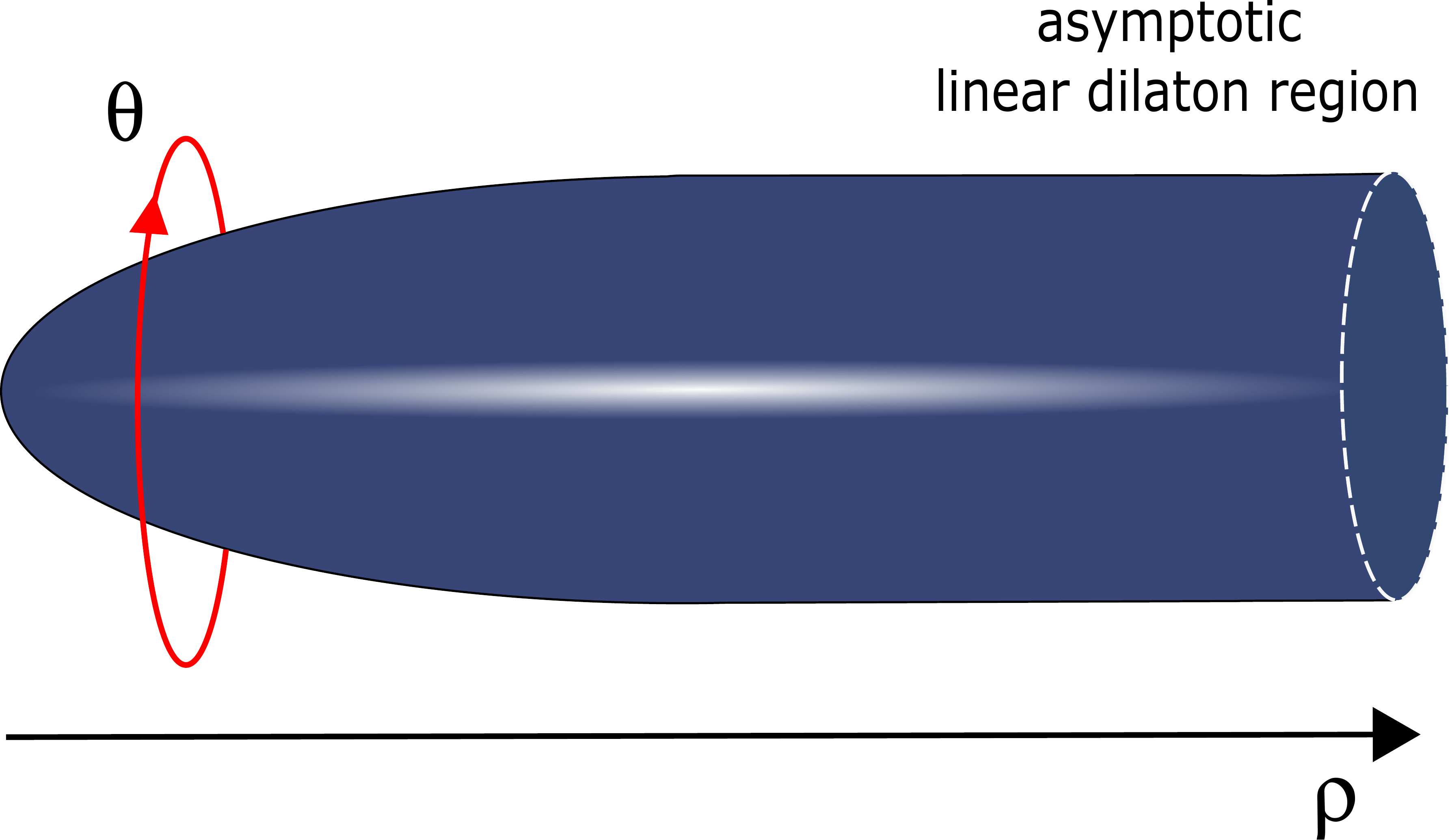}
\caption{Geometry of the $SL(2,\mathbb{R})_k/U(1)$ model.}
\label{geomstr}
\end{figure}

Let us first point out a generic feature of cigar backgrounds. Stringy states in a cigar geometry are labeled by a winding number $w$ around the cigar and a discrete momentum $n$. Not all integer values of $w$ and $n$ are present in the spectrum, due to the fact that the angular circle is not topologically stable. However, given that some state ($w$, $n$) is present, the $\mathbb{Z}_2$ inversion isometry of the angular coordinate requires that the state ($-w$, $-n$) is also present (provided no preferred direction along the angular circle is imposed by some external gauge field). \\
For instance, for type II strings on the $SL(2,\mathbb{R})_k/U(1)$ coset background, one has the (perturbatively exact in $\alpha'$) geometry:
\begin{align}
\label{metricn}
ds^2 &= d\rho^2 + \frac{\alpha' k}{\coth^2\left(\frac{\rho}{\sqrt{\alpha' k}}\right)}d\theta^2,\\
\label{diln}
\Phi &= \Phi_0-\ln\left(\cosh\left(\frac{\rho}{\sqrt{\alpha' k}}\right)\right),
\end{align}
where $\theta \sim \theta + 2\pi$ is an angular coordinate along the cigar and $\rho$ is the radial coordinate. Very far from the tip (at $\rho\to+\infty$), the geometry asymptotes to 
\begin{align}
ds^2 &\approx d\rho^2 + \alpha' k d\theta^2,\\
\Phi &\approx - \frac{\rho}{\sqrt{\alpha' k}},
\end{align}
which is a linear dilaton space (see figure \ref{geomstr}). The only two properties of this asymptotic linear dilaton regime that will be important for us later on, is that firstly the conformal weights in this sector have a mass gap and secondly the linear dilaton space reaches its own critical Hagedorn temperature as one decreases $k$ to $k=1$. \\

\noindent Upon writing the Virasoro zero modes in terms of the Casimirs of the underlying coset manifold, one finds the following differential equation for non-oscillator modes $\phi_{n,w}(\rho)$ \cite{Mertens:2013zya}:
\begin{align}
\label{QFTorig}
&-\frac{\partial_\rho\left(\sinh\left(\sqrt{2/k}\rho\right)\partial_{\rho}\phi_{n,w}(\rho)\right)}{\sinh\left(\sqrt{2/k}\rho\right)} \nonumber \\
&\quad\quad\quad+ \left[-1 + n^2\frac{1}{2k}\coth^2\left(\rho/\sqrt{2k}\right) + w^2\frac{k}{2}\tanh^2\left(\rho/\sqrt{2k}\right) \right]\phi_{n,w}(\rho)= (h+\bar{h}) \phi_{n,w}(\rho).
\end{align}
This is basically simply a geometrization of the Casimir operator $L_0 + \bar{L}_0$. Clearly, the symmetry $w\to-w$ and $n\to-n$ is respected here and both states have the same conformal weight and spatial profile. Obviously, this symmetry extends to the flat $k\to\infty$ limit as well. However, this oppositely wound companion state will turn out to be actually quite well hidden in the partition function and one of the main motivations in the first few sections will be to precisely pinpoint how it is encoded. \\

\noindent Before continuing, let us sum up some more important properties of this field theory perspective. For type II strings, one has three additional sectors in the game (NS-R, R-NS and R-R). The wavefunctions however are of precisely the same form in the flat limit of interest to us. The reason is very well known of course: in flat space, fermions obey the same wave equations as bosons. Up to a possible additive shift of the eigenvalue (coming from the zero-mode fermionic oscillators), the wavefunctions should hence be completely the same. Secondaries of the underlying Kac-Moody algebra also obey the same wave equations, as is determined by the relation $\left[L_0,J^{b}_{-p}\right] = pJ^{b}_{-p}$. \\

\noindent Taking the large $k$ (flat) limit in the above field theory eigenvalue equation (keeping $\rho$ fixed), one obtains:
\begin{align}
\label{QFT}
&-\frac{\partial_\rho\left(\rho\partial_{\rho}\phi_{n,w}(\rho)\right)}{\rho} + \left[-1 +\frac{ n^2}{\rho^2} + \frac{w^2\rho^2}{4}\right]\phi_{n,w}(\rho)= (h+\bar{h}) \phi_{n,w}(\rho),
\end{align}
An immediate feature, is that if $w\neq0$, no continuous part exists in the spectrum at all. The reason is the quadratic potential $+\rho^2$. Conversely, in the large $k$ limit, states with $w=0$ cannot be discrete. There is hence a clean separation between continuous states and discrete states in the large $k$ limit. \\

\noindent These field theoretic guidelines will be important further on, when we try to interpret some of the relevant states contained in the partition function.

\subsection{The large $\tau_2$ limit of the type II partition function on $\mathbb{C}/\mathbb{Z}_N$}
\label{larget}
Having understood the basic necessity for $w<0$ states, let us first analyze how these are encoded in a geometrically flat case: the $\mathbb{C}/\mathbb{Z}_N$ orbifold model.
Thereto, it is interesting to look at the most dominant part as $\tau_2\to+\infty$ of the type II partition function $Z(\tau) $ on the $\mathbb{C}/\mathbb{Z}_N$ orbifold.\footnote{A (bosonic) CFT partition function is of the form $\text{Tr}q^{L_0-c/24}\bar{q}^{\bar{L}_0-c/24}$ where $q=e^{2\pi i \tau}$ and $\tau= \tau_1+i\tau_2$, the torus modulus.} The main reason is to find out precisely how the negatively wound $w=-1$ state is actually realized in the partition function. This will steer us in the right direction in the remainder of this paper. \\

\begin{figure}[h]
\centering
\begin{minipage}{0.4\textwidth}
\centering
\includegraphics[width=0.8\textwidth]{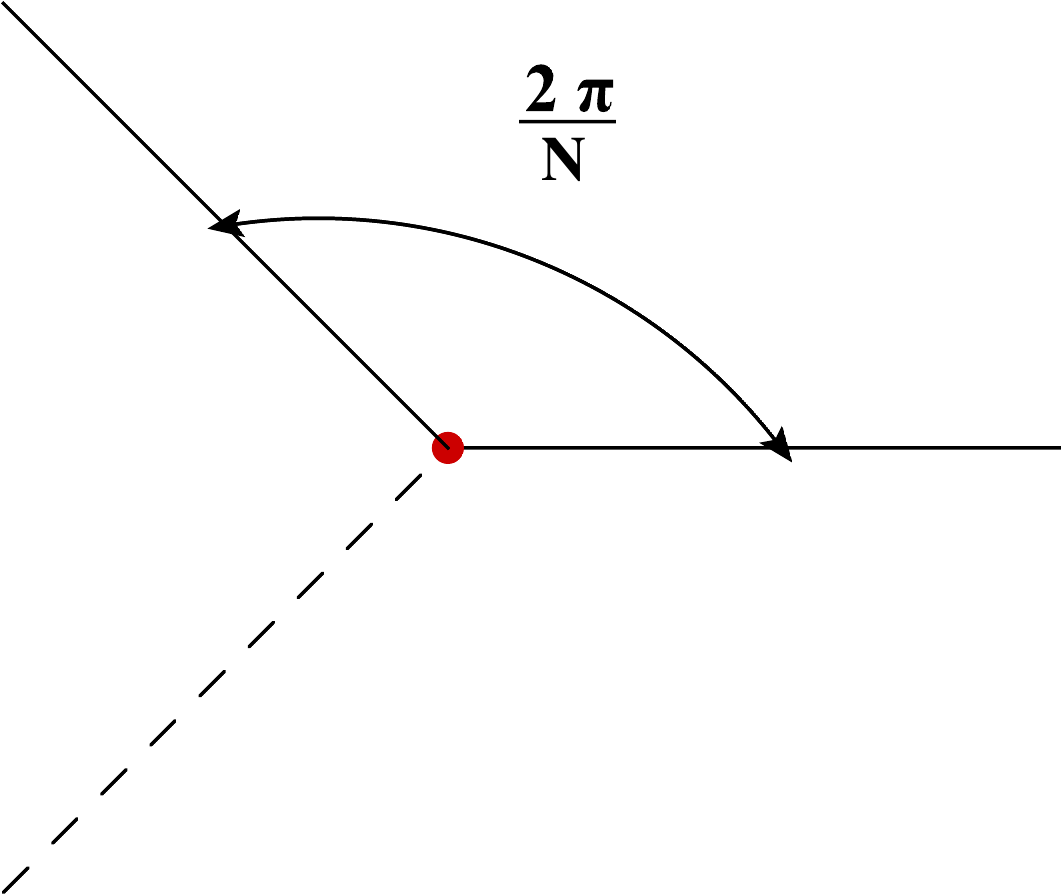}
\end{minipage}
\begin{minipage}{0.4\textwidth}
\centering
\includegraphics[width=0.8\textwidth]{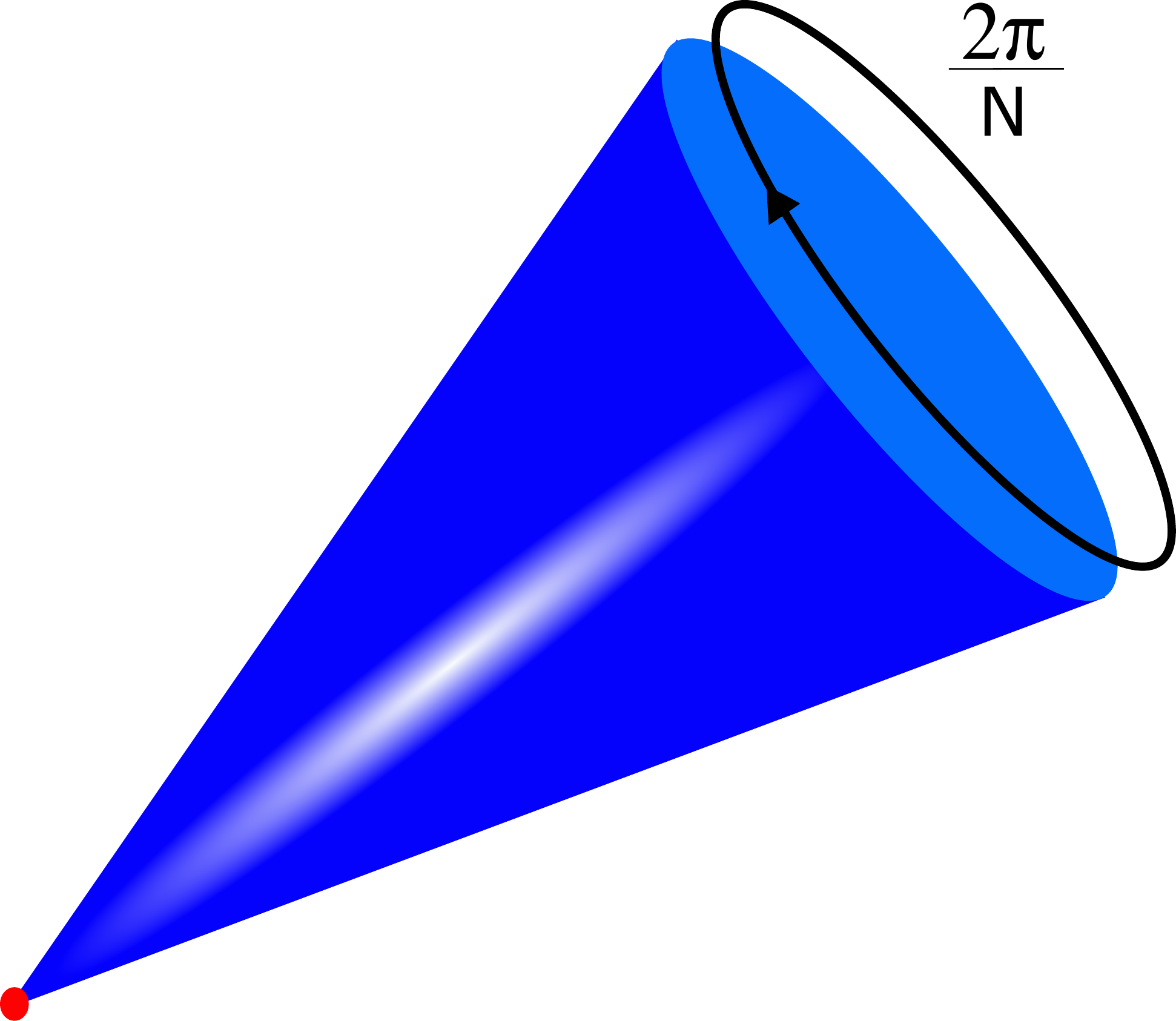}
\end{minipage}
\caption{Left figure: plane identified under a $\mathbb{Z}_N$ rotation ($N=3$ in this example). Right figure: resulting flat cone with angular periodicity $2\pi/N$.}
\label{florbif}
\end{figure}

\noindent The $\mathbb{C}/\mathbb{Z}_N$ model is obtained by identifying the 2d plane under a $\mathbb{Z}_N$ rotation subgroup of $SO(2)$. This identification in string theory amounts to an orbifolding procedure, where one projects onto the invariant sector $\frac{1}{N}\sum_{m=0}^{N-1}$ and includes a summation over twisted states $\sum_{w=0}^{N-1}$. The resulting space is a geometrically flat cone with a conical singularity at the origin (see figure \ref{florbif}). \\

\noindent Let us first very briefly state the situation for bosonic strings 	\cite{Dabholkar:1994ai}\cite{Lowe:1994ah}\cite{Mertens:2013zya}. Taking $\tau_2$ large, one finds the most dominant contribution being provided by $w=1$ and $w=N-1$. These states are to be interpreted as the thermal scalar states (positively and negatively wound) for $\beta=2\pi/N$. Taking $N\to1$, these states would transform into the thermal scalar state for the on-shell (i.e. $N=1$) black hole. This demonstrates that the negatively wound state is actually naturally encoded in the $w=0$ sector of the on-shell black hole. We will further on demonstrate in full detail that this \emph{must} be the case to have a double degeneracy of all states (as $w\to-w$ and $n\to-n$ is a symmetry of the geometry). \\

\noindent We now perform the analogous computation for type II superstrings. The partition function itself for the $\mathbb{C}/\mathbb{Z}_N$ orbifold was written down some time ago \cite{Lowe:1994ah}:\footnote{The prefactor of $1/4$ has its origin in the GSO projection. $N$ needs to be odd here to have an orbifold interpretation.}
\begin{align}
\label{flatpt}
Z(\tau) = \frac{1}{4N}\left(\frac{1}{\left|\eta\right|^2\sqrt{4\pi^2\alpha'\tau_2}}\right)^{6}\sum_{w,m=0}^{N-1}&\sum_{\alpha,\beta,\gamma,\delta}\omega'_{\alpha\beta}(w,m)\bar{\omega}'_{\gamma\delta}(w,m) \nonumber \\
&\times \frac{\vartheta\left[
\begin{array}{c}
\alpha \\
\beta  \end{array} 
\right]^3 \vartheta\left[
\begin{array}{c}
\alpha+\frac{w}{N} \\
\beta+\frac{m}{N}  \end{array} 
\right]\bar{\vartheta}\left[
\begin{array}{c}
\gamma \\
\delta  \end{array} 
\right]^3 \bar{\vartheta}\left[
\begin{array}{c}
\gamma+\frac{w}{N} \\
\delta+\frac{m}{N}  \end{array} 
\right]}{\left|\vartheta\left[
\begin{array}{c}
\frac{1}{2}+\frac{w}{N} \\
\frac{1}{2}+\frac{m}{N}  \end{array} 
\right]\eta^3\right|^2}.
\end{align}
The $\omega'$ prefactors are given as follows
\begin{alignat}{2}
\omega'_{00}(w,m) &= 1 , \quad
&&\omega'_{0\frac{1}{2}}(w,m) = e^{-\frac{\pi i w}{N}}(-1)^{w+1}, \\
\omega'_{\frac{1}{2}0}(w,m) &= (-1)^{m+1}, \quad
&&\omega'_{\frac{1}{2}\frac{1}{2}}(w,m) = \pm e^{-\frac{\pi i w}{N}}(-1)^{w+m}.
\end{alignat}
One needs the following behavior of the modular functions as $\tau_2\to\infty$:
\begin{align}
\vartheta\left[
\begin{array}{c}
0 \\
0  \end{array} 
\right] \to 1 &, \quad 
\vartheta\left[
\begin{array}{c}
1/2 \\
0  \end{array} 
\right] \to 2e^{\pi i \tau/4} , \quad 
\vartheta\left[
\begin{array}{c}
0 \\
1/2  \end{array} 
\right] \to 1 , \\
\label{nontrivtheta}
\vartheta\left[
\begin{array}{c}
\frac{w}{N} \\
b \end{array} 
\right] &\to  e^{\pi i \frac{w^2}{N^2}\tau + 2\pi i \frac{w}{N}b}\underbrace{e^{-2\pi i (\frac{w}{N}\tau+b)}e^{\pi i \tau}}_{\text{if }\frac{w}{N}>1/2}.
\end{align}
Using these limits, one finds that for $\frac{w}{N}<\frac{1}{2}$ the most dominant contribution to the theta function combination in the numerator comes
from the NS and $\widetilde{\text{NS}}$ sectors where $w=1$ is projected in.
It leads to (upon including all remaining contributions and including the antiholomorphic sector):
\begin{equation}
\label{dom1}
Z(\tau) \sim e^{2\pi \tau_2\left(1-\frac{1}{N}\right)}.
\end{equation}
If on the other hand, $\frac{w}{N}>\frac{1}{2}$, one needs to include additional factors coming from the infinite products in the theta functions. 
Note that this is not to be interpreted as an oscillator state, even though the infinite product in the theta function (\ref{nontrivtheta}) contributes non-trivially here. The dominant sector here has $w=N-1$ which is even, and is projected in because of an additional sign in the projector. 
Upon taking the modulus squared again, one finds the most dominant contribution to be
\begin{equation}
\label{dom2}
Z(\tau) \sim e^{2\pi \tau_2\left(1-\left(1-\frac{N-1}{N}\right)\right)} = e^{2\pi \tau_2\left(1-\frac{1}{N}\right)}.
\end{equation}
Both of these most dominant sectors (\ref{dom1}) and (\ref{dom2}) hence make the same contribution.\footnote{ To make the large $\tau_2$ limit in full detail, we include the remaining factors and obtain:
\begin{equation}
Z \approx \frac{V_T}{4N}\int\frac{d\tau_1d\tau_2}{2\tau_2}8Ne^{2\pi \tau_2\left(1-\frac{1}{N}\right)} \approx  V_T\int^{+\infty}\frac{d\tau_2}{\tau_2}\left(\frac{1}{\sqrt{4\pi^2\alpha'\tau_2}}\right)^{6}e^{2\pi \tau_2\left(1-\frac{1}{N}\right)},
\end{equation}
where the factor of $N$ in the first equality arises from the sum over $m$. Clearly the factor of 2 (from $w=1$ and $w=N-1$) is necessary and corresponds to both winding $\pm1$ sectors.} \\

\noindent This computation demonstrates that the negatively wound states actually are encoded in the $w=N-1$ sector with naively one oscillator activated. Taking $N\to1$, one would interpret this sector in the path integral language as the $w=0$ sector with one oscillator excited.

\section{Bosonic spectrum on the cigar}
\label{boso}
In this section, we will investigate some of the properties of the bosonic string on the $SL(2,\mathbb{R})_k/U(1)$ coset CFT in more detail. The next section will be devoted to the analogous study for superstrings. As is well-known, the spectrum on this cigar-shaped background decomposes into a discrete part and a continuous part. The continuous part is more or less standard and corresponds to the modes capable of reaching the asymptotic geometry far from the tip of the cigar. Our interest here will be the discrete sector, which is far more mysterious. \\
Since the next few sections are quite technical, the reader only interested in the results can skip these sections in a first reading and immediately continue to section \ref{largek}, where the spectrum of discrete marginal states in polar coordinates is written down.
\subsection{Partition function}
The (discrete part of the) bosonic partition function $Z_D$ on the $SL(2,\mathbb{R})_k/U(1)$ cigar was written down in \cite{Hanany:2002ev}\cite{Israel:2004ir}\cite{Eguchi:2004yi} and is proportional to
\begin{equation}
Z_D \sim \sum_{n,w,r,\bar{r}\in\mathbb{Z}}\int_{1/2}^{(k-1)/2}dj \lambda_r^j(q)\lambda_{\bar{r}}^j(\bar{q})\delta(2j+r+\bar{r}-kw)\delta_{r-\bar{r},n},
\end{equation}
where
\begin{equation}
\label{bosonicc}
\lambda_r^{j}(q) = q^{-\frac{(j-1/2)^2}{k-2}+\frac{(j+r)^2}{k}}\frac{S_{r}(q)}{\eta^2} = q^{-\frac{(j-1/2)^2}{k-2}+\frac{(j+r)^2}{k}}\frac{\sum_{s=0}^{+\infty}(-)^sq^{s(s+2r+1)/2}}{\eta^2}
\end{equation}
is the bosonic coset discrete character. If there exist values of $j$ in the spectrum for which $j=1/2$ or $j=(k-1)/2$, it is understood that an extra factor of $1/2$ is included.\footnote{This is to be traced back to the way the discrete representations are encoded in the full partition function as discrete poles in the complex plane obtained by a contour shift \cite{Hanany:2002ev}\cite{Israel:2004ir}\cite{Eguchi:2004yi}.} \\

\noindent The problem with this decomposition that we wish to highlight is the sums over $r$ and $\bar{r}$ over $\mathbb{Z}$. For the $SL(2,\mathbb{R})$ principal discrete representations, one needs $m=\pm(j+r)$ for $r\in\mathbb{N}$, not $r\in\mathbb{Z}$ for respectively the lowest and highest weight representations. Here $m$ is the $J_0^3$ eigenvalue as usual. This subtlety is related to the fact that the unitarity constraints in the model:
\begin{equation}
\frac{1}{2}\leq j \leq \frac{k-1}{2},
\end{equation}
with 
\begin{equation}
j=\frac{kw}{2}+ \frac{n}{2}-r = \frac{kw}{2}- \frac{n}{2}-\bar{r} ,\quad r,\bar{r}=0,1,2,\hdots
\end{equation} 
seem to treat the $w>0$ and the $w<0$ sectors in an asymmetric fashion if only positive values of $r$ are retained. For instance, in the large $k$ limit only $w=0$ and $w=1$ is allowed. \\

\noindent Of course, either by employing a twisted vertex operator method or using the LSZ formula of the tree-level amplitudes \cite{Aharony:2004xn}, one finds that actually one should write:
\begin{equation}
j = \frac{k\left|w\right|}{2} - \frac{\left|n\right|}{2} - l,  \quad l=0,1,2,\hdots
\end{equation}
which removes this asymmetry. \\

\noindent The technicality that remains to be resolved then, is to find how the negatively wound states are found in the path integral character decomposition as mentioned above. \\
It turns out that the sums over $r$ in the end will restore this symmetry. \\

\noindent The total sum over $r$ and $\bar{r}$ splits into 4 sectors:
\begin{equation}
\sum_{r,\bar{r}} = \sum_{r\geq 0, \bar{r}\geq 0} + \sum_{r < 0, \bar{r}\geq 0} + \sum_{r < 0, \bar{r}\geq 0} + \sum_{r < 0, \bar{r} < 0},
\end{equation}
which are analyzed in full detail in appendix \ref{detailsbosonic}.\footnote{If one takes a closer look at some of the factors present in the bosonic character (\ref{bosonicc}), it can be checked explicitly that
\begin{equation}
\frac{S_r}{\eta^2} = \sum_{n=0}^{+\infty}\alpha_n(r) q^n,
\end{equation}
for positive $\alpha_n(r)$. This solves an initial worry one might have in that $S_r$ includes a factor of $(-)^s$. Secondly, one finds that 
\begin{equation}
\frac{S_{-r}}{\eta^2} = q^{r}\sum_{n=0}^{+\infty}\alpha_n(r) q^n,
\end{equation}
suggesting strongly that the states with $r\longleftrightarrow -r$ are directly related.} \\

\begin{figure}[h]
\centering
\includegraphics[width=0.5\textwidth]{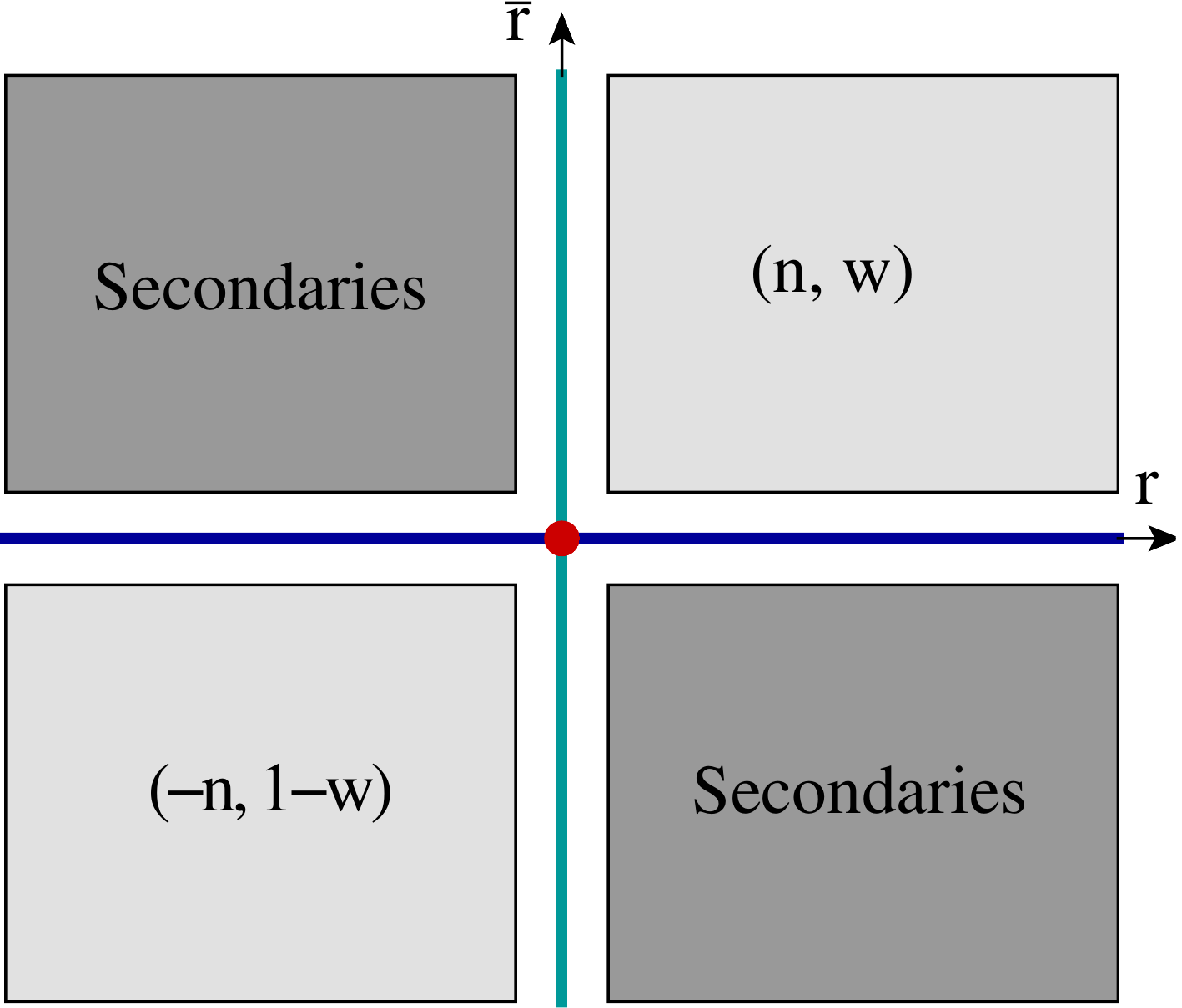}
\caption{Table of states. Equally colored regions correspond to the same conformal weights in the spectrum. The $r=\bar{r}=0$ sector is absent, and the off-diagonal blocks correspond to secondaries of the Virasoro algebra.}
\label{table}
\end{figure}

\noindent The upshot is that states with both negative $r$ and $\bar{r}$ can be shown to have the same conformal weight as the positive $r$ and $\bar{r}$ states, hence effectively doubling the spectrum. Such a state with $n$ and $w$ has the same weight as one with positive values of both $r$ and $\bar{r}$ but with momentum $-n$ and winding $1-w$. Each such state has precisely one partner state and both are allowed by the unitarity constraints. \\
States that combine positive $r$ with negative $\bar{r}$ (or vice-versa) are to be interpreted as secondaries of the conformal algebra. The full table of states in the ($r,\bar{r}$) plane is as shown in figure \ref{table}. \\
We propose to interpret these states with negative $r$ as the negatively wound counterparts of their partner states. The $w$ quantum number does not match since it is shifted by 1 unit, but $w$ is only a summation variable in the partition function; the conformal weight is in fact all that matters.\footnote{We remark that such reinterpretations are inherent to the path integral character decomposition in this model, as was done in \cite{Hanany:2002ev} in the continuous sector.}

\subsection{Concise description of the spectrum}
The spectrum is of the form (for $r,\bar{r}>0$):
\begin{align}
h &= -\frac{\left(\frac{kw}{2}+\frac{n}{2}-r\right)\left(\frac{kw}{2}+\frac{n}{2}-r-1\right)}{k-2} + \frac{\left(\frac{kw}{2}+\frac{n}{2}\right)^2}{k}, \\
\bar{h} &= -\frac{\left(\frac{kw}{2}-\frac{n}{2}-\bar{r}\right)\left(\frac{kw}{2}-\frac{n}{2}-\bar{r}-1\right)}{k-2} + \frac{\left(\frac{kw}{2}-\frac{n}{2}\right)^2}{k}.
\end{align}
The unitarity conditions in both the holomorphic and antiholomorphic part imply that we can write:
\begin{equation}
j = \frac{kw}{2} - \frac{\left|n\right|}{2} - \ell,
\end{equation}
for some $\ell = 0,1,2,\hdots$. Next, we know that for each state, we can find a corresponding state that is both normalizable and of positive norm (i.e. satisfying the unitarity constraint) and that can be interpreted as the negative winding counterpart of the given state. Can we impose this on the spectrum directly? 
Changing $w \to \left|w\right|$ in the first term of the conformal weights does the trick: for every state ($n,w$) one has a state ($-n,-w$) in the spectrum as well. Hence
\begin{equation}
j= \frac{k\left|w\right|}{2} - \frac{\left|n\right|}{2} - \ell,
\end{equation}
as was to be shown.

\section{Superstring spectrum on the cigar}
\label{supe}
Superstring theory on the $SL(2,\mathbb{R})_k/U(1)$ coset has the peculiarity to have an accidental $\mathcal{N}=(2,2)$ worldsheet supersymmetry, as it belongs to the Kazama-Suzuki class of coset theories \cite{Kazama:1988qp}. The $\mathcal{N}=2$ worldsheet superconformal algebra allows the classification of all states in terms of their conformal weight and $U(1)$ $\mathcal{R}$-charge, which we will call $Q$. We will pay attention to this $\mathcal{R}$-charge as well as we go along. This $\mathcal{N}=(2,2)$ structure will obviously carry over to the flat $k\to\infty$ limit.

\subsection{Partition functions}
The type 0B superstring partition function in the discrete sector can be written as \cite{Israel:2004ir}\cite{Eguchi:2004yi}\cite{Sugawara:2012ag}:
\begin{align}
\label{t00}
&Z_D = \frac{1}{2\left|\eta\right|^2}\sum_{a,b=0}^{1}\sum_{n,w,r,\bar{r},f,\bar{f}\in\mathbb{Z}}\int_{1/2}^{(k+1)/2}dj e^{i\pi b (f-\bar{f})} \nonumber \\
&\times\lambda_r^j(q)\lambda_{\bar{r}}^j(\bar{q})q^{\frac{k}{2(k+2)}\left(f+\frac{a}{2}+\frac{2m}{k}\right)^2}\bar{q}^{\frac{k}{2(k+2)}\left(\bar{f}+\frac{a}{2}+\frac{2\bar{m}}{k}\right)^2} \delta(2j+r+\bar{r}-kw+f+\bar{f}+a)\delta_{r-\bar{r}+f-\bar{f},n},
\end{align}
The quantum numbers $m$ and $\bar{m}$ are, just like in the bosonic case, given by
\begin{equation}
m = \frac{kw+n}{2}, \quad \bar{m} = \frac{kw-n}{2}.
\end{equation}
Compared to the bosonic partition function, the major new addition is two additional integers $f$ and $\bar{f}$, which can be roughly identified with fermionic oscillator numbers and the $\mathbb{Z}_2$-valued numbers $a$ and $b$, which implement the different spin structures on the torus. For type 0B string theory, only the diagonal spin sector is present, and hence no separate $a$ and $b$ numbers are required for holomorphic and antiholomorphic sectors. \\
The sum over $b$ realizes the projection
\begin{equation}
(-)^f = (-)^{\bar{f}},
\end{equation}
which is the correct GSO projection for (thermal and non-thermal) type 0B string theory. \\

\noindent For type 0A, one adds a factor of $(-)^{ab}$. The projection changes into
\begin{equation}
(-)^f = (-)^{\bar{f}+a},
\end{equation}
which changes the projection in the R-R sector. This is indeed the correct type 0A GSO projection. \\

\noindent The delta-constraints in (\ref{t00}) can be solved into
\begin{equation}
j = \frac{kw}{2}+ \frac{n}{2} - r - f - \frac{a}{2} =  \frac{kw}{2}+ \frac{n}{2} - \bar{r} - \bar{f} - \frac{a}{2}.
\end{equation}
Fixing $w$, $n$, $f$, $\bar{f}$ and $a$ to arbitrary values, this equation can only be solved for a finite set of $j$ in the interval $\left[\frac{1}{2}, \frac{k+1}{2}\right]$ corresponding to a choice of $r$ and $\bar{r}$. Let us call this set of possible $j$-values, $I$. The sum over $r$ and $\bar{r}$ can then be dispensed by solely taking $j\in I$. We can rewrite
\begin{align}
&Z_D = \frac{1}{2\left|\eta\right|^2}\sum_{a,b=0}^{1}\sum_{n,w,f,\bar{f}\in\mathbb{Z}}\sum_{j\in I} e^{i\pi b (f-\bar{f})} \nonumber \\
&\times\lambda_{m-f-j-a/2}^j(q)\lambda_{\bar{m}-\bar{f}-j-a/2}^j(\bar{q})q^{\frac{k}{2(k+2)}\left(f+\frac{a}{2}+\frac{2m}{k}\right)^2}\bar{q}^{\frac{k}{2(k+2)}\left(\bar{f}+\frac{a}{2}+\frac{2\bar{m}}{k}\right)^2}.
\end{align}
In each chiral sector and for a fixed $a$, $b$, $n$, $w$ and $j$, we can rewrite the contribution as\footnote{We have corrected here a typo that appeared in \cite{Israel:2004ir}. One of the consequences is that we will correctly identify the states contributing to the Witten index later on, opposed to the discussion in \cite{Israel:2004ir}.}
\begin{align}
\label{charform}
\sum_{f\in\mathbb{Z}}e^{i\pi b f} \lambda_{m-f-j-a/2}^j(q)\frac{q^{\frac{k}{2(k+2)}\left(f+\frac{a}{2}+\frac{2m}{k}\right)^2}}{\eta} &=\frac{1}{i^{ab}\eta^3}q^{-\frac{(j-1/2)^2}{k}}q^{\frac{m^2}{k}}\frac{\vartheta_{ab}(\tau)}{1+(-)^bq^{m-j+1/2}} \nonumber \\
&= \frac{1}{i^{ab}}ch_D\left(j,m-j-\frac{a}{2}\right) \left[
\begin{array}{c}
a \\
b  \end{array}
\right] (\tau),
\end{align}
as the discrete (unextended) $\mathcal{N}=2$ character, such that\footnote{The prefactor of $\frac{1}{2}$ is due to the GSO projection.}
\begin{align}
\label{partchar}
Z_D = \frac{1}{2}\sum_{a,b=0}^{1}\sum_{n,w,\in\mathbb{Z}}\sum_{j\in I}
ch_D\left(j,m-j-\frac{a}{2}\right) \left[
\begin{array}{c}
a \\
b  \end{array} 
\right] (\tau) \, \bar{ch}_D\left(j,\bar{m}-j-\frac{a}{2}\right) \left[
\begin{array}{c}
a \\
b  \end{array} 
\right] (\tau).
\end{align}

\noindent For type II superstrings, one should include some further factors into the (discrete part of the) partition function as:
\begin{align}
&Z_D = \frac{1}{4\left|\eta\right|^2}\sum_{a,b=0}^{1}\sum_{\bar{a},\bar{b}=0}^{1}\sum_{n\in\mathbb{Z} + \frac{a+\bar{a}}{2}}\sum_{w,r,\bar{r},f,\bar{f}\in\mathbb{Z}}\int_{1/2}^{(k+1)/2}dj \nonumber \\
&\times(-)^{a+b+ab}(-)^{bw}e^{i\pi bf}(-)^{\bar{a}+\bar{b}+\epsilon \bar{a}\bar{b}}(-)^{\bar{b}w}e^{i\pi\bar{f}\bar{b}}
\lambda_r^j(q)\lambda_{\bar{r}}^j(\bar{q})q^{\frac{k}{2(k+2)}\left(f+\frac{a}{2}+\frac{2m}{k}\right)^2}\bar{q}^{\frac{k}{2(k+2)}\left(\bar{f}+\frac{\bar{a}}{2}+\frac{2\bar{m}}{k}\right)^2} \nonumber \\
&\times\delta\left(2j+r+\bar{r}-kw+f+\bar{f}+\frac{a+\bar{a}}{2}\right)\delta_{r-\bar{r}+f-\bar{f}+\frac{a-\bar{a}}{2},n},
\end{align}
where $\epsilon = 1$ for type IIB and $\epsilon=0$ for type IIA. These sign factors implement the usual GSO projection conditions, combined with what remains from the thermal sign factors \cite{Atick:1988si}: the $(-)^{ma}(-)^{m\bar{a}}$ factors, where $m$ is Poisson dual to $n$, are already absorbed in going from $m$ to $n$ by imposing $n$ to be a half-integer when $a+\bar{a}=1 \text{ mod } 2$ (the spacetime fermionic sectors) \cite{Sugawara:2012ag}. The GSO projection itself is given by $(-)^{f+w+a}=(-)^{\bar{f}+w+\epsilon \bar{a}}=-1$ where $\epsilon=1$ in type IIB and $\epsilon=0$ in type IIA.\\

\noindent The adjustments needed to get to the analogue of equation (\ref{partchar}) are trivial and lead to:\footnote{It is implicitly understood that the $a=b=1$ and $\bar{a}=\bar{b}=1$ characters contains an additional $-i$ factor, compared to their definition above. In more technical detail, one uses $-i \vartheta_{11} = i \vartheta_1$ in equation (\ref{charform}) instead. The prefactor of $\frac{1}{4}$ is due to the GSO projection.}
\begin{align}
&Z_D = \frac{1}{4}\sum_{a,b=0}^{1}\sum_{\bar{a},\bar{b}=0}^{1}\sum_{n\in\mathbb{Z} + \frac{a+\bar{a}}{2}}\sum_{w\in\mathbb{Z}}\sum_{j\in I} (-)^{a+b+ab}(-)^{bw}(-)^{\bar{a}+\bar{b}+\epsilon \bar{a}\bar{b}}(-)^{\bar{b}w} \nonumber \\
&\times ch_D\left(j,m-j-\frac{a}{2}\right) \left[
\begin{array}{c}
a \\
b  \end{array} 
\right] (\tau) \, \bar{ch}_D\left(j,\bar{m}-j-\frac{\bar{a}}{2}\right) \left[
\begin{array}{c}
\bar{a} \\
\bar{b}  \end{array} 
\right] (\tau). 
\end{align}

\noindent From the above expressions, it is clear that in any of these theories, $j$ is restricted to the interval
\begin{equation}
\frac{1}{2} \leq j \leq\frac{k+1}{2}.
\end{equation}
In a vertex operator approach, these constraints are explained due to both normalizability $\left (j>\frac{1}{2}\right)$ and unitarity $\left(j < \frac{k+1}{2}\right)$.\footnote{We note that this is a manner in which string theory differs from field theory: some states are perfectly normalizable primaries (and hence would be allowed in a field theory context), but conformal invariance allows one to construct an entire family of secondary states on top of this, whose unitarity imposes additional constraints which are invisible when analyzing the primary on its own.} States that saturate either of these bounds contribute only with half ``weight" to the partition sum. \\

\noindent The decomposition of these partition functions into the $\mathcal{N}=2$ characters as highlighted above has been the subject of intense study during the past decade \cite{Israel:2004ir}\cite{Eguchi:2004yi}\cite{Eguchi:2010cb}\cite{Sugawara:2011vg}\cite{Troost:2010ud}\cite{Sugawara:2012ag}\cite{Eguchi:2014vaa}. Our focus in what follows, on the other hand, is on the lowest mass states themselves.

\subsection{Description of the spectrum}
For any type of these string theories (0A, 0B, IIA or IIB), one finds for the conformal weights, analogously as for the bosonic string:
\begin{align}
h &= -\frac{j(j-1)}{k} + \frac{m^2}{k} + \frac{\left(f+\frac{a}{2}\right)^2}{2}, \quad r\geq0, \\
h &= -\frac{j(j-1)}{k} + \frac{m^2}{k} + \frac{\left(f+\frac{a}{2}\right)^2}{2} - r, \quad r<0,
\end{align}
where $j= m - r - f - \frac{a}{2} = \bar{m} - \bar{r} - \bar{f} - \frac{\bar{a}}{2}$.\footnote{For type 0 theories, one identifies $a=\bar{a}$ and $b=\bar{b}$. A vertex operator analysis of type 0 strings in this geometry was given in \cite{Giveon:2003wn}.} This leads to an expression for $n$ in terms of the oscillator numbers:
\begin{equation}
\label{momconstraint}
n = l - \bar{l} + f - \bar{f} + \frac{a-\bar{a}}{2}.
\end{equation}
The quantum number $a$ can be zero or one, for NS or R states respectively. \\
Just like before, the $r<0$ sector can be mapped onto the $r>0$ sector upon reinterpreting the quantum numbers. In this case, one can write:
\begin{align}
h &= -\frac{\left(\frac{kw}{2}+\frac{n}{2}-r-\frac{a}{2}-f\right)\left(\frac{kw}{2}+\frac{n}{2}-r-\frac{a}{2}-f-1\right)}{k}+ \frac{\left(\frac{kw}{2}+\frac{n}{2}\right)^2}{k} +\frac{\left(f+\frac{a}{2}\right)^2}{2}-r \nonumber \\
&= -\frac{\left(\frac{k(1-w)}{2}-\frac{n}{2}+r+\frac{a}{2}+f+1\right)\left(\frac{k(1-w)}{2}-\frac{n}{2}+r+\frac{a}{2}+f+1-1\right)}{k}+ \frac{\left(\frac{k(1-w)}{2}-\frac{n}{2}\right)^2}{k} \nonumber \\ 
&\quad\quad+\frac{\left(-f-1-\frac{a}{2}\right)^2}{2},
\end{align}
so one can execute the map
\begin{equation}
\label{invvsy}
r<0,\, n,\, w,\, f \quad \to \quad -r>0,\,  -n,\, 1-w,\, -1-f-a
\end{equation}
to get rid of negative values of $r$. Just like for the bosonic string, for each state present (i.e. whose $j$ satisfies the unitarity constraints), the image state of this map also satisfies the unitarity constraints, where $j \to \frac{k}{2}+1-j$ in the superstring case. \\

\noindent With this redefinition of the quantum numbers, the $\mathcal{R}$-charge gets transformed as
\begin{equation}
Q = \frac{2m}{k} +f +\frac{a}{2} \to  -Q.
\end{equation}
Hence these states have opposite $\mathcal{R}$-charges: the negatively wound cousin of each state is the $\mathcal{R}$-conjugate. \\

\noindent Of course, all of this was just for one chiral sector, and one should combine the results of the holomorphic and anti-holomorphic sectors. We refrain from providing a detailed analysis here, as the results are actually the same as for the bosonic string discussed in section \ref{boso} and we do not wish to make this section even more technical than it already is. \\

\noindent While in the general case, this transformation (\ref{invvsy}) simply shows how to deal with the unphysical $r<0$ sectors, there does exist a physical symmetry of the spectrum that is unveiled by the above procedure. If $r=0$, the transformation becomes
\begin{equation}
\label{involution}
\boxed{
n,\, w,\, f \quad \to \quad -n,\, 1-w,\, -1-f-a}
\end{equation}
which hence is an involution symmetry of the spectrum. We will demonstrate further on that the thermal scalar state and a discrete dilaton state form a pair under this involution. This is also important physically, as we will also show that such pairs are actually to be interpreted as opposite winding cousins of the same underlying state. For bosonic strings, this involution symmetry is absent because the $r=\bar{r}=0$ sector is absent as demonstrated in appendix \ref{detailsbosonic}. \\

\noindent From here on out, we will assume $r,\bar{r}\geq 0$ (using the above mapping to understand the $r,\bar{r}<0$ states) and we will call $r=l\in\mathbb{N}$ and $\bar{r}=\bar{l}\in\mathbb{N}$.

\section{Classification of all lowest-lying states in type II theory}
\label{classification}
Our main interest is the complete classification of all states in the type II theory that are marginal (massless) or tachyonic (which is of course only possible in the NS-NS sector). As is well-known, NS states in a $\mathcal{N}=2$ theory need to satisfy $h\geq \left|Q\right|/2$. States saturating this inequality are called chiral ($c$) primaries for $h=Q/2$ and anti-chiral ($a$)  primaries for $h=-Q/2$. These play a special role in the theory; in particular, when combining left- and right-movers, the theory contains 4 rings of states: the ($c$, $c$), ($c$, $a$), ($a$, $c$) and ($a$, $a$) rings. \\
A priori, the marginality of a state and its chiral property are unrelated, but in practice they will be closely related as we will demonstrate in this and future sections. \\
To find the marginal states in an $\mathcal{N}=2$ worldsheet theory, several options are available. \\
Firstly, one could classify all Ramond ground states which, upon spectral flow, give us the complete set of NS chiral primaries. Alternatively, one could try to construct the full family of NS ground states directly. In any case, it is instructive to find all R ground states first and clarify the $\mathcal{N}=2$ spectral flow transformation in this specific model. \\
In this section, we will in fact find only the marginal chiral primaries in the NS-NS sector, not necessarily all marginal states. We will come back to this later on. 

\subsection{All Ramond ground states}
Let us first catalog all Ramond ground states in the left-moving (holomorphic) sector. Constraints from level-matching left- and right-movers will be discussed below. \\
The conformal weight of an R-ground state is given by
\begin{equation}
h = -\frac{\left(\frac{kw}{2}+\frac{n}{2}-l-f-\frac{1}{2}\right)\left(\frac{kw}{2}+\frac{n}{2}-l-f-\frac{3}{2}\right)}{k}+ \frac{m^2}{k} +\frac{\left(f+\frac{1}{2}\right)^2}{2} = \frac{c}{24} = \frac{1}{8} + \frac{1}{4k}.
\end{equation}
We will look for states that exist for generic $k$ (say, irrational). Then the above condition requires
\begin{align}
\label{firsteqR}
wl + (f+1)\left(w+\frac{f}{2}\right) &= 0, \\
\label{secondeqR}
nl+n(f+1)-2(l+f)-(l+f)^2&=1.
\end{align}
A solution of these equations is $f=-1$, $l=0$ and $w$ and $n$ generic. We prove in appendix \ref{Rdetails} that this is the only solution of this set of equations. Hence the only Ramond ground states that are present in the spectrum are characterized by
\begin{equation}
f=-1 , \quad l=0
\end{equation}
and arbitrary $w$ and $n$.

\subsection{Ramond-Ramond ground states}
Let us next combine left- and right-movers. We have $f=\bar{f}=-1$ and $l=\bar{l}=0$. Also,
\begin{equation}
j = \frac{kw}{2} + \frac{n}{2} +\frac{1}{2} = \frac{kw}{2} - \frac{n}{2} +\frac{1}{2},
\end{equation}
leading to $n=0$. To satisfy the required constraint on $j$, only $w=0$ (with $j=1/2$) and $w=1$ (with $j=k/2+1/2$) are present in the spectrum. Such states hence have $h = \bar{h} = c/24$, and are massive (irrelevant) for finite $k$, upon adding $3/8$ to the weight in each R-sector to account for the additional zero-modes of the worldsheet fermions as discussed in the Introduction. \\
As a check on this, it is known that the Witten index is only sensitive to the Ramond-Ramond ground states. And indeed, the Witten index has been computed and equals 1 \cite{Israel:2004ir}\cite{Sugawara:2012ag}, precisely the sum of both of these states, since both are weighted with a factor $1/2$ for saturating the unitarity bounds. This is done in more detail in appendix \ref{WittenI}.

\subsection{Spectral flow}
On either holomorphic or antiholomorphic sector, spectral flow (in the $\mathcal{N}=2$ algebra) by $\theta$ units acts as \cite{Lerche:1989uy}\footnote{We denoted $L_0 = h$, the conformal weight and $J_0=Q$, the $\mathcal{R}$-charge.}
\begin{align}
L_n' &= L_n + \theta J_n + \frac{c}{6}\theta^2 \delta_{n,0}, \\
J_n' &= J_n + \frac{c}{3}\theta \delta_{n,0}, \\
G_{r}^{+'} &= G_{r+\theta}^{+}, \quad G_{r}^{-'} = G_{r-\theta}^{-}.
\end{align}
One readily shows that, in terms of the quantum numbers, spectral flow acts as
\begin{align}
n_f &\to n_f + \theta, \\
m &\to m + \theta,
\end{align}
where $n_f = f + a/2$. An analogous expression with $\tilde{\theta}$ holds for the right-movers. Note that $j$ is kept invariant under this transformation. Upon combining left- and right-movers, this can only be compatible with 
\begin{align}
m = \frac{kw}{2} + \frac{n}{2}, \quad  \bar{m} = \frac{kw}{2} - \frac{n}{2},
\end{align}
if we flow in \emph{opposite} directions for left- and right-movers, but with the same magnitude. Only in this case, can we interpret this as a shift of the physical discrete momentum (provided $\theta\in\mathbb{Z}/2$).\footnote{$\theta$ can be a quarter integer if we flow from spacetime fermions (with half-integer momentum) to spacetime bosons or vice versa, but this will not be considered here.} Note that this again assumes that $k$ is generic and in general an integer shift of $kw$ cannot be interpreted as a shift in $w$. \\

\noindent All R-ground states flow to NS (anti)chiral primaries when $\theta=\pm 1/2$, but requiring that they be marginal puts a constraint on the $\mathcal{R}$-charge of the initial R-ground state:
\begin{equation}
Q = \frac{2m}{k} + f +\frac{1}{2} = 1 \pm c/6.
\end{equation}
With $c=3+6/k$, this condition entails
\begin{alignat}{3}
n&=1, \quad &&w=2, \quad &&(+\,\text{-sign}), \\
n&=-1, \quad &&w = 1, \quad &&(-\,\text{-sign}).
\end{alignat}
The first candidate state would have $j= k+\frac{1}{2}+1-\frac{1}{2} = k+1$ which is obviously outside the unitarity bound in the first place. The second candidate state is important. \\
Our interest is first in finding all marginal ($c$, $c$) states in the theory. Let us start w.l.g. with a R-R state whose left-moving part is in the ground state. The right-moving part is not (in fact cannot) since in spectral flowing, we should choose the other direction to flow. So the R-R state we start with is characterized by\footnote{$\bar{l}$ and $\bar{f}$ can be found by considering the value of $j$ and the initial $\bar{Q}$ required.}
\begin{align}
w=1, \quad n=-1, \quad l=\bar{l}=0, \quad f=-1, \quad \bar{f}=0,
\end{align}
with weights and charges
\begin{alignat}{2}
h &= c/24, \quad &&Q = 1/2-1/k, \\
\bar{h} &= c/24 +1, \quad &&\bar{Q} = 3/2+1/k.
\end{alignat}
Note that $h-\bar{h} = 1 \in \mathbb{Z}$ and $j=k/2$. This state is hence present in the R-R spectrum of the theory. Next, we spectral flow with $\theta=+1/2$ in the holomorphic sector and with $\theta=-1/2$ in the antiholomorphic sector. One obtains
\begin{align}
h = 1/2&, \quad Q = +1, \\
\bar{h} = 1/2&, \quad \bar{Q} = +1.
\end{align}
The resulting NS-NS state is necessarily a chiral primary and has $w=1$, $n=0$, $f=\bar{f}=0$ and $j=k/2$. This is the thermal scalar zero-mode. \\
This is hence the only marginal ($c$, $c$) state in the NS-NS sector of the theory. \\

\noindent Analogously, one can study the anti-chiral marginal primaries. In this case, the original R-ground state should have
\begin{equation}
Q = \frac{2m}{k} + f +\frac{1}{2} = -1 \pm c/6.
\end{equation}
With $c=3+6/k$ and utilizing the unitarity bounds, this condition gives the initial state (whose left-moving part is chosen to be in the R-ground state):
\begin{align}
w=0, \quad n=+1, \quad l=\bar{l}=0, \quad f=-1, \quad \bar{f}=-2,
\end{align}
with weights and charges
\begin{alignat}{2}
h &= c/24, \quad &&Q = 1/k-1/2, \\
\bar{h} &= c/24 +1, \quad &&\bar{Q} = -1/k-3/2.
\end{alignat}
We spectral flow with $\theta=-1/2$ in the holomorphic sector and with $\theta=+1/2$ in the antiholomorphic sector and obtain a state with
\begin{align}
h &= 1/2, \quad Q = -1, \\
\bar{h} &= 1/2, \quad \bar{Q} = -1.
\end{align}
The resulting NS-NS state is necessarily an ($a$, $a$) primary and has $w=0$, $n=0$, $f=\bar{f}=-1$ and $j=1$, which was interpreted in \cite{Sugawara:2012ag}\cite{Giveon:2013ica} as the dilaton zero-mode, corresponding to an isometry of the metric. It is the only marginal ($a$, $a$) state in the spectrum. \\
However, we argued in several ways already, that this state should actually be interpreted as the negatively wound thermal scalar, with identically the same wavefunctions as a consequence. \\

\noindent Note that both of these (anti)chiral primary states indeed have $h,\bar{h} \leq c/6$, as required by (anti)chiral primaries. This inequality gets saturated at $k\to\infty$. \\

\noindent We know from the general construction of spectral flow that these states must exhaust the available (marginal) ($c$, $c$) and ($a$, $a$) primaries of the theory. It can be shown analogously that no marginal ($c$, $a$) or ($a$, $c$) states can be constructed in this way.

\subsection{All marginal chiral primaries}
As an alternative, one can construct directly all marginal NS-NS chiral primaries and find agreement with the spectral flow analysis presented above.
The $\mathcal{R}$-charge equals
\begin{equation}
Q = \frac{2m}{k} + f +\frac{a}{2}.
\end{equation}
In the NS-sector: $a=0$. The conformal weight equals
\begin{equation}
h = -\frac{\left(\frac{kw}{2}+\frac{n}{2}-r-f\right)\left(\frac{kw}{2}+\frac{n}{2}-r-f-1\right)}{k}+ \frac{\left(\frac{kw}{2}+\frac{n}{2}\right)^2}{k} +\frac{f^2}{2}.
\end{equation}
We can try to find all states which have $Q= \pm1$ and $h=1/2$. These are the marginal (anti)chiral primaries of the theory. \\
The computation is again relatively straightforward and is presented in appendix \ref{chdetails}. In the end, one finds agreement with the states constructed above.

\subsection{Spacetime Fermions: the NS-R and R-NS sectors}
The discrete R-NS states can be found as
\begin{align}
h &= \frac{-j(j-1)}{k} + \frac{m^2}{k} + \frac{\left(f+1/2\right)^2}{2},\\
\bar{h} &= \frac{-j(j-1)}{k} + \frac{\bar{m}^2}{k} + \frac{\bar{f}^2}{2}.
\end{align}
The $j$ values can be found in two ways as 
\begin{equation}
j = \frac{kw+n}{2} - l - f - \frac{1}{2} = \frac{kw-n}{2} - \bar{f} -\bar{l},
\end{equation}
leading to $n =  l -\bar{l} + f - \bar{f} + \frac{1}{2}$, demonstrating clearly that half-integer momenta are needed and that this sector is a spacetime antiperiodic fermion. This is actually a very important conclusion: in field theory it is known that when considering a Dirac fermion in polar coordinates, antiperiodicity of the fermion field around the tip of the cone emerges naturally \cite{Fursaev:1996uz}\cite{Kabat:1995eq}\cite{Dowker:1987pk}. Here spacetime fermions in the discrete sectors can only be present in the spectrum if and only if they are antiperiodic around the tip of the cone. This is a direct demonstration of the thermality of black hole horizons within string theory. \\

\noindent Half-integer values of $n$ are indeed realized for type II strings. Level-matching requires\footnote{As mentioned in the Introduction, we add $3/8$ for each R-sector present to account for the zero-modes of the worldsheet fermions in the additional CFT that we do not specify.}
\begin{equation}
h-\bar{h} = nw + \frac{\left(f+1/2\right)^2}{2} + 3/8 - \frac{\bar{f}^2}{2} \in \mathbb{Z},
\end{equation}
and this is not a trivial condition. The lowest-lying state can be found by level-matching the left-moving Ramond-vacuum: $f=-1$, $l=0$. \\

\noindent On the right-moving side, we choose $\bar{f}=-1$. The level-matching condition then becomes
\begin{equation}
h-\bar{h} = nw \in \mathbb{Z},
\end{equation}
requiring us to choose $w$ even since $n$ is \emph{half}-integer valued. We also need $n = -\bar{l} + 1/2$. Choosing $\bar{l} = 0$, one obtains $n=1/2$. With these values, we have $j=\frac{kw}{2} + \frac{1}{4} + 1 - \frac{1}{2} = \frac{kw}{2} + \frac{3}{4}$, requiring us to choose $w=0$ to satisfy the unitarity constraints. The weight becomes
\begin{align}
h =  -\frac{\left(\frac{3}{4}\right)\left(-\frac{1}{4}\right)}{k} + \frac{1}{16k} + \frac{1}{2} = \frac{1}{4k} + \frac{1}{2},
\end{align}
which is indeed $c/24+3/8$. For completeness, the $\mathcal{R}$-charges of this state are
\begin{align}
Q &= \frac{1}{2k}-\frac{1}{2}, \\
\bar{Q} &= -\frac{1}{2k} -1. 
\end{align}

\noindent Alternatively, on the right-moving side, we can choose $\bar{f}=0$. An analogous computation shows that this state has $w=1$, $n=-1/2$, $\bar{l}=0$ and $j=\frac{kw}{2} + \frac{1}{4}$. Just as above, this state has spin zero and the same conformal weight. The $\mathcal{R}$-charges are
\begin{align}
Q &= -\frac{1}{2k}+\frac{1}{2}, \\
\bar{Q} &= \frac{1}{2k} +1. 
\end{align}

\noindent The NS-R sector of fermions can be treated analogously and leads to states with the same conformal weights.  \\\
The conclusion is that the lowest-lying states in these sectors are massive for any finite $k$.

\section{The large $k$ limit}
\label{largek}
\begin{figure}[h]
\centering
\includegraphics[width=0.55\textwidth]{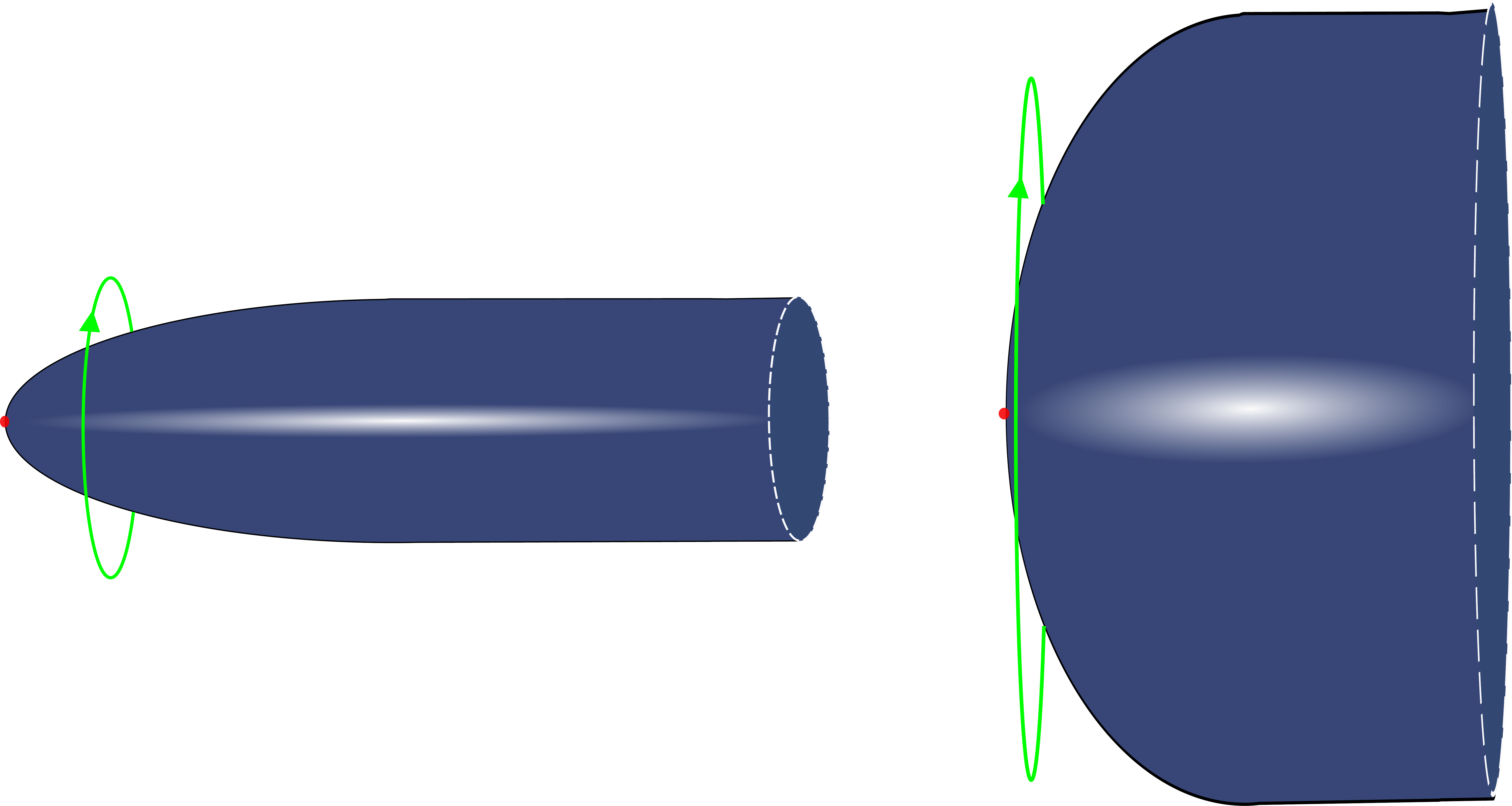}
\caption{The cigar geometry with its winding along the angular coordinate. The asymptotic geometry specifies the winding number number and the winding-dependent thermal GSO projection. As $k$ increases, the cigar flattens out and the tip approaches flat space, but described in polar coordinates. This procedure splits the thermal cigar modes in several classes, depending on their radial spread. Modes that are always sensitive to the asymptotic geometry should be scaled out in the large $k$ limit, if one is interested in the theory at the tip. The winding number transforms into winding around the polar origin.}
\label{flatlmi}
\end{figure}

Our main interest lies in the flat limit ($k\to\infty$) where the cigar flattens out and approaches flat space in polar coordinates: the flat limit of the cigar CFT yields flat space near the tip of the cigar. Moreover, the dilaton field of the cigar model becomes constant.\footnote{One takes $k\to+\infty$ and $\rho$ fixed in equations (\ref{metricn}) and (\ref{diln}) to obtain:
\begin{align}
ds^2 &= d\rho^2 + \rho^2 d\theta^2,\\
\Phi &= \Phi_0.
\end{align}
} For more details on this limit, one can consult e.g. \cite{Giveon:2013ica}\cite{Giveon:2014hfa}\cite{Mertens:2013zya}. \\
A schematic of this flattening of the tip of the cigar is shown in figure \ref{flatlmi}. \\

\noindent Of course, some care must be taken to deal with the asymptotic modes, as for any finite $k$ there is still a region that is asymptotically cylinder-shaped. This issue was inspected in \cite{Mertens:2013zya}, where the continuous sector of modes indeed reflects this property. We refer the reader to that work (and in particular appendix F therein) for details on how the large $k$ limit is taken for the wavefunctions of the modes in the NS-NS sector. \\

\noindent The states that are relevant for our purposes are those that exist for generic $k$, since else they might be an artifact of the large $k$ limit and would not be present near generic black holes (as of course a real black hole is never truly Rindler). In this section, we will find all marginal states in the flat limit and we will demonstrate that they are all generic states. The large $k$ limit also yields a heavy degeneracy of marginal states, required to agree with the vanishing spacetime supersymmetric type II partition function. \\
An immediate simplification in the large $k$ limit is that the winding is restricted to $w=0$ and $w=1$ due to the unitarity constraints. Again we defer the detailed computations to appendix \ref{polarstates} to streamline the story here.

\subsection{All marginal states in flat space in polar coordinates}
Some simple computations show that all the discrete marginal states are given as
\begin{align}
\begin{array}{|c|c|c|c|c|c|c|c|c|c|c|c|c|c|c|c|c|}
\hline
$Sector$ & w & n & l & \bar{l} & f & \bar{f} & Q & \bar{Q} & j & $Comment$ & (-)^{f+w} & (-)^{\bar{f}+w} & $IIA$ & $IIB$ & $0A$ & $0B$\\
\hline
${\color{blue}NS-NS}$ & 0 & 0 & 0 & 0 & -1 & -1 & -1 & -1 & 1 & (a,a) & -1 & -1 & \checkmark & \checkmark & \checkmark& \checkmark\\
${\color{blue}NS-NS}$ & 1 & 0 & 0 & 0 & 0 & 0 & 1 & 1 & \frac{k}{2} & (c,c) & -1 & -1 & \checkmark & \checkmark & \checkmark& \checkmark\\
\hline
$NS-NS$  & 0 & 1 & 1 & 0 & -1 & -1 & -1 & -1 & \frac{1}{2} & (a,a), \,$hw$ & -1 & -1 & \checkmark & \checkmark & \checkmark& \checkmark\\
$NS-NS$  & 0 & -1 & 0 & 1 & -1 & -1 & -1 & -1 & \frac{1}{2} & (a,a), \,$hw$ & -1 & -1 & \checkmark & \checkmark & \checkmark& \checkmark\\
$NS-NS$  & 1 & 1 & 1 & 0 & -1 & -1 & 0 & 0 & \frac{k}{2}+\frac{1}{2} & $hw,\,s=1$ & +1 & +1 & - & - & \checkmark& \checkmark\\
$NS-NS$  & 1 & -1 & 0 & 1 & -1 & -1 & 0 & 0 & \frac{k}{2}+\frac{1}{2} & $hw,\,s=1$ & +1 & +1 & - & - & \checkmark & \checkmark\\
\hline
\hline
$R-R$ & 0 & 0 & 0 & 0 & -1 & -1 & -\frac{1}{2} & -\frac{1}{2} & \frac{1}{2} & $hw$& -1& -1 & - &  -& -& \checkmark\\
$R-R$ & 1 & 0 & 0 & 0 & -1 & -1 & \frac{1}{2} & \frac{1}{2} & \frac{k}{2}+\frac{1}{2} & $hw$& +1& +1 & - &  \checkmark& - &\checkmark\\
\hline
\hline
$R-NS$ & 0 & \frac{1}{2} & 0 & 0 & -1 & -1 & -\frac{1}{2} & -1 & \frac{3}{4}&& -1& -1 &  -&  -& -& -\\
$R-NS$ & 1 & -\frac{1}{2} & 0 & 0 & -1 & 0 & \frac{1}{2} & 1 & \frac{k}{2}+\frac{1}{4} && +1& -1 &  \checkmark&  \checkmark& -& -\\
\hline
$NS-R$ & 0 & -\frac{1}{2} & 0 & 0 & -1 & -1 & -1 & -\frac{1}{2} & \frac{3}{4}&& -1& -1 & \checkmark & -& -& -\\
$NS-R$ & 1 & \frac{1}{2} & 0 & 0 & 0 & -1 & 1 & \frac{1}{2} & \frac{k}{2}+\frac{1}{4} & & -1& +1&  -&  \checkmark& -& -\\
\hline
\end{array} \nonumber
\end{align}
The denotation ``hw" implies that these states have a value of $j$ that saturates the unitarity bounds, and hence their contribution to the partition function is only with half weight. Two of the NS-NS states are of non-zero spin: $s = \left|h-\bar{h}\right| = 1$. We have also denoted for the different superstring theories, whether each state is present in the spectrum ($\checkmark$) or not ($-$).\footnote{We note that the R-R, NS-R and R-NS sectors each also include multiple additional marginal states obtained by turning on the zero-mode worldsheet fermions associated to the additional internal CFT (that we chose not to specify here). This is identically the same as in flat space, and these considerations will not be discussed here.}
Some remarks are in order here.
\begin{itemize}
\item 
All of these states exist for generic $k$, and are hence not an artifact of the large $k$ limit. 
\item 
The first two states are the only states that remain marginal for any finite value of $k$. These two states hence are the only generically marginal states on the cigar.\footnote{We have not explicitly proven this statement in the above, as we only constructed the marginal \emph{(anti)chiral} states, which leaves open the possibility of marginal but non-chiral states. However, if these states are to be marginal for generic $k$, then they should be marginal for $k\to+\infty$ as well, and no such states are found here. Hence no additional generically marginal states exist.}
\item 
Each state here has a partner state obtained by using the symmetry (\ref{invvsy}) described previously. For the R-R, R-NS and NS-R states, both of the states displayed here form a doublet under the involution symmetry (\ref{involution}). In the NS-NS sector, this applies to the first two states as well. The four other NS-NS states get mapped into states with negative $r$ (or $\bar{r}$) which should be interpreted as the negatively wound partner states.\footnote{This resolves a possible worry in that these four displayed states do not contain any states with $Q=\bar{Q}=+1$, which is however required since the $\mathcal{R}$-conjugate states should be present as well.} \\
In particular, the 2 additional ($a$, $a$) states in the NS-NS sector have a partner state with opposite $\mathcal{R}$-charge; there are hence 2 additional half-weight ($c$, $c$) states as well.
\item 
In the infinite $k$ limit, it is straightforward to also find \emph{all} (anti)chiral primaries, marginal or not. We will not show the calculations themselves. The gist of this analysis is that there are no additional (anti)chiral primaries at all beyond those displayed here. In particular, no ($c$, $a$) nor ($a$, $c$) states exist at all. \\
Hence all (anti)chiral primaries are marginal. This means that on the finite $k$ cigar, there are no generically (i.e. for generic $k$) (anti)chiral primaries beyond the thermal scalar and discrete dilaton mode; the 2 additional (spurious) $w=0$ states are ($a$, $a$) states only in the infinite $k$ limit, as are their ($c$, $c$) partners. The scheme of the entire chiral ring in this theory is hence as follows:
\begin{align}
\begin{array}{|c|c|}
\hline
\text{finite } k & k\to \infty \\
\hline
\text{(}c\text{, }c\text{)} \text{ thermal scalar} & \text{(}c\text{, }c\text{)} \text{ thermal scalar} \\
\text{(}a\text{, }a\text{)} \text{ discrete dilaton} & \text{(}a\text{, }a\text{)} \text{ discrete dilaton} \\
- & \text{(}a\text{, }a\text{)}\,\, 2 \, w=0 \text{ modes} \\
- & \text{(}c\text{, }c\text{)}\,\, 2 \, \text{partner modes} \\
\hline
\end{array} \nonumber
\end{align}
where all of these states are marginal as well.
\item
The first four NS-NS states satisfy $h=\bar{h}=c/6=1/2$ (at infinite $k$ only), implying the ($c$, $c$) thermal scalar state (and the 2 spurious half-weight states) satisfy $G^{+}_{-3/2}\left|\psi\right\rangle = \bar{G}^{+}_{-3/2}\left|\psi\right\rangle =0$, whereas the ($a$, $a$) states (the discrete dilaton and the two half-weight $w=0$ states) satisfy $G^{-}_{-3/2}\left|\psi\right\rangle = \bar{G}^{-}_{-3/2}\left|\psi\right\rangle =0$.
\item
For type 0B strings, one requires the GSO projection: $(-)^f=(-)^{\bar{f}}$. Analogously for type 0A strings, where one imposes $(-)^f=(-)^{\bar{f}+\bar{a}}$. \\
The criterion for the GSO projections for type II superstrings in the various string theories\footnote{The GSO projection is $(-)^{f+w+a}=(-)^{\bar{f}+w+\epsilon \bar{a}}=-1$ where $\epsilon=1$ in type IIB and $\epsilon=0$ in type IIA.} always involves the combination $(-1)^{f+w}$ (and its antiholomorphic cousin), which in the NS-sector is actually invariant under the above automorphism (\ref{involution}) of the state spectrum:
\begin{equation}
(-)^{f+w} \to (-)^{-1-f+1-w} = (-)^{f+w},
\end{equation}
implying states are present in the spectrum in pairs. In R-sectors, this factor reverses sign in this procedure.
\item
Just as proven in earlier work for bosonic and type II superstrings \cite{Giveon:2014hfa}\cite{Mertens:2014saa}, the equality between the large $k$ partition function of the cigar and the flat plane holds as well for type 0 string theory, again in spite of the presence of a GSO projection which seems to appoint special significance to the origin (as the thermal GSO projection utilizes the winding number of strings around the origin). Hence once again, the GSO projection does not break the coordinate invariance of the partition function.
\end{itemize}

\section{Continuous states}
\label{continuous}
Up to this point, we only discussed the discrete part of the spectrum, as this contains the most subtle and interesting features. To complete our study of the spectrum, we have to say a few words about the continuous sector as well. \\
The continuous part of the primaries on the cigar (i.e. for finite $k$) is of the form
\begin{align}
h &= \frac{s^2+1/4}{k} + \frac{m^2}{k} + \frac{\left(f+\frac{a}{2}\right)^2}{2}, \\
\bar{h} &= \frac{s^2+1/4}{k} + \frac{\bar{m}^2}{k} + \frac{\left(\bar{f}+\frac{\bar{a}}{2}\right)^2}{2},
\end{align}
where $s\in\mathbb{R}^{+}$, $m = \frac{kw+n}{2}$, $\bar{m} = \frac{kw-n}{2}$. As usual, the quantity $a=0$ in the NS-sector and $a=1$ in the R-sector. \\
In this case, the quantity $f$ really has the interpretation of a fermionic oscillator number. \\
In the following we will immediately focus on the $k\to\infty$ limit and discuss which marginal states are possible. \\
We remark in advance that spectral flow acts in precisely the same way for the continuous states and we will hence not discuss it again here. \\

\noindent NS-NS states should satisfy
\begin{equation}
h-\bar{h} = nw +\frac{f^2-\bar{f}^2}{2} \in \mathbb{Z}.
\end{equation}
Hence $f$ and $\bar{f}$ should be even or odd simultaneously. The weight becomes in the large $k$ limit
\begin{equation}
h+\bar{h} \to \frac{2s^2}{k} + \frac{kw^2}{2} + \frac{f^2+\bar{f^2}}{2}.
\end{equation}
The weight of such a state becomes infinitely large unless $w=0$, implying only discrete momentum states can be continuous in the flat limit. \\
It can be proven that the continuous quantum number $s$ makes no contribution to the convergence or divergence of a state \cite{Mertens:2014nca}.  This leads to
\begin{equation}
h+\bar{h} \to \frac{f^2+\bar{f^2}}{2},
\end{equation}
which is minimized by $f=\bar{f}=0$. This state is the closed string tachyon, which is projected out by GSO in type II theory.\footnote{It is present for type 0 strings, making these tachyonic as usual.} The next state has a single oscillator excited and has $f=\pm1$, $\bar{f}=\pm1$ and $h+\bar{h}=1$, with a quadruple degeneracy. The $\mathcal{R}$-charges of these states are $Q=\pm1$ and $\bar{Q}=\pm1$. Hence one can identify them as a ($c$, $c$), ($c$, $a$), ($a$, $c$) and ($a$, $a$) state.\\
Much like in the discrete sector, these four states are the \emph{only} generically (i.e. for generic $k$) (anti)chiral primaries in the continuous spectrum of the (finite $k$) cigar CFT. \\

\noindent R-R states need to obey the level-matching condition
\begin{equation}
h-\bar{h} = nw +\frac{f^2+f-\bar{f}^2-\bar{f}}{2} \in \mathbb{Z},
\end{equation}
which is always satisfied.
The weight asymptotes to\footnote{Where as usual we added $3/8$ for each R-sector present.}
\begin{equation}
h+\bar{h} \to \frac{2s^2}{k} + \frac{kw^2}{2} + \frac{f^2+f+\bar{f^2}+\bar{f}}{2} + \frac{1}{4} + \frac{6}{8}.
\end{equation}
Hence the criterion for convergence ($h+\bar{h}\geq 1$) becomes as $k\to+\infty$:
\begin{equation}
\frac{f^2+f+\bar{f^2}+\bar{f}}{2} \geq \frac{3}{4}-\frac{6}{8}=0,
\end{equation}
which is always satisfied. For $f=0$ or $f=-1$ (and analogously for $\bar{f}$) these states are marginal and are hence the Ramond ground states. Again we find four states satisfying this property. \\

\noindent The R-NS states are obtained by choosing $a=1$ and $\bar{a}=0$. Level-matching leads to
\begin{equation}
nw +\frac{f^2+f-\bar{f}^2}{2} + \frac{1}{2} \in \mathbb{Z}.
\end{equation}
Again only $w=0$ is allowed to have a finite weight in the end. Hence $\bar{f}$ needs to be odd. The total weight asymptotes to
\begin{equation}
h+\bar{h} \to  \frac{f^2+f+\bar{f^2}}{2} + \frac{1}{8} + \frac{3}{8}.
\end{equation}
This is minimized by setting $f=0$ or $f=-1$ and $\bar{f} = \pm1$, which leads to
\begin{equation}
h+\bar{h} \to  1.
\end{equation}
The state hence becomes marginal in this limit. The state is the combination of the left-moving Ramond vacuum and a right-moving singly excited state.\\
A completely analogous discussion can be made for the NS-R sector states. \\
We find that in both of these sectors again four states (for each $n$) are marginal. This is a manifestation of the restoration of spacetime supersymmetry at infinite $k$.

\subsection{Summary}
\begin{figure}[h]
\centering
\includegraphics[width=0.5\textwidth]{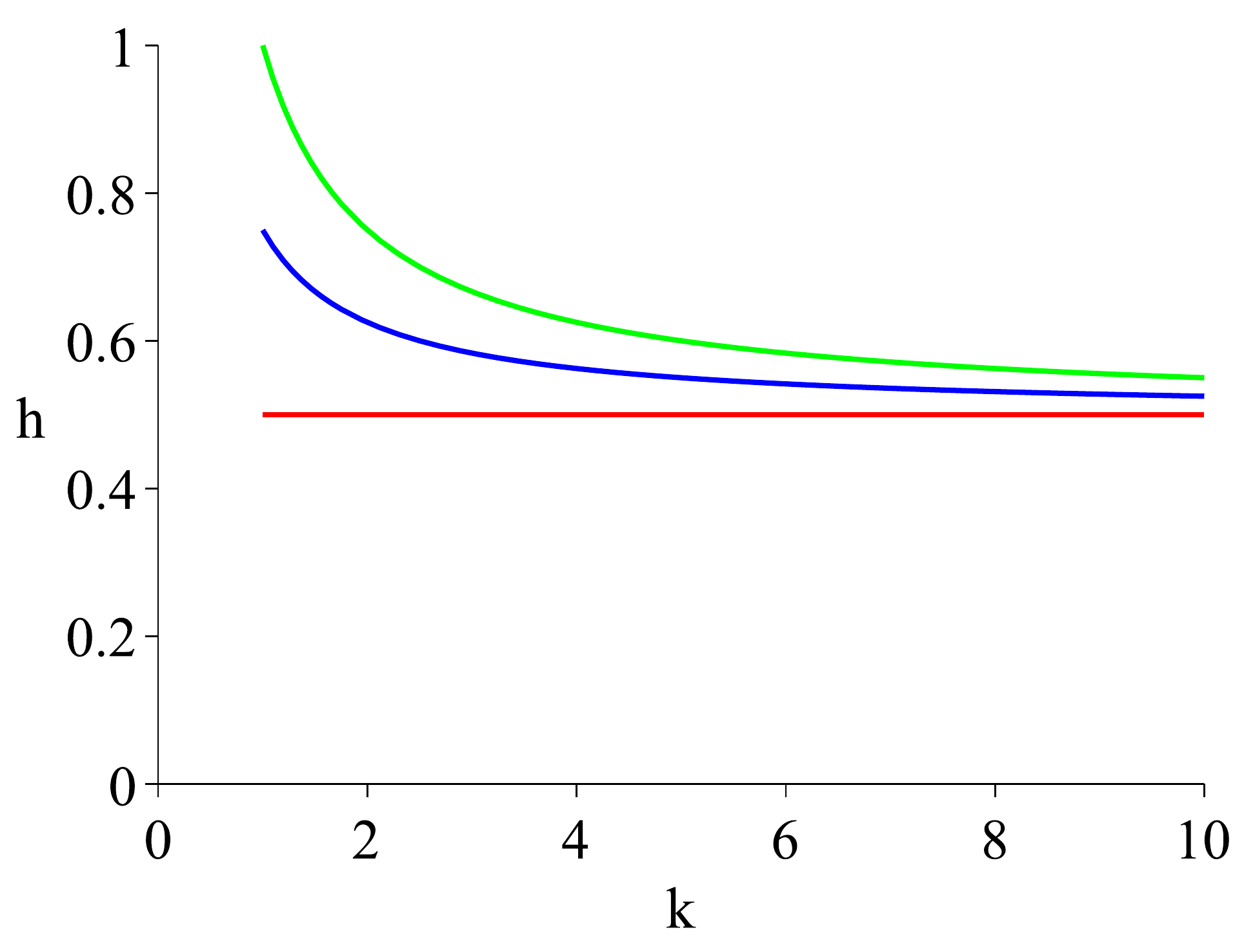}
\caption{Evolution of the important states with $k$. The red curve represents the thermal scalar and discrete dilaton state: the only marginal states for any $k$. The blue line represents all the continuous modes discussed above, as well as the R-NS, NS-R and R-R discrete sectors. Finally, the green curve represents the two additional $w=0$ NS-NS states. These have higher weight for any $k$ but become marginal nonetheless at infinite $k$.}
\label{largeK}
\end{figure}

The behavior of the conformal weights of all of the states discussed in this and the previous section with $k$ is shown in figure \ref{largeK}. The red (bottom) curve represents the two NS-NS states that are always marginal with $h=1/2$. The blue (middle) curve represents most of the other states (the continuous sectors, the R-NS, NS-R and R-R sectors) with $h=\frac{1}{4k}+\frac{1}{2}$. The green (top) curve represents the two remaining NS-NS sectors with $w=0$: $h=\frac{1}{2k}+\frac{1}{2}$.

\subsection{Adding additional dimensions}
Up to this point, we only discussed the 2d space of the cigar itself. Of course, one needs to add additional dimensions in general in order to have a valid string background. Can one generate additional massless states in this way? And if so, which? \\
For the NS-R, R-NS and R-R sectors, we already implicitly took this into account when we added $1/8$ for each additional (complex) fermion contribution. \\
The NS-NS sector is the remaining and most interesting sector to look at. \\
For the discrete sector, one can convince oneself that one cannot make the weight $h+\bar{h}$ in the 2d cigar CFT less than 1.\footnote{This is so even when neglecting the GSO projection on this sector. This is important as one can try the $w=f=\bar{f}=0$ sector and excite one oscillator in the additional dimensions to satisfy GSO in the end.} Thus there is no way in combining a cigar CFT state with an additional oscillator contribution from the additional CFT to obtain an overall marginal, GSO-preserving state. \\
For the continuous sector, this is even more clear: $h+\bar{h}$ on the cigar is at least $1/(2k)$ and hence it is impossible to combine this with an additional oscillator contribution from the other dimensions and obtain a marginal state. This is the mass gap of the linear dilaton space. \\
However, in the $k\to\infty$ limit, this does becomes possible. The natural additional geometry to take here is $\mathbb{R}^8$ and hence obtain 10d flat space with two dimensions described in polar coordinates. One obtains additional massless states by exciting oscillators in the $\mathbb{R}^8$ part. These are precisely the graviton excitations one has in flat space. In the polar plane, their (radial) momentum is described by $\frac{2s}{\sqrt{\alpha' k}}$, whereas the conventional $\alpha' p^2/4$ contribution is present in the other flat dimensions.

\section{Flat $\mathbb{C}/\mathbb{Z}_N$ cones}
\label{cones}
Of course, when studying thermodynamics, we want to find the behavior of the states as the temperature varies. In string theory, this is complicated by the fact that good modular invariant partition functions are only known for $\beta = \frac{2\pi}{N}$ for integer $N$. Nonetheless, it is interesting to try to track the different states as $N$ jumps in a discrete fashion to at least have some idea on what might be going on for a general temperature dependence. The study of the flat $\mathbb{C}/\mathbb{Z}_N$ cones is the purpose of this section. \\

\noindent The modification required to study the $\mathbb{C}/\mathbb{Z}_N$ cones is simply to perform the substitution
\begin{align}
w &\to \frac{w}{N}, \\
n &\to nN,
\end{align}
while keeping the same unitarity constraint on $j$. Since this is the only change, the transformation of the weights (\ref{invvsy}) is altered into
\begin{equation}
r,\, nN,\, \frac{w}{N},\, f \quad \longrightarrow \quad -r,\, -nN,\, 1-\frac{w}{N},\, -1-f-a
\end{equation}
which again entails an involution symmetry when $r=0$. \\

\noindent The main difference is that when $N \neq 1$, there is a non-trivial such symmetry even for bosonic strings. For bosonic strings, one includes an array of tachyonic states in the spectrum with weights
\begin{equation}
h + \bar{h} = -\frac{w^2}{N^2} + \frac{w}{N} - 2.
\end{equation}
Sectors with $w$ and $N-w$ have the same weight. This is the generalization of the involution symmetry (\ref{involution}) to the bosonic flat cones. \\

\noindent Next we will analyze type II superstrings more closely, first in the discrete sector and in the end in the continuous sector.

\subsection{NS-NS sector}
\begin{figure}[h]
\centering
\includegraphics[width=0.5\textwidth]{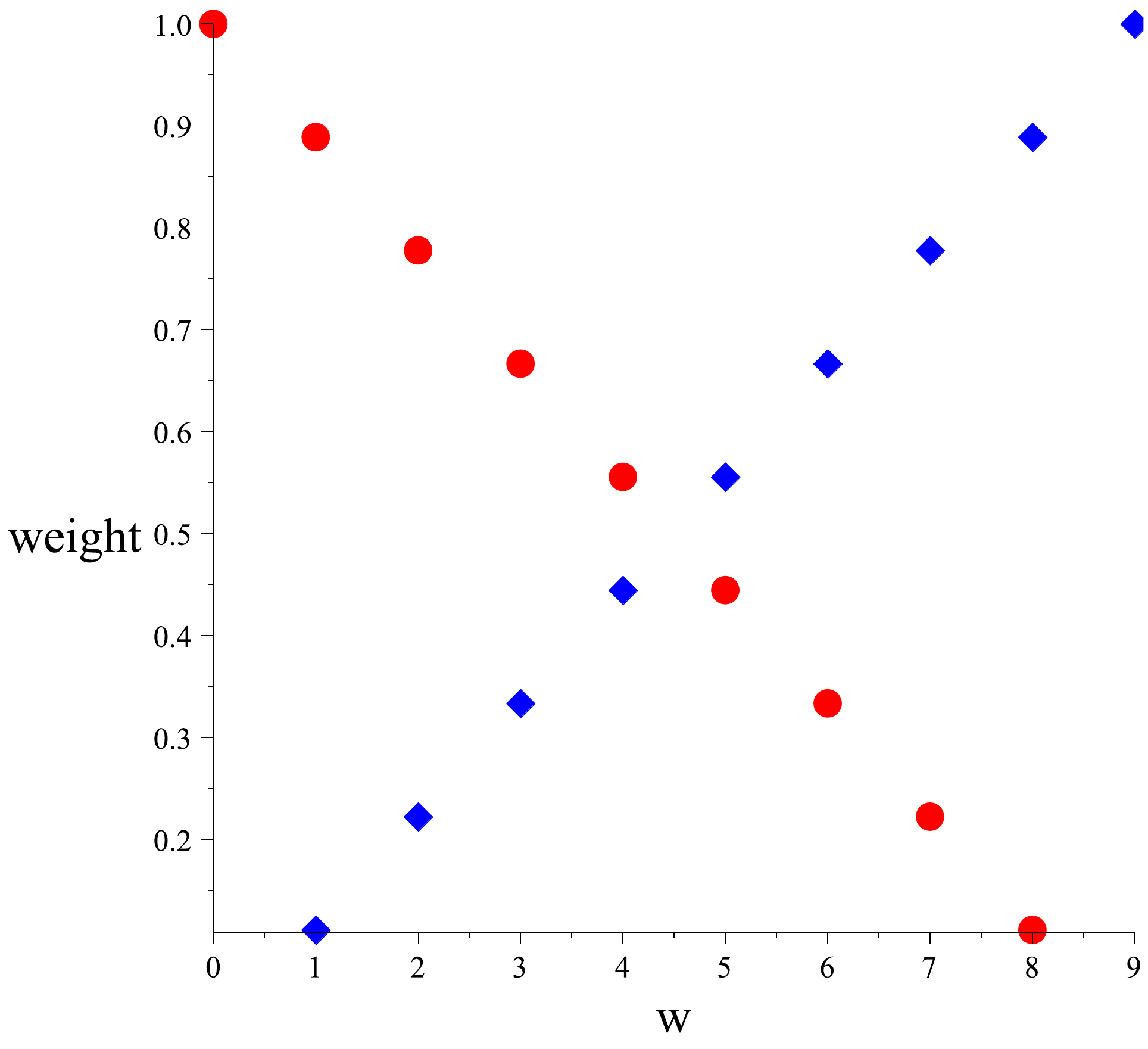}
\caption{Weights $h+\bar{h}$ of NS-NS most tachyonic primaries for the case $N=9$. All states with weight less than 1 are tachyonic. Every state is twofold degenerate. GSO projects half of these out of the spectrum (not shown): the $2^{nd}$, $4^{th}$, $6^{th}$ and $8^{th}$ horizontal lines of states are absent in the spectrum.}
\label{NSNScone}
\end{figure}

There are twisted tachyonic states appearing in both the thermal scalar sector ($f=\bar{f}=0$) and the dilaton sector ($f=\bar{f}=-1$) (and nowhere else). We focus on the case where $l=\bar{l}=0$ as these are the most tachyonic states for fixed $w$.\footnote{We want to remark that there exist other (tachyonic) states on top of these states by having $l$ or $\bar{l}$ nonzero.} The winding states have weight $h+\bar{h} = w/N$ with $j=\frac{kw}{2N}$ and $w=1,\hdots N$ and are hence all tachyonic. The dilaton state remains marginal for any $N$. In addition, there are twisted modes in the dilaton sector having $h+\bar{h}=1-w/N$ with $j=\frac{kw}{2N}+1$ and $w=0,1,\hdots N-1$. Both of these sectors of states are related by the involution symmetry mentioned earlier. Indeed, setting $j' = k/2+1-j$ requires us to take $w'=N-w$. These two sectors of tachyonic states are shown for the case $N=9$ in figure \ref{NSNScone}.

\noindent The most tachyonic states can be found by setting $w=1$ in the thermal scalar family of states or setting $w=N-1$ in the dilaton family of states. Letting $N\to1$, the $w=1$ state will become the thermal scalar in polar coordinates. The $w=N-1$ state in the dilaton sector on the other hand, becomes a state of equal weight but with $w=0$: this is the dilaton state itself.\\

\noindent Hence in a rather precise way, the dilaton state can be identified with the negatively wound thermal scalar state. These results are in agreement with the analysis presented in subsection \ref{larget}. \\

\noindent Finding suitable discrete cousin states for the two $w=0$ half-weight NS-NS states is not possible.\footnote{There do exist marginal states with either $w=0$, $n=1$, $f=\bar{f}=-1$, $l=0$ and $\bar{l}=N$ or  $w=0$, $n=-1$, $f=\bar{f}=-1$, $l=N$ and $\bar{l}=0$, but these fail to satisfy the unitarity constraints ($j=1-\frac{N}{2}$) and are hence absorbed in the continuum when $N>1$.}

\subsection{R-R sector}
It can be shown that the lowest weight states in the R-R sector are characterized by
\begin{equation}
f=\bar{f}=-1
\end{equation}
and have weight and spin
\begin{align}
h+\bar{h} &= 1 + \frac{w(l+\bar{l})}{N}, \\
h-\bar{h} &= nw + \frac{f^2+f}{2} - \frac{\bar{f}^2+\bar{f}}{2}.
\end{align}
Discrete momentum is given in this case by
\begin{equation}
nN = l-\bar{l} + f - \bar{f},
\end{equation}
with $n\in\mathbb{Z}$ and is a more stringent condition on the quantum numbers of a candidate state than when $N=1$, ensuring a $2\pi/N$ rotation is periodic for these R-R states. Clearly, the lowest weight state has $l=\bar{l}=0$ and is marginal for any $N$. In this case, $w$ is allowed to vary from $0$ to $N$. All of these states have the same weight $h+\bar{h}=1$ and spin $0$ (because $n=0$). They have
\begin{equation}
j=\frac{kw}{2N} - \frac{l}{2} - \frac{\bar{l}}{2} + \frac{1}{2} = \frac{kw}{2N} + \frac{1}{2},
\end{equation}
which clearly satisfies the unitarity constraints in this range of $w$.

\subsection{R-NS sector}
The most stringent condition here is found from the momentum constraint (\ref{momconstraint}):
\begin{equation}
nN = l-\bar{l}+f-\bar{f}+1/2,
\end{equation}
for half-integer $n$. This condition ensures that the discrete momentum $nN$ is of the form needed such that a $2\pi/N$ rotation generates anti-periodic fermions. We remark here that the only way to have a solution is when $nN$ is half-integer, which requires $N$ to be odd, a condition that was previously encovered from other perspectives as well. \\
For instance, if we focus on a state with fixed $l$, $\bar{l}$, $f$ and $\bar{f}$, then this state is only present in the spectrum for a subset of all $N$. \\
Let us first focus on the two states we constructed in section \ref{largek} in the R-NS sector and find out which state should be identified with these when $N\neq 1$.
The state with $l=\bar{l}=0$ and $f=\bar{f}=-1$ requires $N=1$ and hence $w=0$. This state is absent for any other conical angle. \\
The other state with $l=\bar{l}=0$ and $f=-1$ and $\bar{f}=0$ requires $N=1$ and hence $w=1$. Again it is absent for any other conical angle.\\

\noindent There is however another state that is massless when $w=1$, $f=-1$ and $\bar{f}=0$. Setting $l=0$ and $\bar{l}=\frac{N-1}{2}$, one finds a massless state for any $N$, satisfying the unitarity constraints. \\
Some trial-and-error shows that all other states with the  same $f$ and $\bar{f}$ are massive for any $N \neq 1$.\footnote{A formal proof goes as follows. If $f=\bar{f}=-1$, the weight of the state equals
\begin{equation}
h+\bar{h} = 1 - \frac{w}{2N} + \frac{w(l+\bar{l})}{N}.
\end{equation}
If $l=\bar{l}$, then we require $N=1$ and hence $w=0$. This leads to weight $1$. Otherwise, if $l-\bar{l}\neq0$, the final term in the weight contributes and outweighs the second term, leading to a massive state. An exception is when $w=0$ and $l=\frac{N-1}{2}$ and $\bar{l}=0$. This state is precisely massless, but it is excluded by the unitarity bounds for $N$ larger than 1.\\
In the other case, $f=-1$ and $\bar{f}=0$, the weight is given by
\begin{equation}
h+\bar{h} = \frac{1}{2} + \frac{w}{2N} + \frac{w(l+\bar{l})}{N}.
\end{equation}
With $n=-\frac{1}{2}$, one readily shows that there exists precisely one massless state with $l=0$ and $\bar{l}=\frac{N-1}{2}$.
}  It can be shown that any other combination of $f$ and $\bar{f}$ also automatically leads to higher weights. This demonstrates that these sectors are quite subtle when analytically continuing in $N$, as no immediate ``sibling'' states are found. \\

\noindent When considering flat space string theory at a finite temperature, all states in the thermal spectrum are continuously deformed by changing the temperature: no states are created or destroyed. This is not true in the polar coordinate description nor for the conical orbifolds. States can appear and disappear abruptly when changing $N$. This is not a sickness of the theory, as it is only the total thermal partition function that matters, not its description in terms of thermal modes, but this does foil a direct thermodynamic interpretation of an individual mode. \\
Some thermal modes however, such as the thermal scalar, are not really affected by this as they exhibit a clean continuation as one varies $N$; we will pick up this line of thought further on when contemplating the entropy. 

\subsection{Continuous states}
Finally, let us look at the continuous states. Studying the continuous states on a cone clearly does not do much: $w=0$ in the continuous sector and $n$ is scaled out in the weights of the states. Hence their weights are kept constant upon changing the temperature (or $N$).  This is expected to some extent, as the continuous states are less sensitive to what is happening at the tip of the cigar or cone.

\section{Large $\tau_2$ limit directly in the cigar partition function} 
\label{directlim}
While the above analysis is quite rigorous, it is somewhat involved to obtain the marginal states: we had to first rely on the (known) character decomposition after which the discrete marginal states were found. It is interesting to note that one can see all of the above constructed most dominant states (both on the cigar and for infinite $k$) directly in the cigar orbifold partition function. 
The most dominant behavior of the partition functions themselves ($\tau_2\to\infty$) can be obtained directly without first going through the character decomposition. \\
In the flat (large $k$) limit, this approach was studied in the earlier literature for the $\mathbb{C}/\mathbb{Z}_N$ orbifolds and everything indeed works out \cite{Dabholkar:1994ai}\cite{Lowe:1994ah}. The study for $N=1$ was not done (due to a lack of technical machinery to tackle this question). \\
On the finite $k$ cigar, we encounter additional technical complications that need to be sorted out. To make this discussion a bit more concrete, consider the type II cigar partition function \cite{Israel:2004ir}\cite{Eguchi:2004yi}\cite{Sugawara:2012ag}:\footnote{We cannot resist making a comment on the overall normalization of these cigar and cigar orbifold partition functions. One of the best ways to determine the normalization is through the elliptic genus, as is done for instance in \cite{Eguchi:2004yi}. This however requires $\mathcal{N}=2$ worldsheet supersymmetry, which is absent in the bosonic case. An alternative method is to use the $k\to\infty$ limit of the cigar orbifolds, as we did in \cite{Mertens:2014saa}. These should reduce to the flat $\mathbb{C}/\mathbb{Z}_N$ models, whose overall normalization is known.}
\begin{align}
\label{largetpartfu}
Z = \frac{k}{4}\int_{\mathcal{F}}\frac{d\tau d\bar{\tau}}{4\tau_2}\sum_{m,w \in \mathbb{Z}}\int_{0}^{1}ds_1 ds_2 &\frac{e^{-\frac{\pi k}{\tau_2}\left|\left(s_1 - w\right)\tau + \left(s_2 - m\right)\right|^2}}{\left|\vartheta_1(u,\tau)\eta^3\right|^2} e^{4\pi\tau_2\left(\frac{1}{4} - \frac{1}{4k}\right)}\sum_i q^{h_i}\bar{q}^{\bar{h}_i}\nonumber \\
&\times \left\{\vartheta_3(u,\tau) \vartheta_3^3 + (-)^{w+1} \vartheta_4(u,\tau) \vartheta_4^3 + (-)^{m+1}\vartheta_2(u,\tau) \vartheta_2^3\right\} \nonumber\\
&\times \left\{\bar\vartheta_3(u,\tau) \bar\vartheta_3^3 + (-)^{w+1} \bar\vartheta_4(u,\tau) \bar\vartheta_4^3 + (-)^{m+1}\bar\vartheta_2(u,\tau) \bar\vartheta_2^3\right\}.
\end{align}
We have included here an arbitrary, unitary and compact internal CFT with weights $h_i$. The only thing needed is that its contribution approaches 1 in the large $\tau_2$ limit. \\
Taking $k$ large amounts to a saddle point approximation \cite{Giveon:2014hfa}\cite{Mertens:2014saa} and yields in the end 
\begin{equation}
Z \sim \left|\vartheta_3^4-\vartheta_4^4-\vartheta_2^4\right| =0,
\end{equation}
demonstrating that indeed the flat space result is obtained. However, this sheds no light on how this cancellation actually occurs using the polar coordinate description. \\
Taking $\tau_2$ large \emph{before} taking the large $k$ limit allows us to follow the most dominant states as the flat limit is taken, and illustrates how the previously constructed marginal states conspire to give a zero net result. \\
Using saddle point methods for the $s_1$-integral in (\ref{largetpartfu}), one can obtain the large $\tau_2$ expansion of the partition function on the cigar itself. It is interesting to see the appearance of the unitarity constraints (through the criterion on the presence of a saddle point) and the appearance of continuous modes (through the absence of any saddle point) appear in this direct way. 
The details will not be presented here. For bosonic strings, the reader is refered to appendix \ref{mostdom}, and the extension to type II superstrings is relatively straightforward (but tedious) and is spelled out in detail in appendix \ref{mostdomII}. \\

\noindent One of the more salient features is that this procedure demonstrates that it is indeed the R-NS and NS-R spacetime fermion sectors that compensate for the thermal scalar and dilaton state and in the end cause the partition function to vanish.

\section{Approaches to the one-loop entropy in Rindler space}
\label{entropy}
Up to this point, we only looked at the partition functions themselves and their spectral content. Within thermodynamics, this partition function in polar coordinates and its orbifold cousins can be identified with the free energy of the string gas in Rindler space (upon subtraction of the temperature-independent Rindler vacuum energy) as $\beta F= -Z$ with $\beta=2\pi$, the inverse Rindler temperature. The thermodynamic entropy $S$ can then be computed as $S= -(\beta\partial_{\beta}-1)Z = \partial_N(NZ)$ for $\beta=2\pi/N$. Note that this derivative requires knowledge of the partition function $Z$ as a function of real $N$. Hence some form of continuation of these partition functions to non-integer $N$ is required. This will unavoidably lead to some ambiguity as we will explain further on. \\
This work was originally motivated by the fact that in \cite{He:2014gva} it was argued that the one-loop entropy for type II superstrings in Rindler space actually vanishes, which is only possible if the contribution of the dominant thermal scalar gets cancelled by some other field. The above considerations demonstrate that there exist indeed other marginal states capable of providing this cancellation. The remainder of this work concerns a deeper study of this question. In this section, we will display three separate approaches towards understanding the one-loop entropy in this space. Each approach is not without reservation however. In the end, we will find a vanishing entropy as well here. The next main section \ref{UV} then finally wraps up the story by highlighting what we have learned regarding the way UV divergences are mitigated within string theory.

\subsection{Negative contributions to the entropy}
As a first topic, we will discuss that it is at least plausible for the entropy to vanish. In order for this to happen, a cancellation has to happen at each mass level on the thermal manifold. \\
The reader might wonder about the fact that we are looking for a mode on the thermal manifold to cancel a contribution from the entropy. The entropy of any non-interacting field (boson or fermion) is always positive, so how can this occur? Actually, the thermal string path integral (whose spectral decomposition we have been studying here) does \emph{not} correspond to a non-interacting Hamiltonian thermal trace $\text{Tr}e^{-\beta H}$ where one sums over all string states in the Lorentzian spectrum; exotic open-closed interactions are already taken into account. The intuition for this \cite{Susskind:1994sm}\cite{Mertens:2014saa}\cite{Mertens:2015adr} comes from the fact that if one draws a torus diagram on flat space, the possibility exists for the origin to be \emph{inside} the worldsheet of the torus. Such string embeddings cannot be interpreted as a non-interacting closed string performing a loop around the Euclidean time direction.\footnote{For any non-interacting field, one always has $F<0$ and $S>0$, provided the zero-point energy is excluded from $F$. These non-interacting properties \emph{are} fulfilled for the partition functions studied in \cite{Mertens:2015adr}.} \\

\noindent This illustrates that the theory cannot be written as the sum of complete (i.e. all wrappings present) free fields (this was done in \cite{Mertens:2015adr}), and as such is not restricted to have a positive entropy after all. It is possible for severe cancellations to occur on the thermal manifold. \\
And indeed, we already saw above that for the partition function, the spacetime fermions contribute with opposite sign than the spacetime bosons, in the end causing a perfect cancellation to occur. Thus one can expect negative contributions to the entropy as well, coming from these same spacetime fermionic sectors. This latter point will be explicitly proven in section \ref{meth3}.

\subsection{Dominance of massless modes}
We have already said several times that we expect the singly wound (thermal scalar) state to be the dominant contribution to the free energy and entropy. But we discovered a host of additional marginal states in polar coordinates. It is hence crucial to first demonstrate in what sense different thermal modes can contribute to the entropy.
We present an analysis on how and when a specific thermal mode can dominate the free energy and entropy: \emph{any} massless mode makes a dominant contribution to thermal quantities, but their functional dependence on the temperature is quite different. This even holds for the R-R modes that have a temperature-independent conformal weight. \\

\noindent Let's look at the situation in flat space and ask in what sense the thermal scalar dominates over other massless states in the spectrum. \\
The free energy of a massive bosonic field in flat space $\mathbb{R}^{d-1,1}$ is of the form
\begin{align}
F &= \frac{V_{d-1}}{\beta} \int \frac{d^{d-1}k}{(2\pi)^{d-1}}\ln\left(1-\exp\left(- \beta \sqrt{m^2+k^2}\right)\right).
\end{align}
Clearly, if all dimensions are compact, the integral becomes a sum, and if the single-particle energy eigenvalue $E=0$ is in the spectrum, the free energy diverges logarithmically. This can also be seen when we rewrite
\begin{equation}
F = -V_{d-1} \int_{0}^{+\infty}\frac{ds}{s(2\pi s)^{d/2}} \sum_{r=1}^{+\infty}\exp\left(-\frac{m^2 s}{2} - \frac{r^2\beta^2}{2s}\right).
\end{equation}
For a massless field, this diverges logarithmically if $d=1$ due to the large $s$ IR behavior of the integral.\footnote{One Poisson resums the series, after which the integral over $s$ diverges as $\int^{+\infty}\frac{ds}{s}$.} 
Within string theory, one sums this over the string spectrum and writes\footnote{The prime means one excludes the $r=0$ contribution from the series.}
\begin{align}
F &= -V_{d-1} {\sum_{r=-\infty}^{+\infty}}' \int_{0}^{+\infty}\frac{d\tau_2}{2\tau_2} \int_{-1/2}^{1/2}d\tau_1\frac{1}{(4\pi^2 \alpha'\tau_2)^{d/2}}\left|\eta(\tau)\right|^{-48}\exp\left( - \frac{r^2\beta^2}{4\pi\alpha'\tau_2}\right),
\end{align}
Clearly the previous possible IR divergence ($\tau_2\to+\infty$) is still present for each massless string mode, as can be seen by expanding the Dedekind $\eta$ function to second order. On top of this, the well-known UV Hagedorn divergence ($\tau_2\to0$) is present as well. Upon transferring from the modular strip $\mathcal{E}$ to the modular fundamental domain $\mathcal{F}$ and Poisson resummation $m\to n$, these divergences are completely mixed. The IR divergences are carried by states whose weight does not depend on $\beta$, whereas the UV divergence is carried by the thermal scalar. If one is asked to write down the most dominant contribution to the entropy, one has to include both of these states. This is not so strange, since we do expect massless states to contribute heavily to the string gas. \\
However, the above IR divergence has a characteristic signature: the free energy of this contribution scales as $F\sim \frac{1}{\beta}$ and has no exponential contribution. Nonetheless, these potentially divergent modes of the massless states contribute to the entropy as well as $S = C$ if $F=-\frac{C}{\beta}$ for some (positive) constant $C$.\footnote{Taking even more $\beta$-derivatives gives a vanishing result though.} Their main difference with the Hagedorn divergence is that the latter causes a divergence as
\begin{equation}
\beta F \propto 
\left\{
    \begin{array}{ll}
        \left(\beta - \beta_{H}\right)^{D/2}\ln\left(\beta - \beta_{H}\right), \quad D\text{ even}, \\
        \left(\beta - \beta_{H}\right)^{D/2}, \quad D\text{ odd},
    \end{array}\right.
\end{equation}
for $D$ non-compact dimensions and hence leads to \emph{non-analytic} behavior as a function of $\beta$. \\
It is somewhat of a surprise to find that the free energy on the full (finite $k$) cigar simply has no additional massless states (due to the mass gap of the linear dilaton asymptotics) and this situation cannot occur. So in this sense, the cigar is actually simpler even than flat space. \\

\noindent A major difference with flat space however, is that as one changes $N$ (or $\beta$) some modes might appear or disappear from the spectrum. We have seen this above already in attempting to find the correct $N\neq1$ continuation of the $w=0$ marginal state in the R-NS sector. This does not happen in flat space, and this obscures whether it makes sense to compute the entropy associated to that mode alone. So we will not attempt to compute the entropy of all the most dominant modes to prove a cancellation. The only reasonable way to proceed seems to be to prove this vanishing on the total entropy expression itself (where all thermal modes have been summed over). \\
What is vital though, is that there are at least modes on the $N=1$ space that have the same most dominant conformal weight as the thermal scalar (i.e. marginal) and this is indeed the case.\\
Some modes however seem to be more robust upon changing $N$, such as the thermal scalar mode. In appendix \ref{toy} we include some simple analogous summations that demonstrate that for such dominant modes, one can compute its entropy directly and claim that this is the dominant contribution to the entropy. This is directly relevant for the bosonic string on the cigar (and its flat limit), where the thermal scalar mode is tachyonic and dominant, with no other mode contributing. \\
Another case for which this is relevant is all of the orbifold models (keeping $N \neq 1$ in the end). Both for bosonic and for type II superstrings, a tachyonic dominant contribution is present that is impossible to compensate by other modes. \\
A much more subtle case is type II superstrings in the flat limit, where as shown extensively above, additional massless states exist, some of which contribute with opposite sign to thermodynamic quantities. Hence a cancellation is possible in this case, and we will need to analyze the full expression for the entropy. \\

\noindent In the next three subsections, we examine three possible lines of attack at obtaining the one-loop entropy for type II superstrings in the flat limit. None are ideal and a fully satisfying construction of the one-loop entropy will have to wait until we fully understand how to deal with string theory on cones of arbitrary deficit.

\subsection{Method 1: The one-loop entropy from sum-over-fields perspective}
\label{metho1}
As a first approach, we approximate string theory as a sum over the fields in its spectrum. The expressions are however a bit unwieldy. We consider here the $\mathbb{C}/\mathbb{Z}_N$ flat cones, for which the partition function for an arbitrary (higher) spin field was written down in  \cite{He:2014gva}. One can then sum these over the full string spectrum. \\
Note that this approach is very different from our perspective up to this point, as no use is made of the thermal spectrum. One takes all of the states in the Lorentzian Rindler spectrum, computes the heat kernel for each individual state and then finally sums these over the spectrum. \\

\noindent For open strings, it is known that this procedure precisely agrees with the actual stringy $\mathbb{C}/\mathbb{Z}_N$ orbifold computation \cite{Mertens:2015adr}. The open superstring sum-over-fields approach yields the entropy ($d=10$):
\begin{align}
\label{nonintpffo}
S &= V_{d-2}  \int_{0}^{+\infty}\frac{ds}{2s}(4\pi s)^{-4} \frac{1}{N}\frac{1}{\eta\left(\frac{is}{2\pi\alpha'}\right)^{9}}\frac{1}{4}\frac{e^{\frac{s}{8\alpha'}}}{\prod_{n=1}^{+\infty}(1-q^n)}\sum_{j=1}^{N-1}\frac{1}{\sin^2\left(\frac{2 \pi j}{N}\right)} \nonumber \\
&\times \left[ \vartheta_3^3\left(0,\frac{is}{2\pi\alpha'}\right) \sum_{m\in\mathbb{Z}}\sum_{p_n,q_n=0}^{+\infty}\left(\frac{2}{3}-2\left|\sum_n (p_n-q_n) + m\right|\right)e^{- \frac{s}{\alpha'}\left(\sum_nn(p_n+q_n)+m^2/2\right)}\right. \nonumber \\
&\left.\quad - \vartheta_4^3\left(0,\frac{is}{2\pi\alpha'}\right) \sum_{m\in\mathbb{Z}}\sum_{p_n,q_n=0}^{+\infty}(-)^m \left(\frac{2}{3}-2\left|\sum_n (p_n-q_n) + m\right|\right)e^{- \frac{s}{\alpha'}\left(\sum_nn(p_n+q_n)+m^2/2\right)} \right. \nonumber \\
&\left.\quad + \vartheta_2^3\left(0,\frac{is}{2\pi\alpha'}\right) \sum_{m\in\mathbb{Z}}\sum_{p_n,q_n=0}^{+\infty}\frac{1}{3}e^{- \frac{s}{\alpha'}\left(\sum_nn(p_n+q_n)+(m-1/2)^2/2\right)}\right].
\end{align}
The function in brackets does \emph{not} vanish. Its large $\tau_2$ expansion is given by
\begin{equation}
8q^{1/2}-64q^{3/2}-296q^{5/2}-1328q^{7/2} + \mathcal{O}(q^{9/2}).
\end{equation}
The leading coefficient can be found as
\begin{equation}
8\left(\underbrace{\frac{8}{6}-1}_{\text{vector}} + \underbrace{\frac{8}{12}}_{\text{gaugino}}\right),
\end{equation}
agreeing with the entropy contribution of a vector field and a gaugino field, the massless sector of open superstrings. \\

\noindent For closed type II superstrings, the partition function as a sum over fields is given by \cite{He:2014gva}\cite{Mertens:2015adr}:
\begin{align}
\label{nonintpff}
Z &= \frac{1}{N}V_{d-2} \int_{\mathcal{E}}\frac{d^2\tau}{4\tau_2}(4\pi^2\alpha' \tau_2)^{-4}\sum_{j=1}^{N-1}\frac{\left|\vartheta_1\left(\frac{j}{N},\tau\right)\right|^8}{\left|\eta(\tau)\right|^{18}\left|\vartheta_1\left(\frac{2j}{N},\tau\right)\right|^2} \nonumber \\
&= \frac{1}{N}V_{d-2}  \int_{\mathcal{E}}\frac{d^2\tau}{2\tau_2}(4\pi^2\alpha' \tau_2)^{-4} \frac{1}{\left|\eta\left(\tau\right)\right|^{18}}\frac{1}{16}\frac{e^{\frac{\pi\tau_2}{2}}}{\left|\prod_{n=1}^{+\infty}(1-q^n)\right|^2}\sum_{j=1}^{N-1}\frac{1}{\sin^2\left(\frac{2 \pi j}{N}\right)} \nonumber \\
&\times \left|\vartheta_3^3(0,\tau) \sum_{m\in\mathbb{Z}}\prod_{n=1}^{+\infty}\sum_{p_n,q_n=0}^{+\infty}e^{\frac{4\pi i j}{N} (p_n-q_n+m)}e^{- \frac{\tau_2}{\alpha'}\left(n(p_n+q_n)+m^2/2\right)}\right. \nonumber \\
&\left.\quad - \vartheta_4^3(0,\tau) \sum_{m\in\mathbb{Z}}\prod_{n=1}^{+\infty}\sum_{p_n,q_n=0}^{+\infty}(-)^m e^{\frac{4\pi i j}{N} (p_n-q_n+m)}e^{- \frac{\tau_2}{\alpha'}\left(n(p_n+q_n)+m^2/2\right)} \right. \nonumber \\
&\left.\quad - \vartheta_2^3(0,\tau) \sum_{m\in\mathbb{Z}}\prod_{n=1}^{+\infty}\sum_{p_n,q_n=0}^{+\infty}e^{\frac{4\pi i j}{N} \left(p_n-q_n+m-\frac{1}{2}\right)}e^{- \frac{\tau_2}{\alpha'}\left(n(p_n+q_n)+(m-1/2)^2/2\right)}\right|^2,
\end{align}
where we dropped the irrelevant non-thermal $j=0$ contribution. The main difference with the actual stringy result is the change in modular integration region (the entire strip region $\mathcal{E}$) and the single summation over $j$. Note that this expression is not modular invariant. \\
Next, we need to perform the derivative and compute the entropy. We will not write down the analogous (extremely long) expression. We only mention that the analogous factor in brackets can be computed explicitly and it is non-vanishing. For the type IIA case for instance, one finds that its expansion for large $\tau_2$ starts with $32q$. This coefficient of the prefactor can again be decomposed as
\begin{equation}
16\left(\underbrace{\frac{9*8}{12}-8}_{\text{metric+dilaton}} + \underbrace{\frac{8*7}{12}-6}_{\text{Kalb-Ramond}} + \underbrace{\frac{8*8}{6}}_{\text{2 gravitinos}} + \underbrace{\frac{8*8}{6}-16}_{\text{R-R 1 and 3-form}}\right),
\end{equation}
coming from the full supergravity multiplet. \\

\noindent Hence, treating string theory merely as an infinite tower of fields, one finds a non-vanishing expression for both the open string and closed superstring entropy. The behavior of both is however very different: the open string result (presumably) has a divergence as $\tau_2\to 0$. This is not too surprising in hindsight. The divergence comes fully from summing the divergences of each individual field in the spectrum and is not cancelled in the process. \\
Closed superstrings, when viewed as a sum over their particle spectrum, on the other hand give a finite expression for the entropy as $\tau_2\to0$ \cite{He:2014gva}.\footnote{This was analyzed in the Melvin context \cite{He:2014gva} we present in the next section, although their result is equally valid for the above expression. The main differences are one factor of $\sqrt{\tau_2}$ and the factor $V_8$ here present in the overall prefactor. Both of these have no effect on the final conclusion of the analysis of \cite{He:2014gva}.} It is to be remarked that this limit contains contributions from the entire tower of string states and the finiteness crucially relies on the presence of an \emph{infinite} tower of states. \\
Let us see what this finiteness means for the individual fields in the closed string spectrum. Schematically, a field of mass $m$ and spin $\mathcal{S}$ in the Rindler spectrum makes a contribution to the entropy of the form \cite{Solodukhin:2015hma}
\begin{equation}
\label{pfparticle}
S \sim \frac{V_{d-2}}{2}\left(\frac{D_{\mathcal{S},m}(d)}{6} - c_{\mathcal{S},m}(d)\right)\int_{\epsilon}^{+\infty}\frac{ds}{s^{d/2}}e^{-m^2 s}.
\end{equation}
The first factor contains the contribution of a bulk piece $D_{\mathcal{S},m}(d)$, proportional to the number of degrees of freedom, and a contact term $c_{\mathcal{S},m}(d) \geq 0$. The small $s$-expansion starts as
\begin{equation}
S \sim \frac{V_{d-2}}{2}\left(\frac{D_{\mathcal{S},m}(d)}{6} - c_{\mathcal{S},m}(d)\right)\left(\frac{1}{\epsilon^{\frac{d-2}{2}}} - \frac{m^2}{\epsilon^{\frac{d-4}{2}}} + \hdots \right),
\end{equation}
and is an expansion in $m^2\epsilon$. For non-zero mass, both subdominant (but divergent) and finite terms are present. \\
This contact term and its origin in state-counting has been studied extensively in the past years \cite{Kabat:1995eq}\cite{Kabat:1995jq}\cite{Kabat:2012ns}\cite{Donnelly:2012st}\cite{Donnelly:2014fua}\cite{Donnelly:2015hxa}. \\
The fact that the small $\tau_2$ limit in string theory turns out to be finite, means that the divergent parts of all fields in the string spectrum cancel out as one takes the sum. Note that this requires a cancellation at each fixed (negative) order in $\epsilon$. The finite parts do remain however. \\
For $d=10$, this cancellation at the level of the individual fields requires the non-trivial equalities
\begin{align}
\sum_{i \in \mathcal{H}_{\text{Rindler}}}m_i^{2p}\left(\frac{D_{\mathcal{S}_i,m_i}(d)}{6} - c_{\mathcal{S}_i,m_i}(d)\right) = 0, \quad p=0,1,2,3.
\end{align}
The summation is over the full physical (Lorentzian signature) string Hilbert space $\mathcal{H}_{\text{Rindler}}$.\footnote{The Rindler string Hilbert space has precisely the same oscillator structure as the flat space string Hilbert space \cite{deVega:1987um}. This is also seen by the nice agreement in the open string sector above, where the particle partition functions are summed over the flat space spectrum.} \\

\noindent Even though the closed string result is hence finite, closed string theory is more than just the sum of its constituent fields (unlike open string theory) and hence a more careful analysis is required. The analyses in the following subsections will indicate that the type II result actually vanishes. \\
One way of anticipating this already is that for the partition functions on the flat $\mathbb{Z}_N$ cones, the sum over QFTs is infinitely larger than the stringy result, interpreted geometrically as an infinite overcounting of tori worldsheets \cite{Mertens:2015adr}. This is of course for the partition function and not for the entropy, but one would expect this same feature to be present as well, since it is linked to a geometric (worldsheet) feature. \\

\noindent In \cite{He:2014gva}, it was argued that the one-loop entropy should vanish. This was argued to be in unison with the viewpoint of Susskind and Uglum that sufficiently supersymmetric backgrounds experience no renormalization in Newton's constant and should hence have vanishing quantum corrections to the entropy \cite{Susskind:1994sm}. The equivalence of quantum corrections to the Bekenstein-Hawking entropy and the renormalization of Newton's constant has a long history (see e.g. \cite{Susskind:1994sm}\cite{Fursaev:1994ea}\cite{Solodukhin:1995ak}\cite{Solodukhin:2011gn}\cite{Solodukhin:2015hma}). \\
Let us next explore the approach of \cite{He:2014gva} in greater detail.

\subsection{Method 2: Melvin Regularization}
The Melvin model represents the space $\frac{\mathbb{C}\times S_1}{\mathbb{Z_N}} \times \mathbb{R}^7$. The orbifolding procedure does not have any fixed points because of the simultaneous rotation along the $S_1$. This immediately removes all intricacies of the model and hence provides a well-behaved regularization. \\
It has been shown in \cite{Takayanagi:2001jj} that this model experiences some quite pathological features in the $R\to0$ limit. For instance, it depends crucially on whether $N$ is rational or not. In the latter case, one finds tachyons arbitrarily close to the standard closed string tachyon mass. So even though it is a modular invariant model, this pathology makes us question its relevance in general. \\
The Melvin model partition function (with the radius of the $S_1$ being $NR$) for type II superstrings is given by
\begin{equation}
\label{melv}
Z_R = Z_0 R \int_{\mathcal{F}}\frac{d^2\tau}{\tau_2^5}\sum_{m,w\in\mathbb{Z}}e^{-\frac{\pi R^2}{\alpha'\tau_2}\left|m-w\tau\right|^2}\frac{\left|\vartheta_1\left(\frac{m-w\tau}{N},\tau\right)\right|^8}{\left|\eta(\tau)\right|^{18}\left|\vartheta_1\left(2\frac{m-w\tau}{N},\tau\right)\right|^2},
\end{equation}
with $Z_0$ an $R$- and $N$-independent prefactor: $Z_0 \sim \frac{V_7}{\alpha'^4}$. The two building blocks of this partition function (the $\mathbb{C}/\mathbb{Z}_N$ model and the $S_1$)  are clearly visible and are coupled through the $m$ and $w$ quantum numbers. \\
Given this expression, one can go two ways, one wrong and the other as given in \cite{He:2014gva} and in agreement with the argument we will present in section \ref{meth3} further on. It is instructive to explore both in more detail. \\

\noindent Taking the above expression as a function of $N$, it is very intriguing that this partition function is modular invariant for any real $N$. This in contrast to the $\mathbb{C}/\mathbb{Z}_N$ orbifold models that only exhibit modular invariance for integer $N$. \\
For integer values of $N$, upon taking the limit of $R\to0$, (\ref{melv}) agrees with the $\mathbb{C}/\mathbb{Z}_N \times \mathbb{R}^8$ models, where the additional eighth flat direction is found as the $R\to0$ limit of $L = \frac{\sqrt{\alpha'}}{NR}$, with $V_8 = LV_7$. \\
It is hence very tempting to define the entropy by computing it first for non-zero $R$ directly in the above model, and only in the end take the $R\to0$ limit. This provides an on-shell (i.e. modular invariant) calculation of the entropy. This route however is not the correct one as we will demonstrate in several ways here. \\

\noindent The entropy $S_R=\left.\partial_N(NZ_R)\right|_{N=1}$ can be immediately computed. Note that this is not really an entropy as the topology of the space is different when $R\neq0$, but still we might expect it to turn into the Rindler entropy in the $R\to0$ limit. To obtain it, first notice that upon setting $N=1$ in $Z_R$ itself (equation (\ref{melv})), one finds a vanishing result due to
\begin{equation}
\vartheta_1(m-w\tau,\tau) \sim \vartheta_1(0,\tau) = 0.
\end{equation}
The denominator also vanishes, but this does not alter the fact that the full partition function will vanish as well. This can be appreciated in several ways. The first is that the background for $N=1$ is flat space times a circle, which is spacetime supersymmetric and hence needs to have $Z_R=0$. A second way is to note that even though both numerator and denominator vanish separately, their ratio still vanishes by elementary calculus. A third way to appreciate this is to note that, in the $R\to0$ limit, the $N=1$ limit corresponds to the untwisted contribution, which in principle diverges due to the $\vartheta_1$ in the denominator. However, this divergence is to interpreted as the transverse area of the plane and is hence an IR volume divergence. \\
In any case, this vanishing happens for any $R$ and in particular will happen in the $R\to0$ limit as well. \\

\noindent Upon computing the entropy $S_R$, it is immediate that $S_R$ will vanish as well, simply because the power of $\vartheta_1$ in the numerator is high enough, and this for any value of $R$. \\
Finally, one is tempted to declare that
\begin{equation}
\label{wrongprop}
S \stackrel{?}{=} \lim_{R\to0}S_R
\end{equation}
is the entropy of Rindler space. This however is wrong: the limit is not smooth. \\
A first way of appreciating this is to perform the analogous computation for open superstrings. 
The open superstring partition function on the flat $\mathbb{Z}_N$ orbifold is given by
\begin{equation}
Z \sim \frac{V_8}{N} \int^{+\infty}_0 \frac{dt}{2t} t^{-4} \sum_{j=1}^{N}\frac{\vartheta_1\left(\frac{j}{N},it\right)^4}{N\sin(2\pi j/N)\vartheta_1\left(\frac{2j}{N},it\right)\eta(it)^9}.
\end{equation}
This partition function can be modified into a Melvin model as
\begin{equation}
Z_R \sim V_8 \int^{+\infty}_0 \frac{dt}{2t} t^{-4} \frac{R}{\sqrt{\alpha' t}}\sum_{j\in\mathbb{Z}}e^{-\frac{\pi R^2}{\alpha't}j^2}\frac{\vartheta_1\left(\frac{j}{N},it\right)^4}{N\sin(2\pi j/N)\vartheta_1\left(\frac{2j}{N},it\right)\eta(it)^9}.
\end{equation}
The associated entropy $S_R=\partial_N(NZ_R)$ evaluated at $N=1$ is readily seen to vanish again due to $\vartheta_1(0,it)=0$ for any $R$.
But the open superstring entropy was computed in the previous section (equation (\ref{nonintpffo})) and did not vanish! \\
It is instructive to see this inequivalence in a simple mathematical toy model. Say we want to compute the ``entropy" of the following ``partition function"
\begin{equation}
\label{toyp}
Z = \frac{1}{N}\sum_{j=0}^{N-1}\sin^2\left(\frac{\pi j }{N}\right).
\end{equation}
One finds explicitly that $S=\left.\partial_N(NZ)\right|_{N=1} = \frac{1}{2}$. Alternatively, one constructs a Melvin model and computes
\begin{equation}
Z_R = R\sum_{n\in\mathbb{Z}}e^{-\pi R^2 n^2}\sin^2\left(\frac{\pi n }{N}\right).
\end{equation}
Upon setting $n=j+Nm$ ($j:0\to N-1$), Poisson resumming $m$ and taking the small $R$ limit, one finds agreement with the above ``partition function" (\ref{toyp}). However, computing the ``entropy" directly will give zero, since $\sin(\pi n)=0$ for any $n$. Both results disagree: even though the Melvin model (\ref{melv}) gives the correct $\mathbb{C}/\mathbb{Z}_N$ partition functions in the $R\to0$ limit, the entropy is incorrect when comparing it to the direct evaluation. \\

\noindent Let's go back to the partition function on the Melvin model (\ref{melv}) and use it solely to extend the modular domain as in \cite{He:2014gva}, after which we immediately take the small $R$ limit, before computing the entropy. This provides an alternative approach, and is the one followed in \cite{He:2014gva}. Transforming the modular fundamental domain $\mathcal{F}$ into the strip $\mathcal{E}$ using the standard unfolding theorem \cite{McClain:1986id}\cite{O'Brien:1987pn}, one obtains
\begin{equation}
Z_R = Z_0 R \int_{\mathcal{E}}\frac{d^2\tau}{\tau_2^5}\sum_{n\in\mathbb{Z}}e^{-\frac{\pi R^2}{\alpha'\tau_2}n^2}\frac{\left|\vartheta_1\left(\frac{n}{N},\tau\right)\right|^8}{\left|\eta(\tau)\right|^{18}\left|\vartheta_1\left(\frac{2n}{N},\tau\right)\right|^2}.
\end{equation}
Again splitting the index as $n=j+Nm$ ($j:0\to N-1$) and taking the small $R$ limit, one obtains
\begin{equation}
Z_{\text{strip}} = Z_0 \frac{1}{N}\int_{\mathcal{E}}\frac{d^2\tau}{\tau_2^5}\sqrt{\alpha'\tau_2}\sum_{j=0}^{N-1}\frac{\left|\vartheta_1\left(\frac{j}{N},\tau\right)\right|^8}{\left|\eta(\tau)\right|^{18}\left|\vartheta_1\left(\frac{2j}{N},\tau\right)\right|^2},
\end{equation}
where no $R$-dependence is left in the prefactor. This partition function $Z_{\text{strip}}$ is crucially only proportional to $V_7$, not to $V_8$. This expression is almost precisely the same as that one would write down by summing over fields in the spectrum (as in the previous subsection \ref{metho1} (first line of equation (\ref{nonintpff}))), except for the prefactor $V_8 \to V_7$ and an additional factor of $\sqrt{\alpha'\tau_2}$ in the integrand. \\

\noindent Next, one computes the entropy $S_{\text{strip}} = \left.\partial_N(NZ_{\text{strip}})\right|_{N=1}$ of this partition function.\footnote{For which one can drop the $j=0$ non-thermal contribution.} Note however, that it is not modular invariant anymore for $N$ non-integer. Moreover, the computation of the derivative requires a specific prescription to define the continuation in $N$. \\

\noindent When considering the string partition function on the conical $\mathbb{C}/\mathbb{Z}_N$ orbifolds, we noted in \cite{Mertens:2015adr} that one can unfold the fundamental domain, at the cost of including an overall infinity. We found there that an infinite overcounting of tori is present for the thermal part of the partition functions (and hence the entropies). We can write, very schematically:\footnote{The prime on the summation index means that $m=w=0$ is excluded.}
\begin{equation}
\int_{\mathcal{E}}d^2\tau \sum_{j=1}^{N-1} f(\tau,j) \,\, \sim \,\, \infty \int_{\mathcal{F}}d^2\tau {\sum^{N-1}_{m,w=0}}'f(\tau,m+w\tau),
\end{equation}
transforming a modular expression in the strip domain $\mathcal{E}$ to one in the fundamental domain $\mathcal{F}$. We see here that the above Melvin procedure can be viewed as merely regularizing this transition, where this ``infinity" is given by $L$, the length of the transverse $8$-dimension. Besides this, there is an additional factor of $\sqrt{\alpha'\tau_2}$ present, which is necessary for dimensional reasons. \\
Now, if the strip modular expression would turn out to be finite, then the fundamental domain expression would automatically need to vanish. This would mean the actual entropy $S$ would vanish. \\
And in \cite{He:2014gva}, it was checked very carefully that indeed the above strip entropy $S_{\text{strip}}$ is finite. On a technical level, this was possible only due to the transition to the strip domain, which in turn was possible only by using this Melvin model approach. This was the main reason for considering this method. Hence the one-loop entropy $S$ vanishes by this argument. \\
An alternative way of formulating this \cite{He:2014gva}, is that the finiteness of the strip modular expression demonstrates that $S_{\text{strip}} \sim V_7$, which is not proportional to $V_8$ as one would expect, and hence the entropy $S$ actually vanishes.

\subsubsection{Discussion}
To summarize, we learned that
\begin{equation}
\lim_{R\to 0}S_R \neq S,
\end{equation}
or taking the $R\to0$ limit does not commute with taking the $\frac{\partial}{\partial N}$ derivative. \\
The most important ramification is that one should not be tempted into trusting the modular invariant formula $S_R$ to define the one-loop entropy. \\
One can also use this Melvin model regularization as a trick to perform the unfolding procedure, as highlighted above. This leads to a strip entropy $S_{\text{strip}}$ whose finiteness is the important feature. When treated this way, one finds $S=0$ in the end, albeit in a rather indirect way. \\

\noindent It is important to emphasize though that, although this method is capable of demonstrating the vanishing of the entropy, for systems that do not exhibit enough spacetime supersymmetry where this vanishing does not happen, it is unclear how to obtain a precise numerical value for the entropy. \\
A somewhat unsatisfactory feature of this approach is that the final vanishing of the entropy is supposed to be a consequence of spacetime supersymmetry; yet the employed method does not make this manifest: the finiteness of the ``entropy" in the strip $S_{\text{strip}}$ seems like a happy coincidence. \\
\noindent On a related note, one can readily check that when trying to apply this argument to bosonic strings, one finds a divergent result for both the large $\tau_2$ and small $\tau_2$ regions. However, this divergence is directly related to the presence of the bosonic string tachyon, but one would naively expect the closed string tachyon not to be the reason of the divergence, as it should simply be a consequence of the absence of (a sufficient amount of) spacetime supersymmetry. \\

\noindent As a final remark, one should be careful when trying to apply this argument to different models, as a divergence in $S_{\text{strip}}$ is \emph{not} a pathology of the model: it might get cancelled by the infinite prefactor. A finite strip result leads to a vanishing entropy. A divergent strip result can lead to anything. 

\subsubsection{The Melvin model of the cigar CFT}

\noindent This vanishing of $S$ is supposedly directly linked to the spacetime supersymmetry and the non-renormalization of Newton's constant. It is hence of interest to look at the entropy in a black hole background geometry that is not supersymmetric and find out what the most dominant contribution is in that case. The cigar CFT itself (finite $k$) represents a non-supersymmetric black hole background. \\

\noindent The partition function on the cigar orbifold model $\frac{SL(2,\mathbb{R})_k/U(1)}{\mathbb{Z_N}}$is given by \cite{Israel:2004ir}\cite{Eguchi:2004yi}\cite{Sugawara:2012ag}
\begin{align}
\label{cigarpt}
Z(\tau) = &\frac{k}{4N} \sum_{\sigma_L, \sigma_R}\sum_{w,m =0}^{N-1} \int_{-\infty}^{+\infty}ds_1ds_2 \epsilon(\sigma_L;w,m)\epsilon(\sigma_R;w,m) \nonumber \\
&\times f_{\sigma_L}(s_1\tau+s_2,\tau)f_{\sigma_R}^{*}(s_1\tau+s_2,\tau)e^{-\frac{\pi k}{\tau_2}\left|\left(s_1 - \frac{w}{N}\right)\tau + \left(s_2 - \frac{m}{N}\right)\right|^2}.
\end{align}
where 
\begin{equation}
f_{\sigma}(u,\tau) = \frac{\vartheta_{\sigma}(u,\tau)}{\vartheta_1(u,\tau)}\left(\frac{\vartheta_{\sigma}(0,\tau)}{\eta}\right)^3,
\end{equation}
where $\vartheta_{\sigma} = \vartheta_{1,2,3,4}$ for $\sigma$ = $\widetilde{\text{R}}$, $\text{R}$, $\text{NS}$, $\widetilde{\text{NS}}$ respectively and
$\epsilon = (1, (-1)^{w+1}, (-1)^{m+1}, (-1)^{w+m})$ for ($\text{NS}$, $\widetilde{\text{NS}}$, $\text{R}$, $\widetilde{\text{R}}$) respectively. \\

\noindent The associated Melvin model of the cigar $\frac{SL(2,\mathbb{R})/U(1) \times S_1}{\mathbb{Z_N}}$ can be written as
\begin{align}
\label{cigarpt}
Z(\tau) = & Z_0 R\frac{k}{4N} \sum_{\sigma_L, \sigma_R}\sum_{w,m =-\infty}^{+\infty} \int_{-\infty}^{+\infty}ds_1ds_2 \epsilon(\sigma_L;w,m)\epsilon(\sigma_R;w,m) \nonumber \\
&\times f_{\sigma_L}(s_1\tau+s_2,\tau)f_{\sigma_R}^{*}(s_1\tau+s_2,\tau)e^{-\frac{\pi R^2}{\alpha'\tau_2}\left|m-w\tau\right|^2} e^{-\frac{\pi k}{\tau_2}\left|\left(s_1 - \frac{w}{N}\right)\tau + \left(s_2 - \frac{m}{N}\right)\right|^2}.
\end{align}
Some checks can be performed at this point. The first is that the $R\to0$ limit agrees with the cigar orbifold partition function (up to a prefactor coming from the $S_1$). \\
It is immediate again that this expression is modular invariant for any real $N$. One can also immediately check that the large $k$ limit of this partition function indeed agrees with the $\frac{\mathbb{C}\times S_1}{\mathbb{Z}_N}$ Melvin model, since this reduces the $s_1$ and $s_2$ integrals to just their saddle at $s_1=w/N$ and $s_2=m/N$. \\

\noindent Executing the unfolding procedure, one arrives at the strip partition function:
\begin{align}
\label{cigarptstr}
Z(\tau) = & Z_0 R\frac{k}{4} \sum_{\sigma_L, \sigma_R}\sum_{m =-\infty}^{+\infty} \int_{-\infty}^{+\infty}ds_1ds_2 \epsilon(\sigma_L;m)\epsilon(\sigma_R;m) \nonumber \\
&\times f_{\sigma_L}(s_1\tau+s_2,\tau)f_{\sigma_R}^{*}(s_1\tau+s_2,\tau)e^{-\frac{\pi R^2}{\alpha'\tau_2}m^2} e^{-\frac{\pi k}{\tau_2}\left|s_1\tau + \left(s_2 - \frac{m}{N}\right)\right|^2}.
\end{align}
Again splitting the index into $m=j+Nn$, performing a Poisson summation on $n$ and taking the small $R$ limit, one finds\footnote{The periodicity of the expression in $m\to m+N$ can be seen by shifting $s_2$ by 1 and realizing that $\vartheta_1(\nu+1,\tau) = - \vartheta_1(\nu,\tau)$ and $\vartheta_2(\nu+1,\tau) = - \vartheta_2(\nu,\tau)$ whereas $\vartheta_3$ and $\vartheta_4$ are invariant under this shift of $\nu$.}
\begin{align}
\label{cigarptstr}
Z(\tau) = & Z_0  \frac{k}{4N} \sum_{\sigma_L, \sigma_R}\sum_{j =0}^{N+1}\sqrt{\alpha'\tau_2} \int_{-\infty}^{+\infty}ds_1ds_2 \epsilon(\sigma_L;j)\epsilon(\sigma_R;j) \nonumber \\
&\times f_{\sigma_L}(s_1\tau+s_2,\tau)f_{\sigma_R}^{*}(s_1\tau+s_2,\tau) e^{-\frac{\pi k}{\tau_2}\left|s_1\tau + \left(s_2 - \frac{j}{N}\right)\right|^2}.
\end{align}
The entropy associated to this partition function should be divergent as $\tau_2\to0$. Since only then, can the original partition function in the fundamental domain be non-zero, as it should be due to the absence of spacetime supersymmetry on the cigar. \\
One can easily see that this partition function itself is divergent. Choose $\tau_1=0$ and fix $N$ and $j$ and $s_2=j/N$. Then the small $\tau_2$ asymptotics of the theta-functions can never compensate for the divergence induced by the $\eta$-functions. \\
As the partition function itself is divergent as $\tau_2\to0$ for any $N$, it seems that the strip entropy $S_{\text{strip}}$ should diverge as well, leaving open the possibility indeed of a finite non-zero entropy $S$. A more detailed computation is made difficult due to the quadratic exponential summation over $j$ to be performed. \\
We postpone a more detailed analysis of this model to possible future work.

\subsection{Method 3: Saddle point approach for the entropy}
\label{meth3}
In \cite{Mertens:2014saa} we have demonstrated that the cigar orbifold CFT allows us to write down a candidate formula for $N$ non-integer, and that the large $k$ limit of this partition function then provides a natural candidate for the flat conical backgrounds for \emph{arbitrary} deficit angles. In this section, we will put that formula to the test, by computing the entropy in this way for the $\mathbb{Z}_N$-orbifolds of the cigar and then taking the large $k$ limit. We will find a vanishing one-loop entropy as well, in agreement with \cite{He:2014gva}. \\

\noindent So here we propose the procedure of computing the entropy at finite $k$, and in the end taking the infinite $k$ limit:
\begin{equation}
S = \lim_{k\to+\infty} S_k.
\end{equation}
In comparison to (\ref{wrongprop}), this approach is not sensitive to subtleties about the order of taking the limits. Physically, the reason is that at the very get-go we are already on a cigar-shaped manifold and we are hence computing a legitimate candidate for the entropy of a black hole for any finite $k$. \\ In other words, optimistically put, if the proposed formula is correct, it will be correct for any $k$. \\

\noindent The type II superstring partition function on the cigar orbifold is of the form \cite{Israel:2004ir}\cite{Eguchi:2004yi}\cite{Sugawara:2012ag}
\begin{align}
\label{pfcigar}
Z_N(\tau) = \frac{k}{4N}\sum_{m,w \in \mathbb{Z}}\int_{0}^{1}ds_1 ds_2 &\frac{e^{-\frac{\pi k}{\tau_2}\left|\left(s_1 - \frac{w}{N}\right)\tau + \left(s_2 - \frac{m}{N}\right)\right|^2}}{\left|\vartheta_1(u,\tau)\eta^3\right|^2} \nonumber \\
&\times \left\{\vartheta_3(u,\tau) \vartheta_3^3 + (-)^{w+1} \vartheta_4(u,\tau) \vartheta_4^3 + (-)^{m+1}\vartheta_2(u,\tau) \vartheta_2^3\right\} \nonumber\\
&\times \left\{\bar\vartheta_3(u,\tau) \bar\vartheta_3^3 + (-)^{w+1} \bar\vartheta_4(u,\tau) \bar\vartheta_4^3 + (-)^{m+1}\bar\vartheta_2(u,\tau) \bar\vartheta_2^3\right\},
\end{align}
where $u=s_1\tau + s_2$.\footnote{One still needs to integrate this over the fundamental modular domain $\mathcal{F}$ afterwards, but we will refrain from writing this in this subsection. We also chose not to write the internal CFT at this point: its presence is immaterial for the argument we will present here.} This expression is suitable to generalize to non-integer $N$ \cite{Mertens:2014saa}. In terms of the individual states and the character decomposition at $k\to\infty$, this continuation can be interpreted as continuing each mode in $N$, and providing a sharp truncation on the allowed modes as they enter or exit the allowed physical interval \cite{Mertens:2014saa}. \\
In the past, a continuation in $N$ was proposed as well for the flat $\mathbb{C}/\mathbb{Z}_N$ models \cite{Dabholkar:1994ai}\cite{Lowe:1994ah}, though the continuation itself was not found (mainly due to the integer $N$ appearing directly in the summations for the $\mathbb{C}/\mathbb{Z}_N$ models). With the above expression, at least this step is solved by going to the cigar model instead, and only in the end taking the large $k$ limit. \\

\noindent While it is unclear whether this continuation is the same as the Melvin continuation discussed above, it has equal claim to be valid and is certainly the natural candidate for the entropy on the $SL(2,\mathbb{R})_k/U(1)$ black hole model itself. \\

\noindent Our strategy will be to differentiate this expression with respect to $N$ to obtain the entropy on the cigar CFT, and afterwards take the large $k$ limit. Just as mentioned in the previous subsections, all of these approaches should be taken with a grain of salt, since apparently only string theory on $\mathbb{C}/\mathbb{Z}_N$ with $N$ an integer makes sense. This obscures any attempt at even correctly defining the entropy at the very start. This is a very difficult problem and we will not say anything new about this here. \\ Anyhow, let us proceed and see what we find. The resulting expression for the entropy $S = \left.\partial_N(NZ_N)\right|_{N=1}$ is then equal to \\
\begin{align}
\label{entrrr}
S = &\frac{k}{4} \sum_{m,w \in \mathbb{Z}}\int_{0}^{1}ds_1 ds_2 \frac{e^{-\frac{\pi k}{\tau_2}\left|\left(s_1 - w\right)\tau + \left(s_2 - m\right)\right|^2}}{\left|\vartheta_1(u,\tau)\eta^3\right|^2} \nonumber \\ 
&\times \left[-\frac{\pi k}{\tau_2}\left\{\left((s_1-w)\tau+(s_2-m)\right)\left(w\bar\tau+m\right) + \left((s_1-w)\bar{\tau}+(s_2-m)\right)\left(w\tau+m\right)\right\}\right] \nonumber \\ 
&\times \left|\vartheta_3(u,\tau) \vartheta_3^3 + (-)^{w+1} \vartheta_4(u,\tau) \vartheta_4^3 + (-)^{m+1}\vartheta_2(u,\tau) \vartheta_2^3\right|^2.
\end{align}

\begin{figure}[h]
\centering
\includegraphics[width=0.4\textwidth]{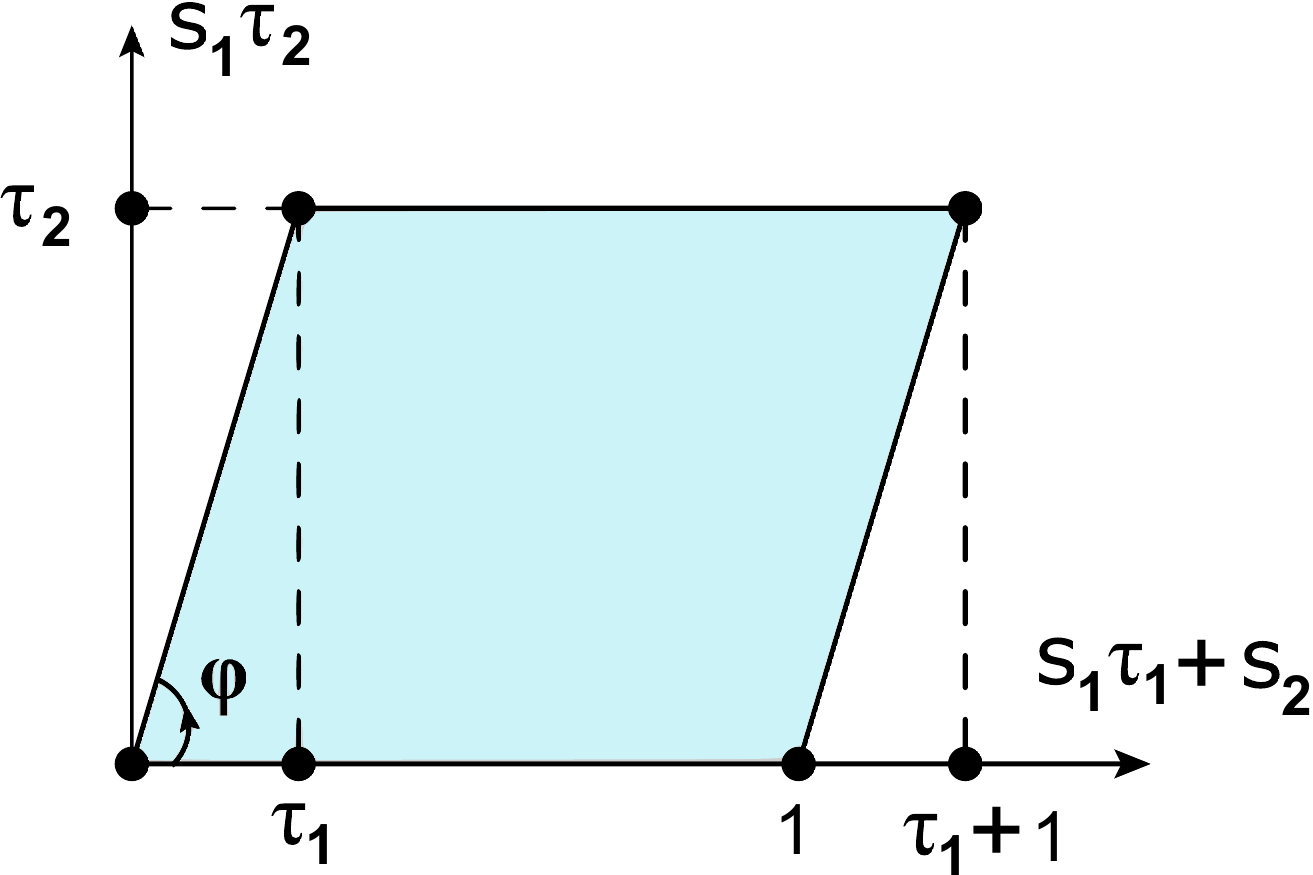}
\caption{The integration region when $w=m=0$ is the parallelogram described in polar coordinates. In particular the opening angle is $\text{Arctan}(\tau_2/\tau_1)$. For the other three regions, the same integration region is used, but the polar coordinate frame is centered around each of the other three vertices.}
\label{s1s2}
\end{figure}

\noindent There are four saddle points of the $s_1$- and $s_2$-integrals as $k\to+\infty$, at $w=m=0$, $w=1, m=0$, $w=0, m=1$ and $w=m=1$. All four of these are on the border of the integration region. For each of these four sectors, we make the transition to polar coordinates as
\begin{align}
\left(s_1 - w\right)\tau_1 + \left(s_2-m\right) &= \rho \cos\varphi , \\
\left(s_1 - w\right)\tau_2 &= \rho \sin\varphi.
\end{align}
Since for each sector, the dominant region is located around $\rho\approx 0$, the integrals get transformed to 
\begin{equation}
\int_{0}^{1}ds_1 ds_2 \to \int_{0} d\rho \frac{\rho}{\tau_2}\int_{I_{w,m}}d\varphi,
\end{equation}
where for each $w$ and $m$, the interval of integration $I_{w,m}$ for $\varphi$ varies and is given by
\begin{align}
w=0, m=0 \quad \Rightarrow \quad &\varphi: 0 \to \text{Arctan}(\tau_2/\tau_1) ,\\
w=0, m=1 \quad \Rightarrow \quad &\varphi: \text{Arctan}(\tau_2/\tau_1) \to \pi, \\
w=1, m=0 \quad \Rightarrow \quad &\varphi: -\pi + \text{Arctan}(\tau_2/\tau_1) \to 0, \\
w=1, m=1 \quad \Rightarrow \quad &\varphi: \pi \to \pi + \text{Arctan}(\tau_2/\tau_1).
\end{align}
In more detail, the integration region for the $w=m=0$ case is described by polar coordinates in the parallelogram in figure \ref{s1s2}.
For the other three cases, one employs polar coordinates around the other three vertices of the parallelogram. \\

\noindent The entropy in any one of these four sectors can then be rewritten as
\begin{align}
S_{w,m} = \frac{k}{4} \iint d\varphi d\rho \left(-\frac{\pi k \rho}{\tau_2^2}\right)&\frac{e^{-\frac{\pi k}{\tau_2}\rho^2}}{\left|\eta\right|^6}\left\{\rho e^{i\varphi}\left(w\bar\tau+m\right) + \rho e^{-i\varphi}\left(w\tau+m\right)\right\} \nonumber \\
&\times \frac{(\text{theta functions})}{\left|\vartheta_1(u,\tau)\eta^3\right|^2}.
\end{align}
The functions in the second line depend on $u=s_1\tau + s_2 = \rho e^{i\varphi} + w\tau + m$ and hence also depend on $\rho$. In the large $k$ limit, the integral is dominated by the saddle point at $\rho=0$. The function on the second line should hence be Taylor-expanded around $\rho=0$ to obtain a series of increasingly subdominant terms. \\
In fact, the theta function in the denominator contains a double pole at $\rho=0$ for all four saddles, and this is the most dominant contribution.  

\subsubsection{Most dominant term}
The most dominant term is obtained by simply setting $u=s_1\tau + s_2 = w\tau + m$.
For any of the four saddle points, the theta function combination vanishes (due to the Jacobi identity) and hence the most dominant term actually vanishes. \\

\noindent For $\rho \approx 0$, the $\varphi$-integral is fixed and gives some $\rho$-independent value.\footnote{The sum of the four sectors yields for the $\varphi$-integral:
\begin{align}
\label{phiint}
&\int_{-\pi + \text{Arctan}(\tau_2/\tau_1)}^{0}d\varphi \left(e^{i\varphi}\bar\tau + e^{-i\varphi}\tau\right) + \int_{\pi}^{\pi + \text{Arctan}(\tau_2/\tau_1)}d\varphi \left(e^{i\varphi}(\bar\tau+1) + e^{-i\varphi}(\tau+1)\right) \nonumber \\
&+\int_{\text{Arctan}(\tau_2/\tau_1)}^{\pi}d\varphi \left(e^{i\varphi} + e^{-i\varphi}\right) = -4\tau_2\left(1+ \frac{1}{\left|\tau\right|}\right).
\end{align}
This integral is manifestly negative. This is reassuring since the entropy, if non-zero, should preferably be positive and this minus sign cancels the one in front of the full expression from the derivative w.r.t. $N$.
}
The $\rho$-integral can be done and gives (upon including all $k$-dependent prefactors):
\begin{equation}
\sim \int_{0}^{}d\rho k\frac{\pi k}{\tau_2} e^{- \frac{\pi k}{\tau_2}\rho^2}\frac{\rho^2}{\tau_2}\frac{1}{\rho^2} = \frac{\pi k^{3/2}}{2 \tau_{2}^{3/2}}.
\end{equation}
But, as said above, this gets multiplied by a theta-function combination that vanishes. \\

\noindent It is instructive to write down the full expression now as:
\begin{equation}
S = \frac{k^{3/2}}{8\tau_2^{1/2}}\left(1+\frac{1}{\left|\tau\right|}\right)\left[\underbrace{\left|\vartheta_3^4-\vartheta_4^4\right|^2}_{\text{NS-NS}} + \underbrace{\left|\vartheta_2^4\right|^2}_{\text{R-R}} - \underbrace{\left(\vartheta_3^4-\vartheta_4^4\right)\bar{\vartheta_2}^4}_{\text{NS-R}}- \underbrace{\vartheta_2^4\left(\bar{\vartheta_3}^4-\bar{\vartheta_4}^4\right)}_{\text{R-NS}}\right] = 0.
\end{equation}
From this it is clear that the cancellation of terms in the entropy happens due to a Bose-Fermi cancellation: spacetime fermionic sectors (NS-R and R-NS) contribute negatively to the entropy, whereas the bosonic sectors NS-NS and R-R contribute positively with precisely the same magnitude such that a cancellation occurs. Note that to get information on the absolute sign of each of these contributions, we did require the explicit result (\ref{phiint}). \\
We do remark that it is quite unorthodox for such a cancellation to occur within the entropy, but as mentioned before it is not ruled out due to the fact that this set-up does not correspond to a non-interacting Hamiltonian trace $\text{Tr}e^{-\beta H}$.

\subsubsection{Subdominant terms}
The next three terms are respectively proportional to $k$, $\sqrt{k}$ and 1. They are the only possible finite contribution. Higher order terms are of order $1/\sqrt{k}$ and vanish in the large $k$ limit. Hence we are interested in only these terms. \\

\noindent Upon contemplating the expansion in $\rho$ of
\begin{align}
\frac{1}{\left|\vartheta_1(u,\tau)\eta^3\right|^2} &\left\{\vartheta_3(u,\tau) \vartheta_3^3 + (-)^{w+1} \vartheta_4(u,\tau) \vartheta_4^3 + (-)^{m+1}\vartheta_2(u,\tau) \vartheta_2^3\right\} \nonumber \\
&\times \left\{\bar\vartheta_3(u,\tau) \bar\vartheta_3^3 + (-)^{w+1} \bar\vartheta_4(u,\tau) \bar\vartheta_4^3 + (-)^{m+1}\bar\vartheta_2(u,\tau) \bar\vartheta_2^3\right\},
\end{align}
as a product of series expansions of the three factors, it is clear that the Jacobi identity severely restricts the most dominant contributions, since
\begin{equation}
\vartheta_3(u,\tau) \vartheta_3^3 + (-)^{w+1} \vartheta_4(u,\tau) \vartheta_4^3 + (-)^{m+1}\vartheta_2(u,\tau) \vartheta_2^3 = 0,
\end{equation}
for $u =w\tau+m$. It turns out that the first derivative of either of the terms in the numerator also vanishes at $u=w\tau+m$:
\begin{equation}
(\partial_\nu\vartheta_3)(u,\tau) \vartheta_3^3 + (-)^{w+1} (\partial_\nu\vartheta_4)(u,\tau) \vartheta_4^3 + (-)^{m+1}(\partial_\nu \vartheta_2)(u,\tau) \vartheta_2^3=0.
\end{equation}
This is non-trivial and is proven explicitly in appendix \ref{appTheta}. To have a finite contribution, we should hence differentiate both factors in the numerator at least two times, making the first possibly non-zero contribution appearing at fourth order in the expansion. But these terms vanish in the large $k$ limit! We hence conclude that no subleading terms survive the large $k$ limit at all. \\
The precise Bose-Fermi cancellation continues to occur at these orders.\footnote{We leave it as an open problem here to check whether bosons continue to contribute positively to $S$ and fermions negatively. This should be straightforward but tedious.} \\

\noindent Since all further subdominant terms vanish as $k\to\infty$, this computation shows that the flat limit has vanishing entropy, in accordance with the Melvin approach followed in \cite{He:2014gva} and in the previous subsection. A vanishing entropy is at odds with the non-interacting thermodynamic interpretation of the entropy, and hence demonstrates once again that negative contributions are present as well; Susskind and Uglum interpret these as exotic open strings with ends stranded on the black hole horizon \cite{Susskind:1994sm}. In \cite{Mertens:2014saa}, we have given an interpretation of this phenomenon from the thermal manifold perspective as that of strings circling the origin but without a definite winding number (a possibility only present in string theory). 

\subsubsection{Comments on uniqueness}
As mentioned above, the continuation to non-integer $N$ is ambiguous and one can readily imagine artificial choices that would make the entropy non-zero in the end. The hidden assumption of course is to follow equation (\ref{pfcigar}) as closely as possible. The above way of defining and computing the entropy seems quite natural, but in fact there is a whole class of equally natural choices (related to the ambiguity in continuing to real $N$). The partition function itself (\ref{pfcigar}) is invariant under the shifts $s_1 \to s_1+1$ and similarly for $s_2$, if $N$ is an integer. Thus we can freely change the integration square in the $(s_1, s_2)$ plane into any $1\times1$ square we like. However, the entropy does depend on this choice! This can be seen directly in equation (\ref{entrrr}) where this shift should normally be compensated by a shift in $w$ and $m$, but the entropy is not invariant under this. Thus one finds in general a different result.\footnote{One can also take any linear combination of these squares where the coefficients sum to 1. \\
As mentioned before, in the large $k$ limit, the continuation proposed in equation (\ref{pfcigar}) corresponds to a sharp truncation on the orbifold sums. The ambiguity that arises then is that for instance
\begin{equation}
\sum_{n=0}^{\left\lfloor N-1\right\rfloor}f(n) \neq  \sum_{n=-\left\lfloor N\right\rfloor}^{-1}f(n), \quad \text{for }f(n+N) = f(n), 
\end{equation}
when $N$ is non-integer. This corresponds precisely to the shifting of the integration square on the cigar model.}
However, in our specific case, the theta-function combination still vanishes by the Jacobi identity. This shows that this vanishing argument will be relatively independent of the choice of continuation in $N$. \\
Perhaps a more sophisticated argument in favor of this, is to insist on having a modular invariant partition function for non-integer $N$ before computing the entropy. One good way of obtaining modular invariants is to start with a given sector that one wants to be included, and then add to this the sum of all of the images under the modular group $PSL(2,\mathbb{Z})$. This is by construction modular invariant. Contemplating what a generic such term would look like, starting from (\ref{pfcigar}) for non-integer $N$, one concludes that the entropy at $N=1$ associated to that particular term also vanishes. Hence even performing a modular completion of (\ref{pfcigar}) leads to a vanishing entropy in the end. \\
In more general language, we can multiply (\ref{pfcigar}) by $(1+f(N))$ where $f(N)$ vanishes for integer $N$ by definition. This leads in general to a different entropy, unless the prefactor already vanishes which it does here.

\subsection{Summary}
\noindent These three separate calculations all include some reservations (mainly related to the ambiguity in considering a continuation to real $N$). However, they each approach the problem from a different angle, and they each shed some light on the problem at hand. Both the Melvin approach and the cigar approach lead to the same result $S=0$. We believe that together they make it indeed plausible that this cancellation indeed occurs, and that the entropy vanishes at one loop. The reader can decide for him-/herself which avenue is preferred. \\ 

\noindent Let us make a small remark here. The entropy of open superstrings diverges due to the sum over the UV divergences of each field in the spectrum, whereas the entropy for type II closed superstrings vanishes. This is in contrast to the partition functions themselves: both vanish due to Jacobi's obscure identity. This difference between open and closed strings (which is related to the polynomial dependence on $\tau_2$ for small $\tau_2$) is technically similar as that which led early research into string thermodynamics in flat space \cite{Alvarez:1986sj} to the conclusion that for open strings, the Hagedorn temperature is limiting, while it is not for closed strings. In that case, the free energy diverges at $T_H$ for open strings, whereas it is finite for closed strings. \\

\noindent So we confirmed here a vanishing genus one entropy for Rindler space. \\
Of course, a real black hole is never truly Rindler, and as we mentioned a few times already, effects that are present only at infinite $k$ should not be taken as being generic for black hole horizons. The appearance of a host of states at the infinite $k$ limit, is a consequence of spacetime supersymmetry in flat space, whereas a generic black hole (no matter how large) breaks these symmetries. Hence the detailed cancellation (that must be present here to have a vanishing entropy) between states will not happen for a generic black hole. The $SL(2,\mathbb{R})_k/U(1)$ black hole indeed demonstrates that for this particular black hole, no cancellation is possible at all and the entropy cannot vanish. The \emph{only} state that is massless for any (large) $k$ is the thermal scalar state (with the identification of the dilaton mode as the negatively wound thermal scalar). It should dominate the entropy and contain the dominant thermodynamic information of the gas of strings around the black hole horizon \cite{Mertens:2013pza}\cite{Mertens:2013zya}\cite{Mertens:2014nca}\cite{Mertens:2014cia}\cite{Mertens:2014dia}\cite{Mertens:2014saa}\cite{Mertens:2015hia}.

\section{UV divergences of QFT and their disappearance in string theory}
\label{UV}
To conclude this work, we would like to come back to some well-established features of black hole physics: within QFT thermodynamics one encounters a UV divergence coming from the region close to the Rindler horizon \cite{'tHooft:1984re}\cite{Susskind:1994sm}. We have explicitly seen that in string theory, this divergence seems absent. However, the reason for its absence could be tracked down due to the presence of (a sufficient amount of) spacetime supersymmetry. In this section, we will try to pinpoint what generically happens to the QFT divergences as one computes one-loop string thermodynamics using Euclidean techniques. We choose to portray our discussion in a more general setting than done up to this point. \\

\noindent This section can be viewed as a synthesis of the results presented in \cite{Mertens:2015adr} on the partition functions one obtains by summing over the fields in the spectrum and the results presented above. It was found there that if one drops all polarization dependence of the higher spin fields in the spectrum and then sums these QFT partition functions over the string spectrum, one obtains a partition function that experiences a thermal maximal acceleration divergence close to the horizon. It was also found there that for closed strings, summing the full contribution for each higher spin field (i.e. including polarization dependence) leads to an expression that is still not equal to the actual stringy result: a mismatch in the modular domain and the thermal summations is present, that cannot be solved by using the unfolding theorems of \cite{McClain:1986id}\cite{O'Brien:1987pn}. \\
This polarization dependence gives rise to an additional \emph{negative} contribution to the entropy, as has been studied extensively for spin 1 gauge fields in Rindler space from a myriad of perspectives \cite{Kabat:1995eq}\cite{Kabat:1995jq}\cite{Kabat:2012ns}\cite{Donnelly:2012st}\cite{Donnelly:2014fua}\cite{Donnelly:2015hxa}. This contribution arises due to the so-called edge states in Lorentzian signature, or due to string worldsheets intersecting the Rindler origin in Euclidean signature \cite{Mertens:2015adr}.

\noindent In Lorentzian Rindler space, each field contributes divergently to the free energy and entropy of the system. The divergence can be regulated with a brick wall at $\rho=\epsilon_B$ and the divergence is ultraviolet since it corresponds to the near-horizon blueshifted region. \\

\noindent We know that string theory has a finite one-loop entropy and hence, as a UV-complete theory, is free of these divergences. We will take a look at what precisely happens to the UV divergences of QFT when these are summed over the spectrum into a stringy expression for the string entropy. \\
The upshot will be that the UV divergence for the entropy and free energy of fields around black holes is removed in the same way as the UV divergence of the cosmological constant at one loop in flat space: by a reduction of the modular domain. \\

\noindent The explanation of this feature directly from the Lorentzian thermal trace $\text{Tr}e^{-\beta H}$ seems much more difficult to make and is left for future work. \\

\noindent Our analysis will be restricted to the Rindler limit and its flat conical $\mathbb{C}/\mathbb{Z}_N$ cousins. \\

\noindent The first step is to realize that the geometric brick wall cut-off $\epsilon_B$ in Lorentzian signature is directly related to a cut-off in the Schwinger proper-time language by regularizing $\int_{0}^{+\infty}\frac{ds}{s} \to \int_{\epsilon_{S}}^{+\infty}\frac{ds}{s}$. \\
This link was studied in $d=4$ in work by Emparan \cite{Emparan:1994qa}.
Emparan's analysis can be readily generalized to any $d$ using a scaling argument, and allows us to link the heat kernel UV cut-off $\epsilon_{S}$ to $\epsilon_{B}$ as\footnote{
In more detail, the brick wall cut-off procedure yields an expression for a massless scalar as
\begin{equation}
F \sim \frac{A}{\epsilon_{B}^{d-2}\beta^d},
\end{equation}
whereas the Euclidean heat kernel approach gives
\begin{equation}
F \sim \frac{A}{\epsilon_{S}^{\frac{d-2}{2}}\beta^2}.
\end{equation}
When $d=2$, extra care is needed. In this case, a logarithmic divergence is present in $\epsilon$ in both regularization procedures and the $\epsilon$'s can be identified.}
\begin{equation}
\epsilon_{S}\sim \epsilon_{B}^2\beta^2
\end{equation}
with a $d$-dependent numerical prefactor (and for $d\neq2$). \\
Finite mass corrections to these expressions contribute always subdominantly. This is particularly transparent in heat kernel language. \\
We take this calculation as an indication that the geometric near-horizon divergence is the same as the small Schwinger proper time divergence in the heat kernel formalism, also for other spins.

\subsection{Divergences in the heat kernel}
Before proceeding, it is illuminating to clarify the different kinds of UV divergences that can appear in the heat kernel formalism.
The cosmological constant in QFT has the form
\begin{equation}
E_0\sim \int_{0}^{+\infty}\frac{ds}{s^{d/2+1}},
\end{equation}
and is UV divergent as is well-known. The flat space free energy in QFT on the other hand is of the form
\begin{equation}
F\sim \int_{0}^{+\infty}\frac{ds}{s^{d/2+1}}e^{-A/s}
\end{equation}
for a positive $A$. This is convergent for $s\to0$ thanks to the exponential damping.
In Rindler space, the free energy in QFT has the form
\begin{equation}
F\sim \int_{0}^{+\infty}\frac{ds}{s^{d/2+1}}
\end{equation}
and has the same type of UV divergence as the cosmological constant, coming from the near-horizon region of black holes. This already strongly suggests the same stringy mechanism is at work here as with the cosmological constant: a reduction in the modular integration domain. \\

\noindent How does string theory behave?
For the cosmological constant, string theory simply uses the fundamental domain $\mathcal{F}$ instead of the strip $\mathcal{E}$, immediately killing the UV divergence. The QFT result is an infinite overcounting of the string result. \\
The free energy in flat space string theory behaves as (for $s\approx0$)
\begin{equation}
F\sim \int_{0}^{+\infty}\frac{ds}{s^{p}}e^{-A/s}e^{+B/s},
\end{equation}
for some $p$ and positive $A$ and $B$. It is usually convergent, unless $T>T_H$, the Hagedorn temperature. At that point, an exponential divergence kicks in.\footnote{The polynomial behavior on its own as $s\to0$ can go either way, which is what lead early literature on string thermodynamics to distinguish between theories where the Hagedorn temperature can be reached in case of convergence at $T=T_H$ (this is the case for closed strings) or cannot be reached (for open strings) \cite{Alvarez:1986sj}. } \\ 
The free energy and entropy in Rindler space, if excluding the polarization contributions of higher spin fields, is badly divergent:
\begin{equation}
F\sim \int_{0}^{+\infty}\frac{ds}{s^{p}}e^{+B/s}, \quad B>0,
\end{equation}
and this is the so-called maximal acceleration divergence \cite{Sakai:1986sc}\cite{Parentani:1989gq}\cite{McGuigan:1994tg}\cite{Barbon:1994ej}\cite{Barbon:1994wa}\cite{Dabholkar:1994gg}\cite{Emparan:1994bt}\cite{Mertens:2015adr}, far worse than the milder UV divergences of each field separately. \\
The proper treatment of string theory in Rindler space involves including the polarization contributions, but even then the result is still an infinite overcounting of the actual stringy computation \cite{Mertens:2015adr}. The latter mismatch in the end is the mechanism responsible for removing the individual UV divergences of each field.

\subsection{Summing over fields}
Consider a fixed higher spin field. Computing its heat kernel on the flat cone, including its polarization contributions, gives a result that has an overall UV divergence: both the non-interacting part and its polarization part have separate UV divergences, as can be seen in equation (\ref{pfparticle}). Both of these terms however contribute with opposite sign to the final outcome. \\
It is possible for cancellations to occur upon summing over the string spectrum, as we saw happens for closed type II superstrings in section \ref{metho1}: the divergences cancel leaving a finite result in the end. \\
This cancellation is however not expected to be true in general, but is a consequence of spacetime supersymmetry. It is not the general mechanism for removing the UV divergences, since if string theory is a UV complete theory, then also SUSY-breaking black hole backgrounds should be free of UV divergences. The way this works can be seen by comparing the flat cone partition functions as a sum over fields to those within string theory. \\
The bosonic closed string partition function on the flat $\mathbb{Z}_N$ orbifold can be written down as 
\begin{align}
\label{fundpf}
Z_{ST} &= V_{d-2}  \int_{\mathcal{F}}\frac{d\tau^2}{4\tau_2}(4\pi^2\alpha' \tau_2)^{-12} \frac{1}{N}{\sum_{m,w=0}^{N-1}}'\frac{\left|\eta(\tau)\right|^{-42}e^{2\pi\tau_2\frac{w^2}{N^2}}}{\left|\vartheta_1\left(\frac{m}{N} + \frac{w}{N}\tau,\tau\right)\right|^2}.
\end{align}
On the other hand, the sum-of-fields approach (including polarization effects of higher spin fields) yields \cite{He:2014gva}\cite{Mertens:2015adr}
\begin{align}
\label{stripf}
Z_{QFT} = V_{d-2}  \int_{\mathcal{E}}\frac{d\tau^2}{4\tau_2}(4\pi^2\alpha' \tau_2)^{-12} \frac{1}{N}\sum_{j=1}^{N-1}\frac{\left|\eta(\tau)\right|^{-42}}{\left|\vartheta_1\left(\frac{j}{N} ,\tau\right)\right|^2}.
\end{align}
One immediately finds that, schematically, $Z_{QFT} = \infty Z_{ST}$ \cite{Mertens:2015adr} as we noted above already. The ratio is an infinite combinatorial prefactor, as the sum-over-fields approach (including polarization effects) gives an infinite overcounting of the actual stringy result. Within string language, the UV region is cut off by utilizing the fundamental domain $\mathcal{F}$ instead of the strip $\mathcal{E}$. Hence the mechanism at work here is precisely the same as that causing the cosmological constant to be finite within any string theory. \\

\noindent To conclude, UV divergences in thermal quantities are avoided in full-fledged string theory by choosing the fundamental modular domain $\mathcal{F}$ instead of the strip domain $\mathcal{E}$, combined with making a suitable choice of the summations. \\

\noindent This resolves an old puzzle: Polchinski in his seminal work on flat space string thermodynamics \cite{Polchinski:1985zf} concludes that the free energy of a string gas is just the sum of the free energies of the fields in the spectrum. If this were all there is to black holes, it would be impossible to get rid of the UV divergence as each term in the sum has this same UV divergence. The divergence can get cancelled when spacetime supersymmetry plays a role, but this cannot be the general mechanism as the UV divergence needs to be dealt with for \emph{any} black hole in string theory, supersymmetric or not. \\ 

\noindent We should note that the string theory part of the story had been largely understood and anticipated in the past: string theory has no UV divergences close to the horizon. The new additional part here, completing the story, is the possibility to compare this to the sum of QFTs constituting the string spectrum. The latter is only made possible through the work of He et al. \cite{He:2014gva}.

\subsection{Summary}
\begin{itemize}
\item If one excludes polarization effects (i.e. contact terms or edge states) in the heat kernel of each field in the spectrum, the string theory experiences a maximal acceleration divergence as analyzed in \cite{Mertens:2015adr}.
\item If one includes polarization effects in the heat kernel for each field in the spectrum, the resulting ``string" theory is free from the maximal acceleration divergence. However, generically, the UV divergences arising from each field cause the string result to have a UV divergence as well. It can happen that the UV divergence itself gets cancelled by summing over the entire tower of string states, but this is not the generic situation. 
\item If one in addition restricts the modular integration region to the fundamental domain $\mathcal{F}$ and doubles the quantum numbers (i.e. include a winding number), then the resulting free energy and entropy are finite. If the background is sufficiently supersymmetric, then the partition function and entropy both vanish, as is the case in Rindler space as argued above. \\
So for Rindler space, two effects are needed to get a vanishing entropy: firstly, the presence of negative surface contributions to the entropy and secondly, an infinite overcounting of field theory compared to string theory. Both are crucial.
\end{itemize}

\subsection{Some thoughts on the fall-to-the-center and tortoise coordinates}
\label{falltoc}
Whereas the above scenario on cancelling UV divergences seems consistent to understand the Euclidean story, its Lorentzian counterpart is far more subtle. To appreciate this, we would like to discuss a bit the Lorentzian counterpart of the field theory equation (\ref{QFT}). For simplicity, consider a massless scalar field in Lorentzian Rindler space $ds^2=-\rho^2dt^2+d\rho^2$ with action:
\begin{equation}
S = \frac{1}{2} \int d\rho \rho\left[(\partial_{\rho}\phi)^2 - \frac{1}{\rho^2}(\partial_{t}\phi)^2\right].
\end{equation}
It is known that a continuum of eigenmodes corresponding to this action exist, related to the UV divergence close to the black hole horizon. \\
It can be transformed to tortoise coordinates as $u=\ln(\rho)$ into
\begin{align}
S &= \frac{1}{2}\int du \rho^2\left[\frac{1}{\rho^2}(\partial_{u}\phi)^2 - \frac{1}{\rho^2}(\partial_{t}\phi)^2\right] \\
&= \frac{1}{2}\int du \left[(\partial_{u}\phi)^2 - (\partial_{t}\phi)^2\right],
\end{align}
which is the standard action on the full real line; the horizon $\rho=0$ has been mapped into $u=-\infty$. From this perspective, the continuum arises quite naturally: a half-infinite space has suddenly appeared. \\
For fixed energy $\omega$, the action then takes the form
\begin{equation}
S = \frac{1}{2}\int d\rho \rho\left[(\partial_{\rho}\phi)^2 - \frac{\omega^2}{\rho^2}\phi^2\right].
\end{equation}
Within a path integral, one can make the transition from this second-quantized perspective to a first-quantized picture.
After absorbing the measure into the wavefunction, one obtains
\begin{equation}
S = \frac{1}{2}\int dx\left[\dot{x}^2 - \frac{\omega^2}{x^2} - \frac{1}{4x^2}\right]
\end{equation}
on the half-line $x\in(0,+\infty)$. For any real $\omega$, this describes a particle in an attractive $1/x^2$ potential that is unstable towards the fall to the center. The critical value for this potential corresponds to $\omega=0$ when the potential has coefficient $1/4$. The operator has a continuum of bound states with negative eigenvalue. Hence in order to better understand perturbative modes around Lorentzian black holes, we need to understand quantum mechanics in the unstable $-1/x^2$ potential. It should be noted here, that it is frequently mentioned in the earlier QM literature,\footnote{See e.g. \S 35 in \cite{Landau:1991wop}.} that the unstable $-1/x^2$ potential is of mere academic interest as no physical system exhibits this behavior. It is however apparent here that the problems of field theory near a black hole horizon are of precisely this kind. Moreover, since string theory in the Lorentzian background is decomposed into its QFT spectrum, it remains a mystery on how precisely string theory resolves this UV divergence from the Lorentzian perspective.\footnote{We suspect the so-called edge modes or surface modes are the cause here: they are completely missed from a Lorentzian QFT perspective, whereas the Euclidean point of view contains these implicitly (as the negative contributions to the heat kernels). Even more so, discarding the surface modes in the Euclidean picture, leads to a maximal acceleration divergence close to the black hole horizon.} The Euclidean perspective however, to which this work is devoted, is much better understood and the finiteness of thermodynamical quantities is explicitly visible. \\

\noindent For imaginary $\omega$, one is considering discrete momentum states on the Euclidean geometry (as above). For nonzero imaginary $\omega=in$, the potential is $\sim +1/x^2$ and is repulsive. A continuum of scattering states is found as discussed previously. \\
Winding on the Euclidean geometry corresponds to adding a $+x^2$ potential. If $w\neq0$, a discrete spectrum is found. When $n=-i\omega=0$, this potential still has the critical $-1/x^2$ piece as well. Nonetheless, no continuum is found. The singly-wound thermal scalar precisely saturates the fall-to-the-center instability. For $n\neq0$, the potential has a shape that resembles $+1/x^2$ for small $x$ and $+x^2$ for large $x$. \\

\begin{figure}[h]
\begin{minipage}{0.32\textwidth}
\centering
\includegraphics[width=0.9\textwidth]{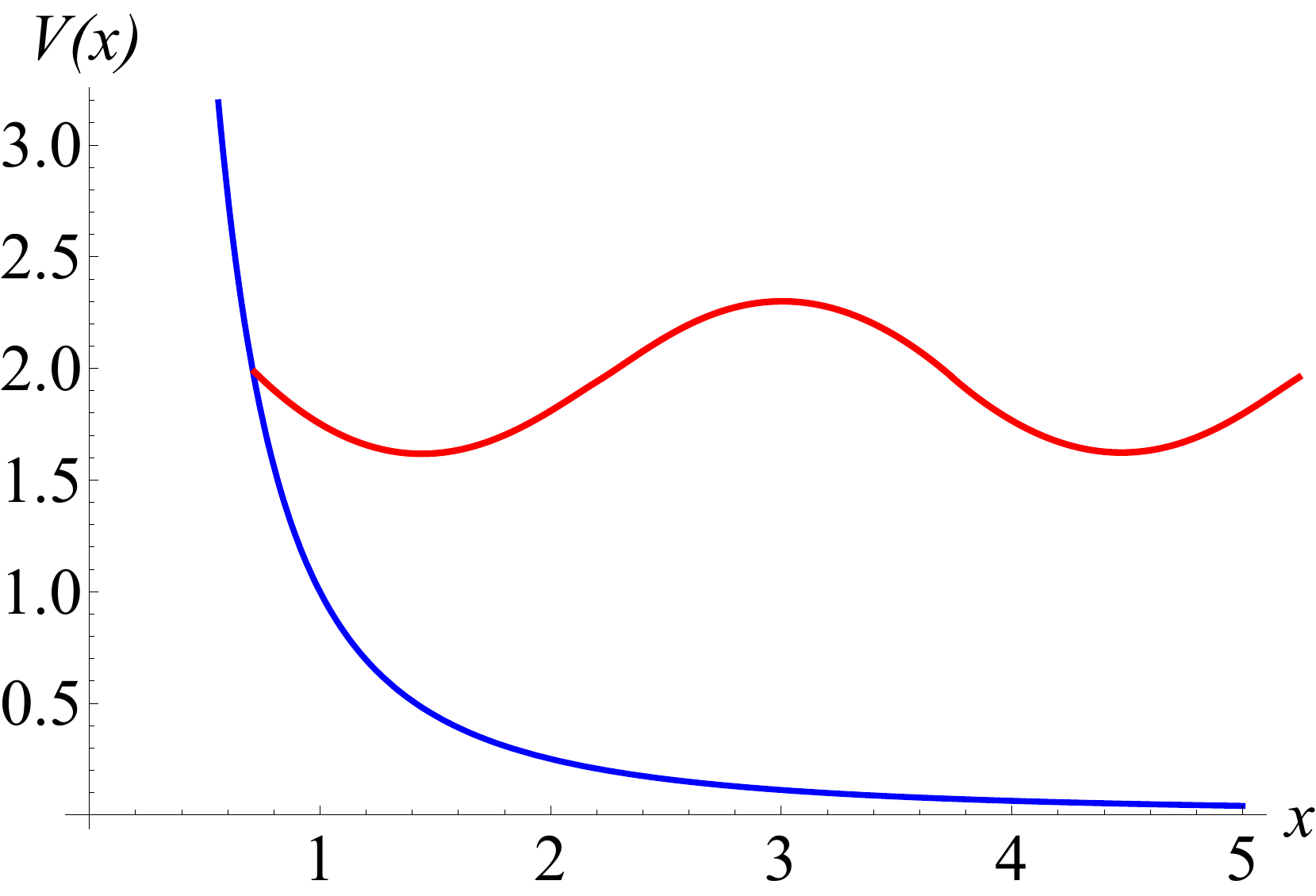}
\end{minipage}
\begin{minipage}{0.32\textwidth}
\centering
\includegraphics[width=0.9\textwidth]{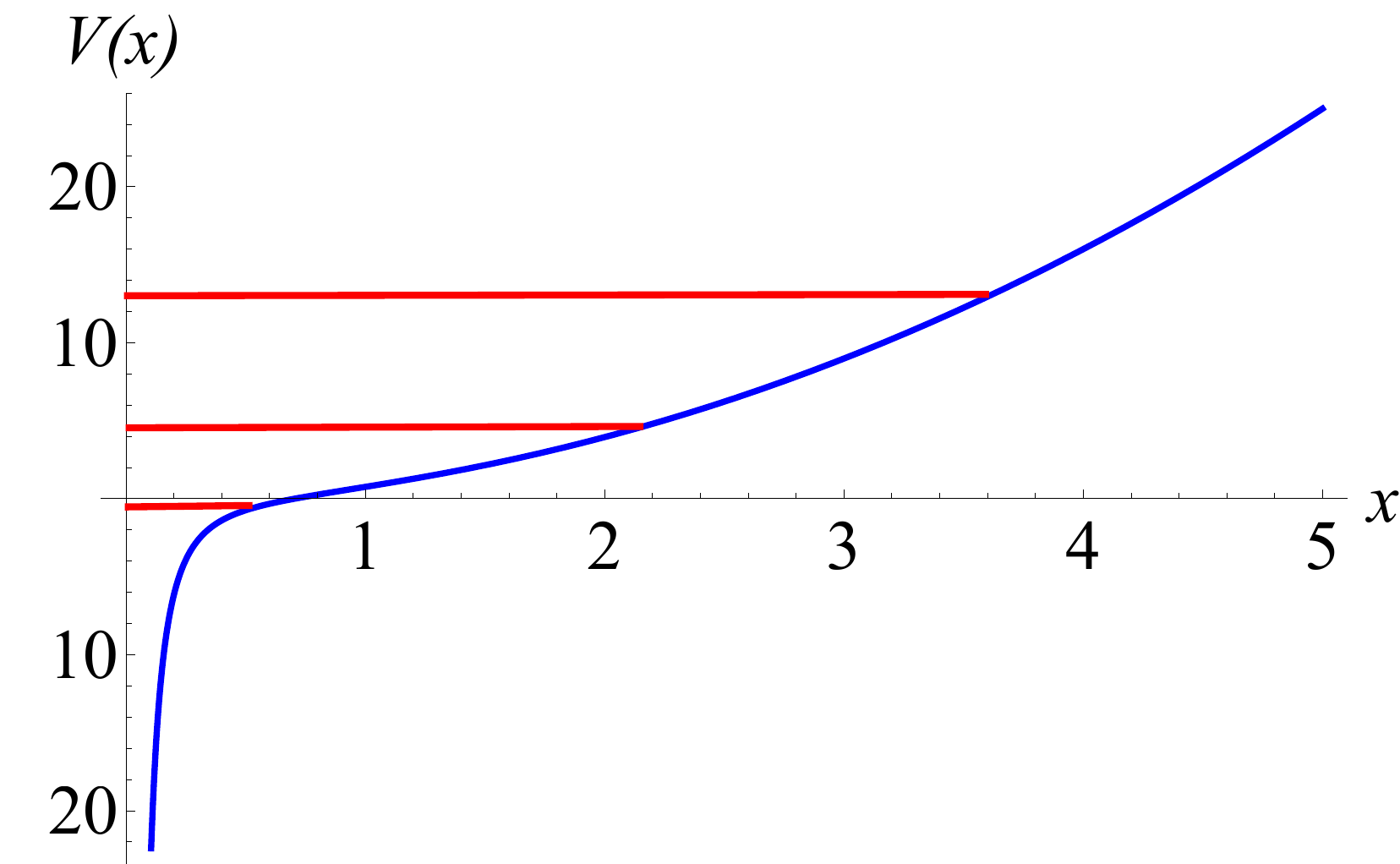}
\end{minipage}
\begin{minipage}{0.32\textwidth}
\centering
\includegraphics[width=0.9\textwidth]{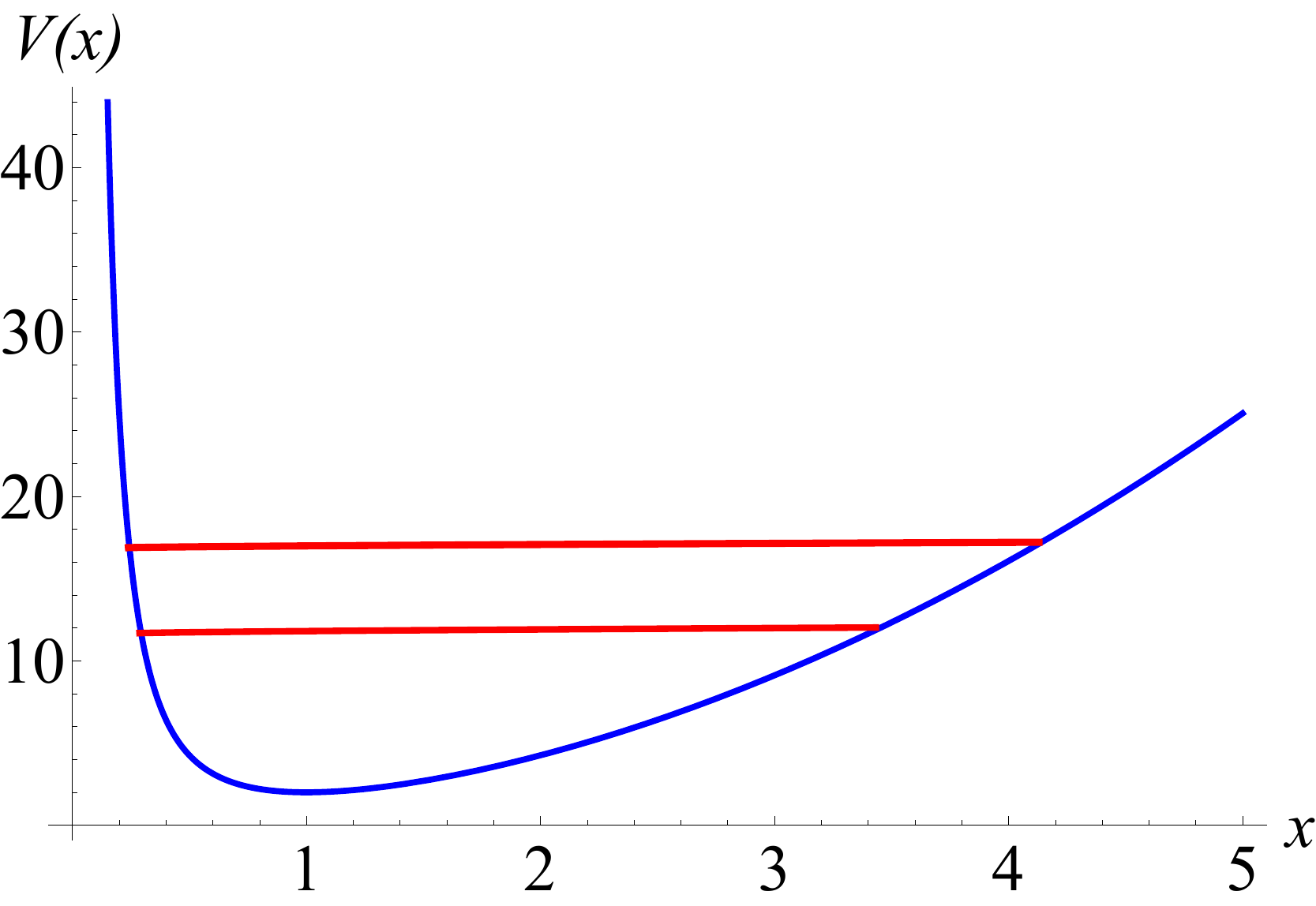}
\end{minipage}
\caption{Illustration of the potential $V(x)$ for the Euclidean case with representative eigenvalues. Left figure: pure momentum modes $n$. Middle figure: pure winding modes $w$. Right figure: mixed modes with $n$ and $w$.}
\label{euclideanpot}
\end{figure}

\begin{figure}[h]
\centering
\includegraphics[width=0.4\textwidth]{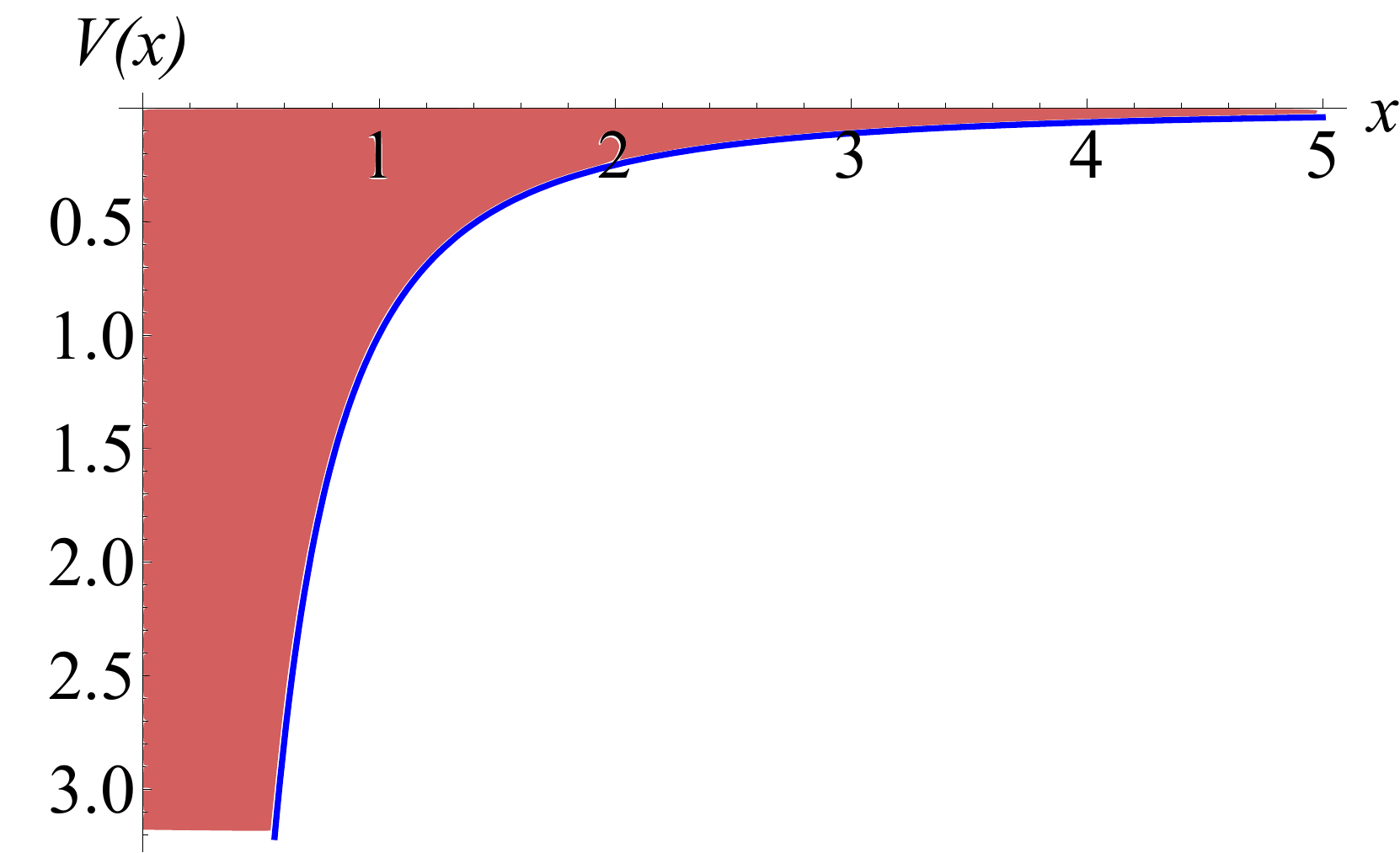}
\caption{Illustration of the potential $V(x)$ for the Lorentzian case. The potential is unstable and a continuous spectrum of bound states is found.}
\label{lorentzianpot}
\end{figure}

\noindent The different behaviors of the potential and the nature of the resulting spectrum is sketched in figure \ref{euclideanpot} for the Euclidean case and in figure \ref{lorentzianpot} for the Lorentzian case.

\noindent To summarize, Lorentzian states have a continuum of bound states coming from the near-horizon region. This can be seen in different equivalent ways. One can transform to tortoise coordinates to see that there is in fact a full halfspace hidden near the horizon. The continuum simply arises from the non-compactness of this dimension. A second way is to solve the eigenvalue problem directly, where an unstable attractive $-1/x^2$ potential is present: this has a continuum of bound states for any real $\omega$. This feature is not present on the Euclidean manifold, since the nature of the potential changes. \\
What this discussion teaches us, is that it is not trivial to make the analytic continuation for this type of background. Solving this side of the story would prove very interesting.

\section{Conclusion and Outlook}
\label{concl}
In this work, we have pointed out several features of the $SL(2,\mathbb{R})_k/U(1)$ cigar CFT and its flat limit.
\begin{itemize}
\item The sums over negative $r$ can be mapped into positive $r$ and hence effectively doubles the degeneracy of all states. This corresponds to the symmetry of $w\to-w$ and $n\to-n$ in the field theory action describing the primaries and hence resolves a mismatch between on the one hand, the path integral character decomposition and on the other hand, the field theory and vertex operator constructions.
\item For a finite value of $k$, the NS-NS sector contains only one chiral-chiral primary (the thermal scalar) and one antichiral-antichiral primary (the dilaton zero-mode). Both states are marginal and they are related by an involution symmetry of the spectrum. There do not exist discrete ($a$, $c$) nor ($c$, $a$) fields in this background. These two states exhaust the discrete marginal states of the theory at a generic value of $k$. \\
At $k\to\infty$, 2 additional ($a$, $a$) and 2 additional ($c$, $c$) states appear, each contributing with half-weight.
\item The R-R, R-NS and NS-R sectors are all massive but become marginal as $k\to\infty$, in correspondence with the restoration of spacetime supersymmetry.
\item The R-R ground states can spectral flow to the chiral (and anti-chiral) primaries, but only when only one of the left- or right-movers is in the ground state. The reason is that spectral flow requires the left- and right-movers to flow in the opposite direction.
\item All marginal states in the large $k$ limit have been classified. We have emphasized the fact that all of these states exist for generic $k$ and are hence not an artifact of the large $k$ limit. However, their marginality on the other hand is an artifact of the infinite $k$ limit, and we hence expect them not to play any major influence on real black holes (that are not truly geometrically flat). This is of major importance, since at large $\tau_2$, the most dominant contribution is of course extremely sensitive to whether a state is precisely marginal or just nearly so.
\item When the temperature changes (the introduction of a conical deficit), the NS-NS states are naturally continued. The dilaton zero-mode can be identified with the negatively wound thermal scalar state and is in this sense a temperature-dependent state. The involution symmetry also gets naturally extended into this regime. The R-R ground states have weights that are independent of the temperature. The R-NS and NS-R fermions are more subtle as no straightforward $N\neq1$ partner state is found. 
\item The continuous states are simpler to describe: $f$ is directly related to the fermionic oscillator level in this case. The NS-NS sector contains for each discrete momentum one of each of the four states of the $\mathcal{N}=2$ superconformal ring. \\
For a general $k$, there are no marginal continuous states however. So for a real black hole, there is not a single state left to cancel the dominant behavior of the thermal scalar, and its contribution will really be the most significant. We indeed interpreted its influence as realizing the long string random walk surrounding a generic (uncharged) black hole in previous work \cite{Mertens:2013pza}\cite{Mertens:2013zya}\cite{Mertens:2014nca}\cite{Mertens:2014cia}\cite{Mertens:2014dia}\cite{Mertens:2014saa}\cite{Mertens:2015hia}.
\item We have found several alternative perspectives on some well-known features of this background. Firstly, discrete spacetime fermions are naturally anti-periodic around the tip of the cigar: periodic fermions do not exist in the discrete sector. Secondly, $N$ must be odd, since else no anti-periodic fermions would exist.
\end{itemize}

\noindent To summarize, whereas for the finite $k$ cigar CFT only 2 marginal states exist, as $k\to\infty$ one finds lots of extra marginal states appearing, required for the restoration of spacetime supersymmetry. Precisely this supersymmetry, in turn, requires the vanishing of the one-loop entropy. This can only happen if the discrete marginal thermal scalar state gets compensated by other marginal states. In this work, we have demonstrated that there indeed exist states with these properties and hence this cancellation might indeed occur. In the end, we presented a direct partition function argument that demonstrates that this cancellation indeed occurs between bosons and fermions on the thermal manifold, even for the entropy. Since the partition function does not correspond to a free Hamiltonian trace $\text{Tr}e^{-\beta H}$, no constraint on the positivity of the entropy, order by order in string perturbation theory, is required a priori.  \\

\noindent In the final section, we gave a complete picture on how one ``builds up" string theory starting with the constituent field theories and how the near horizon UV divergence is averted in string theory. Within that language, the vanishing of the entropy is explained by a combination of two facts: the entropy of higher spin fields carry negative contributions themselves and the modular integration region is shrunk to the fundamental domain, mitigating the infinite overcounting of string worldsheets. \\

\noindent There are several questions left open for future study. \\
We started out by giving a motivation in terms of the field theory of the winding states. This works well for the thermal scalar state itself, but is somewhat ill-understood for all of the other states uncovered in the polar coordinate description. It would be very interesting to precisely pin down the field theory action to be utilized for the other sectors (in particular the spacetime fermions). We consider this the most immediate open question at this point. \\
The vanishing of the one loop entropy for sufficiently supersymmetric backgrounds is very interesting. For instance, if we assume this vanishing to continue to all orders in string perturbation theory (which is plausible), the tree level contribution to the black hole entropy is perturbatively exact. \\
Comparing this to an argument we presented in \cite{Mertens:2015hia} where we obtained the tree level contribution of the black hole entropy by throwing in non-interacting long strings, we see that we have in fact obtained the total answer for the case of Rindler space (or, plausibly, any sufficiently supersymmetric black hole). This means adding interactions of the infalling strings to the story has no net result, and the entropy remains the same. \\
Of course, all given ``derivations" of this vanishing of $S$ are plagued by the same underlying disease: one does not know how to make sense of string theory on spaces with arbitrary conical deficit. This makes performing a derivative in $N$ ambiguous. However, it seems that at least the presented approaches agree that the one-loop entropy should vanish for Rindler space. \\
On an even more ambitious front, the major issue is to get a handle on the negative contact terms present in the string partition functions in terms of the Lorentzian entanglement entropy of edge states, such as was done in \cite{Donnelly:2014fua}\cite{Donnelly:2015hxa} for the spin 1 case. In particular, it was argued there that the negativity of the contact terms is merely a feature of heat kernel regularization and is hence not truly physical. However, perturbative string theory is deeply linked to this type of regularization as the $\tau_2$ parameter can be identified with the Schwinger proper time. \\

\begin{figure}[h]
\centering
\includegraphics[width=0.6\textwidth]{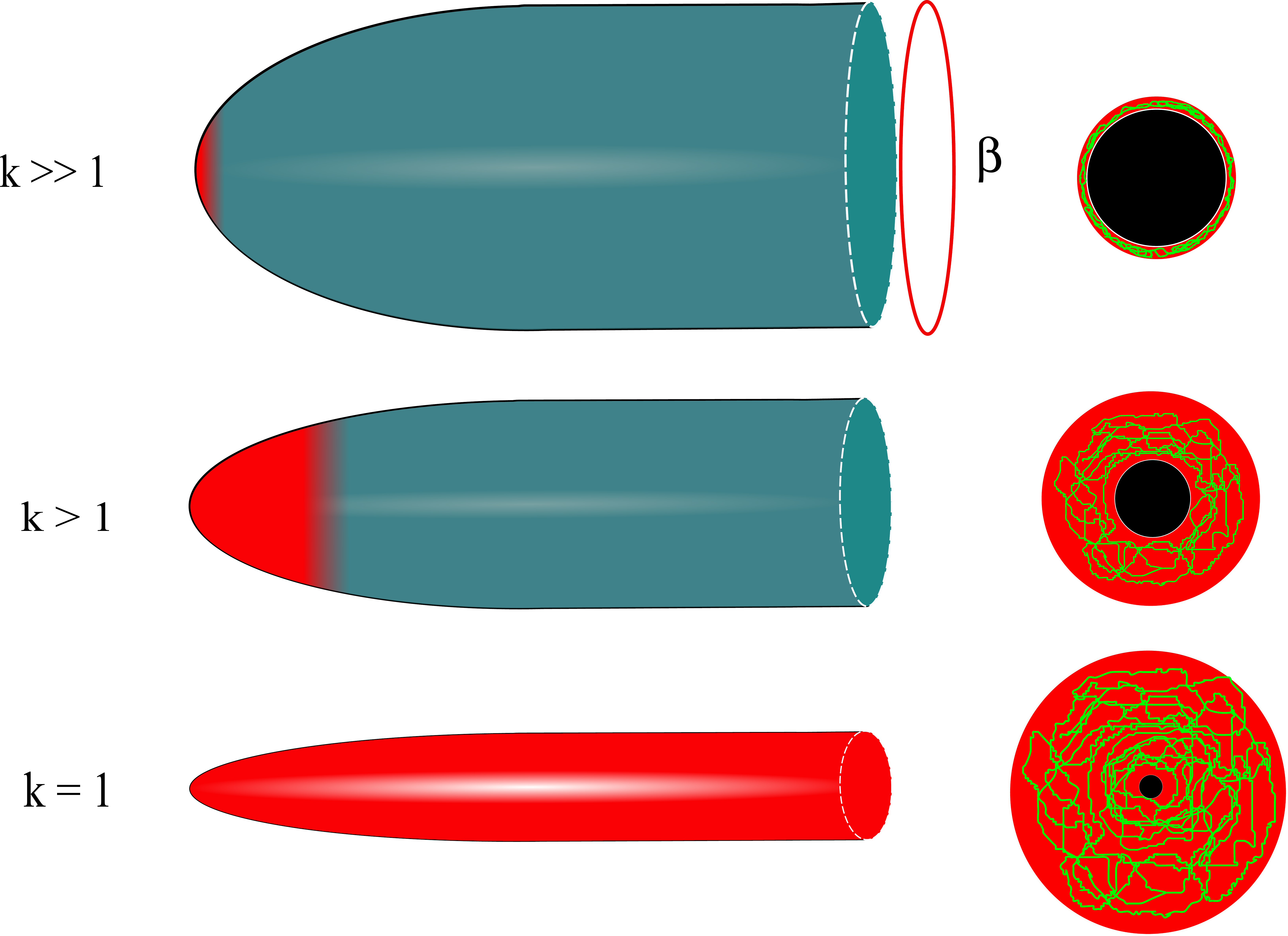}
\caption{Cigar-shaped Euclidean black holes with the location of the dominant thermal scalar state colored in red. As $k$ decreases from infinity, the bound thermal scalar state spreads out over the cap of the cigar, until at $k=1$ this mode becomes part of the continuum and stretches all the way to the asymptotic linear dilaton region, as this region is precisely at its own Hagedorn temperature now. To the right, is the counterpart of this story on the Lorentzian background, where the thermal scalar state determines the location of the long random walking string (colored in green) making up the stretched horizon of the hole. For $k=1$ this stretched horizon diverges to infinity and encompasses the entire space.}
\label{cartoon}
\end{figure}

\noindent We have confirmed here (using an ansatz for the off-shell partition function and entropy) that the one-loop entropy vanishes for Rindler space, but not for the cigar (with finite $k$). This result is in unison with the spectrum of states. Hence most of these additional marginal states are a special feature of Euclidean Rindler space, and are expected to become massive (irrelevant) for any large black hole. \\
\noindent At least the $SL(2,\mathbb{R})_k/U(1)$ black hole provides one example where this is indeed borne out in full detail. This provides a good opportunity to try to convince the sceptical reader, who doubts the thermodynamic random walking interpretation of the thermal scalar state for black hole geometries, that this does provide the most natural interpretation of the facts. The argument we will present now tries to tie together the random walking long string on the cylinder at infinity with that near the black hole horizon. \\
Suppose one starts at the asymptotic linear dilaton Hagedorn temperature: this is at $k=1$. Then the continuous versions of these singly wound modes are marginal and spread over the asymptotic cylindrical geometry.\footnote{Again here we encounter an argument why the correct interpretation for the dilaton zero-mode is the $w=-1$ thermal scalar: the continuous state with the quantum numbers of the dilaton zero-mode is nowhere near marginal at $k=1$.} Indeed, their weight is given by:
\begin{equation}
h = \frac{1}{4k} + \frac{kw^2}{4},
\end{equation}
which becomes $1/2$ for $w=\pm1$ and $k=1$. Far enough away from the tip, no funny business should happen and we have a genuine long string phase, just as in flat space. Indeed, all values of $w$ and $n$ are allowed in the continuous series (for \emph{finite} $k$) and a true thermodynamic interpretation can be given. \\ 
Imagine we now lower the asymptotic temperature (or increase $k$ from its lowest value of 1). These continuous modes become irrelevant. However, the discrete versions of both modes appear in the spectrum and are marginal throughout this process. Using our result that the dilaton zero-mode should actually be interpreted as the negatively wound thermal scalar, both of the zero-modes have the same spatial wavefunction. Obviously, the behavior of the most dominant mode changed from continuous to discrete. It is very natural that the long string phase is confined now to the near-horizon region where the discrete modes live. We find the alternative explanation, that this discrete mode has nothing to do with random walks, much more exotic, since one would then have to explain what it does represent \emph{and} why it transcends so nicely into a continuous mode that \emph{definitely} has a long random walking interpretation. \\
We end with a graphical representation of this story in figure \ref{cartoon}.

\section*{Acknowledgements}
The authors thank A. Lewkowycz for an interesting discussion. TM gratefully acknowledges financial support from Princeton University, the Fulbright program and a Fellowship of the Belgian American Educational Foundation. TM also thanks the UGent Special Research Fund for supporting him in Ghent where this work was started. The work of VIZ was partially supported by the RFBR grant 14-02-01185.

\appendix
\section{Details on the bosonic character decomposition}
\label{detailsbosonic}

We study all the different possibilities for the signs of $r$ and $\bar{r}$.

\subsection*{$r\geq0$, $\bar{r}\geq0$ sector}
The sector for $r\geq0$ and $\bar{r}\geq0$ is the simplest: it simply leads to the expected lowest weight discrete representations $\mathcal{D}_j^+$ of $SL(2,\mathbb{R})$. The unitarity constraints for this case lead to
\begin{align}
\frac{1}{2} &< \frac{kw}{2} + \frac{n}{2} - r < \frac{k-1}{2}, \\
\frac{1}{2} &< \frac{kw}{2} - \frac{n}{2} - \bar{r} < \frac{k-1}{2}.
\end{align}
This can only be satisfied if $w>0$ (strictly positive). The spin of such states is $h-\bar{h} = nw$. \\
The special case when $r=\bar{r}=0$ is actually absent. Indeed, this $r=\bar{r}=0$ sector has
\begin{align}
j = \frac{kw}{2} + \frac{n}{2} = \frac{kw}{2} - \frac{n}{2},
\end{align}
and hence $n=0$. The unitarity constraints then imply
\begin{equation}
\frac{1}{2} < \frac{kw}{2} <\frac{k-1}{2},
\end{equation}
for which no solution is possible. This $r=\bar{r}=0$ sector is hence absent altogether.

\subsection*{$r<0$ and $\bar{r}<0$ sector}
The sector where both $r$ and $\bar{r}$ are negative can be dealt with in the following way.
It can be shown that the characters obey the following relation:
\begin{equation}
\lambda_{r}^{j} = \lambda_{-r}^{k/2-j}.
\end{equation}
Thus changing $j \to k/2-j$ is all one should do. This change in $j$ entails replacing $r'=-r$, $w' = 1-w$ and $n' = -n$. Upon reinterpreting the states in this fashion, this is actually precisely the same as the previous sector. What this means is that every state is doubly degenerate. For instance, the state with $r=-3$, $w=-2$ and $n=-1$ has the same weight as the state with $r=3$, $w=3$ and $n=1$. The former state should be thought of as merely doubling the spectrum; and providing the negatively wound partner of each state of positive $r$. Note that the unitarity constraint is invariant under this transformation. The apparent mismatch in the shift in $w$ should not bother us too much, as $w$ is in the end merely a summation index in the partition function and we can freely interpret it in any way we want: the invariant information is the conformal weight itself, and the degeneration caused by this additional sector is the important point. \\
The spin of these states equals 
\begin{equation}
h-\bar{h} = nw - r +\bar{r} = nw - n = n(w-1) = n'w',
\end{equation}
and the original $w = 0,-1,-2 \hdots$ to satisfy the unitarity constraints.

\subsection*{$r>0$, $\bar{r}<0$ and $r<0$, $\bar{r}>0$ sectors}
Next consider the $r>0$ and $\bar{r}<0$ sector. The weights are given by
\begin{align}
h &= -\frac{j(j-1)}{k-2}+ \frac{m^2}{k}, \\
\bar{h} &= -\frac{j(j-1)}{k-2}+ \frac{\bar{m}^2}{k} - \bar{r},
\end{align}
where $j = \frac{kw}{2}+\frac{n}{2} - r = \frac{kw}{2}+\frac{n}{2} - \bar{r}$ and hence $r-\bar{r} = n$. The spin of such a state equals $h - \bar{h} = nw +\bar{r}$, which unlike before cannot generally be reabsorbed into a redefinition of $n$ or $w$. This state is hence a true secondary (an oscillator state) and $-\bar{r}$ can be interpreted as the anti-holomorphic oscillator number: for a general CFT, one would expect $ h- \bar{h} = N - \bar{N}$. Note that $r-\bar{r} = n > 0$ in this sector. \\

\noindent The other sector $r<0 $, $\bar{r} > 0$ has $r-\bar{r} = n > 0$ and $h - \bar{h} = nw - r$. This represents a state with a holomorphic oscillator. \\
The above mapping maps a sector $r>0,\bar{r}<0, w, n$ into $r<0,\bar{r}>0, 1-w ,-n$ and hence both quadrants in the ($r,\bar{r}$) plane give equal weights.

\subsection*{$r=0$, $\bar{r} <0$ sector}
In the previous paragraph, we demonstrated that the mixed sign sector states should be interpreted as Virasoro secondaries. Not so when either $r$ or $\bar{r}$ vanishes.
The only state in the mixed sign sector that can (and should) be interpreted as a primary is the one where $r=0$. One then has $\bar{r} = -n$. The weights are
\begin{align}
h &= -\frac{\left(\frac{kw}{2}+\frac{n}{2}\right)\left(\frac{kw}{2}+\frac{n}{2}-1\right)}{k-2}+ \frac{\left(\frac{kw}{2}+\frac{n}{2}\right)^2}{k}, \\
\bar{h} &= -\frac{\left(\frac{kw}{2}-\frac{n}{2}-\bar{r}\right)\left(\frac{kw}{2}-\frac{n}{2}-\bar{r}-1\right)}{k-2}+ \frac{\left(\frac{kw}{2}-\frac{n}{2}\right)^2}{k} - \bar{r}.
\end{align}
It readily follows that setting $n'=-n$ and $w'=1-w$, one can reinterpret these weights as
\begin{align}
h &= -\frac{\left(\frac{kw'}{2}+\frac{n'}{2}\right)\left(\frac{kw'}{2}+\frac{n'}{2}-1\right)}{k-2}+ \frac{\left(\frac{kw'}{2}+\frac{n'}{2}\right)^2}{k}, \\
\bar{h} &= -\frac{\left(\frac{kw'}{2}-\frac{n'}{2}+\bar{r}\right)\left(\frac{kw'}{2}-\frac{n'}{2}+\bar{r}-1\right)}{k-2}+ \frac{\left(\frac{kw'}{2}-\frac{n'}{2}\right)^2}{k}.
\end{align}
In this case, the spin can be reinterpreted by a rescaling of $n$ and $w$:
\begin{equation}
h-\bar{h} = nw-n=n(w-1) = n'w'.
\end{equation}
The weights of these primaries are precisely the same as those from $r=0$ and $\bar{r} > 0$.

\subsection*{$r<0$, $\bar{r}=0$ sector}
This sector is completely analogous to the previous sector.

\section{Some details on the Ramond ground states}
\label{Rdetails}
For fixed $w$ and $f$, the equation (\ref{firsteqR}) allows solving for $l$. Since $l$ has to be a positive integer, this severely restricts the allowed values of $f$ to:
\begin{equation}
f=-2w \quad \vee \quad f=-1 \quad \vee \quad f=-w \text{ if $w$ odd} \quad \vee \quad f=-w-1 \text{ if $w$ odd}.
\end{equation}
We will discuss these four options. 
\begin{itemize}
\item $f=-2w$. \\
In this case, one finds
\begin{equation}
l=0, \quad n=1-2w,
\end{equation}
and hence $j=\frac{kw}{2}+w$ which automatically is outside the unitarity bound.

\item $f=-1$. \\
In this case, one finds
\begin{equation}
l=0,
\end{equation}
with arbitrary $n$. These states have $j=\frac{kw}{2} + \frac{n}{2} + \frac{1}{2}$ and they are allowed provided $w$ and $n$ are chosen suitably. In particular, note that $n=w=0$ is included in the spectrum, albeit with half weight.

\item $f=-w$ for $w$ odd. \\
In this case, one finds
\begin{equation}
l=\frac{w-1}{2}, \quad n=-\frac{w-1}{2}
\end{equation}
and $j=\frac{kw}{2} + \frac{w}{4} + \frac{1}{4}$ which again violates the unitarity bound unless $w=1$. But in this case, we are in fact considering the previous situation again.

\item $f=-w-1$ for $w$ odd. \\
In this case, one finds
\begin{equation}
l=\frac{w-1}{2}, \quad n=-\frac{w+1}{2}
\end{equation}
and $j=\frac{kw}{2} + \frac{w}{4} + \frac{3}{4}$ which again violates the unitarity bound.

\end{itemize}

\section{Witten Index}
\label{WittenI}
Here we shortly discuss how the Witten index indeed only obtains contributions from the two R-R states displayed previously, and this for any $k$ (including the $k\to\infty$ limit). \\
The Witten index can be computed by considering the trace over all states in the left and right twisted Ramond sector, as:
\begin{equation}
I_W = \text{Tr}_{RR}(-)^{F+\bar{F}}q^{L_0-c/24}\bar{q}^{\bar{L}_0-c/24}.
\end{equation}
This corresponds to taking only the $a=b=1$ sector in the partition function (irrespective of the final GSO projection one utilizes). The characters are given by the expression (including a non-zero $z=e^{2\pi i \nu}$ as a regulator):
\begin{align}
ch_D\left(j,m-j-\frac{a}{2}\right) \left[
\begin{array}{c}
a \\
b  \end{array} 
\right] (\tau,\nu) 
&= i^{ab}\sum_{f\in\mathbb{Z}}e^{i\pi b f} \lambda_{m-f-j-a/2}^j(q)\frac{q^{\frac{k}{2(k+2)}\left(f+\frac{a}{2}+\frac{2m}{k}\right)^2}}{\eta} z^{f+a/2+\frac{2m}{k}} \nonumber \\
&=\frac{1}{\eta^3}q^{-\frac{(j-1/2)^2}{k}}q^{\frac{m^2}{k}}z^{\frac{2m}{k}}\frac{\vartheta_{ab}(\tau,\nu)}{1+(-)^bzq^{m-j+1/2}}.
\end{align}
This vanishes (almost) always when $a=b=1$. The exception occurs if $m-j+1/2=0$; then the $z\to1$ limit is non-trivial:
\begin{align}
&\sum_{f\in\mathbb{Z}}e^{i\pi f} \lambda_{m-f-j-1/2}^j(q)\frac{q^{\frac{k}{2(k+2)}\left(f+\frac{1}{2}+\frac{2m}{k}\right)^2}}{\eta} z^{f+1/2+\frac{2m}{k}} \nonumber \\
&=-i(-i)q^{-\frac{-m^2}{k}}q^{\frac{m^2}{k}} \nonumber \\
&=-1.
\end{align}
One obtains 
\begin{align}
I_W = \sum_{n,w,\in\mathbb{Z}}\sum_{j\in I} \delta_{m-j+1/2}\delta_{\bar{m}-j+1/2} =1.
\end{align}
This immediately requires $n=0$ and $w=0$ or $w=1$ as the only options, with respectively $j=1/2$ and $j=k/2+1/2$, both of which saturate the unitarity bound. These are precisely the two RR states we constructed earlier. \\

\noindent In \cite{Israel:2004ir}, the state responsible for the Witten index was incorrectly identified as a $j=1$, $m=0$ state, due to a typo in the character decomposition formula.

\section{Details on the chiral primaries}
\label{chdetails}
Setting $Q=1$ implies
\begin{equation}
n=0, \quad w+f=1.
\end{equation}
Setting $h=1/2$ leads to the conditions
\begin{align}
(l+f)^2 +(l+f)&=0, \\
w(l+f)+w/2+f^2/2 &= 1/2.
\end{align}
The total solution is one of the four options:
\begin{alignat}{3}
f&=0, \,\, &&w=1, \,\, &&l=0, \\
f&=1, \,\, &&w=0,\,\, &&l=-1, \\
f&=1, \,\,&&w=0, \,\, &&l=-2, \\
f&=-2, \quad &&w=3, \quad &&l=1.
\end{alignat}
The final one has $j=3k/2-1$ which is excluded by the unitarity bounds. The middle two violate the assumption that $l>0$. The first one is the only one remaining and is the ($c$, $c$) thermal scalar state, upon combining left- and right-movers. \\

\noindent Setting $Q=-1$ for antichiral primaries implies
\begin{equation}
n=0, \quad w+f =-1,
\end{equation}
for which the total solution is one of the four options:
\begin{alignat}{3}
f&=2, \quad &&w=-3, \quad &&l=-2, \\
f&=-1, \quad &&w=0, &&l=1, \\
f&=0, &&w=-1, &&l=-1, \\
f&=-1, &&w=0, &&l=0.
\end{alignat}
The first and third one violate $l>0$ whereas the second one has $j=0$ which is outside the unitarity bounds. The final state is the ($a$, $a$) dilaton zero-mode.\\
No mixed ($c$, $a$) or ($a$, $c$) states can be found.

\section{All marginal states in polar coordinates}
\label{polarstates}
\subsection{The NS-NS sector}
As $k\to\infty$, we find
\begin{equation}
h+\bar{h} \to 2w(l+f)-wn + w + \frac{f^2}{2} + \frac{\bar{f}^2}{2},
\end{equation}
and 
\begin{equation}
h-\bar{h} \to nw  + \frac{f^2}{2} - \frac{\bar{f}^2}{2},
\end{equation}
requiring $f$ and $\bar{f}$ to be odd or even simultaneously. We will examine the two cases $w=0$ and $w=1$ separately. 

\subsection*{$\mathbf{w=0}$}
The lowest state has $f=\bar{f}=0$ and would be the bound version of the closed string tachyon. However, it does not satisfy the unitarity constraints. The states with $f=\bar{f}=-1$ are marginal states. These represent the dilaton zero-mode with $j=1$ and $n=0$. \\
One can obtain additional intuition that this state should \emph{not} be interpreted physically as a no-winding state from that fact that the field theory action (\ref{QFTorig}) we presented earlier, upon taking the large $k$ limit, only allows for discrete states when $w\neq0$. This clearly demonstrates again the discrepancy between the physical interpretation of winding and momentum on the one hand, and the quantum numbers $w$ and $n$ as they appear in the partition function, showing that a reinterpretation is required indeed.

\subsection*{$\mathbf{w=1}$}
The sector with $f=\bar{f}=0$ is minimized by $l=\bar{l}=0$ with weight 1 and spin 0 and $n=0$. This is the thermal scalar state with $j=k/2$. \\ 

\noindent It turns out, that there are actually two more physical states with $w=0$ and two more with $w=1$. These are different in the sense that they saturate the unitarity bounds, and hence only contribute to the partition function with half weight. We do not write them down here, but instead refer the reader to the table of marginal states we display in the main text. Moreover, these states are actually all doubly degenerate, as the above construction on how to deal with negative $r$ yields another four marginal states. \\ 
These states still exist for finite $k$, though they become massive and we have hence not discussed these previously.

\subsection{The R-R sector}
Ramond ground states (in, say, only the holomorphic sector) are characterized for $w=0$ by
\begin{equation}
f=0 \quad \vee \quad f=-1,
\end{equation}
for any $l$ such that the unitarity constraints are satisfied. For $w=1$ one finds instead for the ground states
\begin{equation}
f=-1, l =0 \quad \vee \quad f=-2,l=0.
\end{equation}

\noindent Matching $j$ fixes $n=l-\bar{l}+f-\bar{f}$ and unitarity then restricts these candidate states to only a small subset of these states, namely
\begin{alignat}{2}
w&=0, \,\, l=\bar{l}=0,\,\, f=\bar{f}=-1,\,\,n=0,\,\,j=1/2,\,\,&&Q=\bar{Q}=-1/2, \\
w&=1, \,\, l=\bar{l}=0, \,\,f=\bar{f}=-1,\,\,n=0,\,\,j=(k+1)/2,\,\,&&Q=\bar{Q}=+1/2.
\end{alignat}
Both of these R-R ground states contribute with only half-weight to the partition function as they saturate the unitarity bound. \\
These states fit into the scheme outlined in section \ref{classification} for a generic $k$.

\subsection{The R-NS sector}
The weight of the state has the limiting behavior
\begin{equation}
h+\bar{h} \to \frac{1}{8} + \frac{3}{8} + 2w + 2w(l+f)-wn + \frac{f^2 + f}{2} + \frac{\bar{f}^2}{2},
\end{equation}
where $n=l-\bar{l} + f -\bar{f} +1/2$ and the spin asymptotes to
\begin{equation}
h-\bar{h} \to nw  + \frac{f^2 + f}{2} - \frac{\bar{f}^2}{2} + \frac{1}{2}.
\end{equation}
Since $n$ is a half-integer, requiring integer spin provides a restriction on $\bar{f}$. 
\subsubsection*{$\mathbf{w=0}$}
In this case, $\bar{f}$ needs to be odd to have integer spin. \\
The lowest state that is allowed has $f=\bar{f}=-1$. It has $h+\bar{h} = 1$ and $h-\bar{h}=0$. It has $j=\frac{3}{4}-\frac{l+\bar{l}}{2}$. Clearly, the only state allowed has $l=\bar{l}=0$ and hence $n=\frac{1}{2}$ and $j=\frac{3}{4}$.
\subsubsection*{$\mathbf{w=1}$}
In this case, $\bar{f}$ needs to be even to have integer spin. \\
There are four candidates of lowest weight states which have
\begin{align}
f=-1, \quad \bar{f} &= 0, \\
f=-2, \quad \bar{f} &= 0, \\
f=-1, \quad \bar{f} &= -2, \\
f=-2, \quad \bar{f} &= -2.
\end{align}
All of these sectors have a candidate state with $l=\bar{l}=0$. However, only the first option satisfies the unitarity constraints. The lowest weight state is characterized by $n=-\frac{1}{2}$ and has $j=\frac{k}{2}+\frac{1}{4}$. This state has $h+\bar{h} = 1$ and $h-\bar{h}=0$. \\

\noindent Again, both of these states precisely correspond to the generic classification scheme presented in section \ref{classification}.

\subsection{The NS-R sector}
The only difference is $n=l-\bar{l} + f -\bar{f} -1/2$. The lowest $w=0$ state has $f=\bar{f}=-1$, $h+\bar{h} = 1$ and $h-\bar{h}=0$. Again $l=\bar{l}=0$, $j=\frac{3}{4}$ and $n=-\frac{1}{2}$. This is the state constructed in \cite{Sugawara:2012ag}.

\section{Most dominant states of cigar CFT from the large $\tau_2$ limit: bosonic case}
\label{mostdom}
It turns out one can look at the large $\tau_2$ behavior directly in the partition function (i.e. without first going through the character decomposition), and even see the unitarity constraints pop up. Taking $\tau_2$ large in the $\frac{SL(2,\mathbb{R})_k/U(1)}{\mathbb{Z}_N}$ cigar orbifold partition function:\footnote{We have included here a unitary compact CFT with weights $h_i$. All we need of this CFT is that its large $\tau_2$ limit yields 1 (due to the unit operator).} 
\begin{align}
\label{bosstart}
Z &= \frac{1}{4N}\sqrt{k(k-2)}\int_{\mathcal{F}}\frac{d\tau d\bar{\tau}}{4\tau_2} \int_{0}^{1}ds_1ds_2 \nonumber \\
&\sum_{m,w\in\mathbb{Z}}\sum_i q^{h_i}\bar{q}^{\bar{h}_i}e^{4\pi\tau_2(1-\frac{1}{4(k-2)}) -\frac{k\pi}{\tau_2}\left|(s_1 - \frac{w}{N})\tau +(s_2 - \frac{m}{N})\right|^2+2\pi\tau_2s_1^2} \nonumber \\
&\frac{1}{\left|\sin(\pi(s_1\tau + s_2))\right|^2}\left|\prod_{r=1}^{+\infty}\frac{(1-e^{2\pi i r \tau})^2}{(1-e^{2\pi i r \tau - 2\pi i (s_1\tau +s_2)})(1-e^{2\pi i r \tau + 2\pi i (s_1\tau +s_2)})}\right|^2,
\end{align}
one obtains
\begin{align}
\label{starting}
Z &\approx \frac{1}{N}\sqrt{k(k-2)}\int_{\mathcal{F}}\frac{d\tau d\bar{\tau}}{4\tau_2} \int_{0}^{1}ds_1ds_2 \nonumber \\
&\sum_{m,w\in\mathbb{Z}}\sum_i q^{h_i}\bar{q}^{\bar{h}_i}e^{4\pi\tau_2(1-\frac{1}{4(k-2)}) -\frac{k\pi}{\tau_2}\left|(s_1 - \frac{w}{N})\tau +(s_2 - \frac{m}{N})\right|^2+2\pi\tau_2s_1^2} e^{-2\pi s_1\tau_2}.
\end{align}
In the large $\tau_2$ region, the $s_1$-integral has a saddle point since 
\begin{align}
\label{starting}
Z_{w,m} &\approx \frac{1}{N}\sqrt{k(k-2)}\int_{\mathcal{F}}\frac{d\tau d\bar{\tau}}{4\tau_2} \int_{0}^{1}ds_1ds_2 \nonumber \\
&e^{4\pi\tau_2\left[1 -\frac{(k-2)}{4}\left(s_1 - \frac{w}{N}\frac{k}{k-2}+\frac{1}{k-2}\right)^2 + \frac{w^2}{N^2}\frac{k^2}{4(k-2)}-\frac{kw}{2N(k-2)}-\frac{kw^2}{4N^2}\right] -\frac{k\pi}{\tau_2}\left((s_1 - \frac{w}{N})\tau_1 +(s_2 - \frac{m}{N})\right)^2}.
\end{align}
The saddle integral gives a prefactor of $1/\sqrt{(k-2)\tau_2}$, whereas the remaining $s_2$-integral and sum over $m$ yield an extra factor of $\sqrt{\tau_2/k}N$. To find the latter, one uses
\begin{equation}
\sum_{m\in\mathbb{Z}}\int_{0}^{1}ds_2 e^{-\frac{k\pi}{\tau_2}\left(\left(s_1 - \frac{w}{N}\right)\tau_1 +\left(s_2 - \frac{m}{N}\right)\right)^2} \approx \sqrt{\frac{\tau_2}{k}}N, \quad \tau_2\to\infty.
\end{equation}
Hence, all prefactors of $k$ cancel, as well as the prefactor $N$. These states are hence really discrete, as no prefactors remain.
We notice a very important point here: no $N$ prefactor is present at all anymore for $Z$. The entropy is defined as $S =  \left.\partial_N(NZ)\right|_{N=1}$. Hence the most dominant behavior of the entropy gets contributions from all states (even those whose weight itself does not depend on $N$). Hence all states described in this paper are relevant for thermodynamics. \\ 
The exponential factors that remain combine into
\begin{equation}
h+\bar{h} = -\frac{\frac{kw}{2N}\left(\frac{kw}{2N}-1\right)}{k-2} + \frac{kw^2}{4N^2} -1,
\end{equation}
which is precisely the correct conformal weight of the lowest weight primary. Moreover, a saddle is only found when
\begin{equation}
0< \frac{w}{N}\frac{k}{k-2}-\frac{1}{k-2}<1 \quad \Longleftrightarrow  \quad \frac{1}{2} < \frac{kw}{2N} < \frac{k-1}{2},
\end{equation}
which is the unitarity constraint. Of course, this is only the constraint applied to states with solely $w\neq0$ turned on, but it is curious to find the unitarity constraint so directly. \\
Note that we used the version of the partition function (\ref{bosstart}) that is amenable to continuation to real $N$, since $N$ only appears as a parameter in this expression. Hence even for these partition functions, the most dominant state remains the same. The same will be through upon differentiating with respect to $N$ to obtain the entropy. \\

\section{Most dominant states of cigar CFT from the large $\tau_2$ limit: Type II case}
\label{mostdomII}

In this section, we will look at the large $\tau_2$ limit of the cigar orbifold partition function for type II superstrings ($u=s_1\tau+s_2$):
\begin{align}
Z = \frac{k}{4N}\int_{\mathcal{F}}\frac{d\tau d\bar{\tau}}{4\tau_2}\sum_{m,w \in \mathbb{Z}}&\int_{0}^{1}ds_1 ds_2 \frac{e^{-\frac{\pi k}{\tau_2}\left|\left(s_1 - \frac{w}{N}\right)\tau + \left(s_2 - \frac{m}{N}\right)\right|^2}}{\left|\vartheta_1(u,\tau)\eta^3\right|^2} e^{4\pi\tau_2\left(\frac{1}{4} - \frac{1}{4k}\right)}\nonumber \\
&\times \left\{\vartheta_3(u,\tau) \vartheta_3^3 + (-)^{w+1} \vartheta_4(u,\tau) \vartheta_4^3 + (-)^{m+1}\vartheta_2(u,\tau) \vartheta_2^3\right\} \nonumber\\
&\times \left\{\bar\vartheta_3(u,\tau) \bar\vartheta_3^3 + (-)^{w+1} \bar\vartheta_4(u,\tau) \bar\vartheta_4^3 + (-)^{m+1}\bar\vartheta_2(u,\tau) \bar\vartheta_2^3\right\}.
\end{align}
In this expression, we have already taken the large $\tau_2$ limit of the internal CFT and the $bc$ ghost CFT. The written remaining parts contain the cigar CFT, the superconformal partners, and the $\beta \gamma$ CFT. We will study where all the different states are located that turn out to become marginal in the flat limit. This will help clarify how a cancellation can occur. \\
In the following computations, we keep $N$ explicit, even though for our interests here $N=1$. We are mainly focusing on the discrete marginal states discussed extensively above.  We will see indications of continuous states at various places, but will not perform an extensive and exhaustive discussion of these. \\

\noindent We have the expansions for large $\tau_2$ (for $u=s_1\tau+s_2$):
\begin{align}
\vartheta_3(u,\tau) &\approx 1 - q + q^{s_1+1/2}e^{2\pi i s_2} + q^{-s_1 + 1/2}e^{-2\pi i s_2} + \hdots, \\
\vartheta_4(u,\tau) &\approx 1 - q - q^{s_1+1/2}e^{2\pi i s_2} - q^{-s_1 + 1/2}e^{-2\pi i s_2} + \hdots, \\
\vartheta_2(u,\tau) &\approx q^{1/8}q^{-s_1/2}e^{-i\pi s_2} + \hdots
\end{align}

\subsection*{NS-NS sector}
The NS-NS sector arises from the $\vartheta_3$ and $\vartheta_4$ contributions.
\subsubsection*{w=1}
The most dominant state comes from the $+1$ for $\vartheta_3$ and $\vartheta_4$. One finds
\begin{align}
Z_{w,m} &\approx \frac{1}{N}k\int_{\mathcal{F}}\frac{d\tau d\bar{\tau}}{4\tau_2} \int_{0}^{1}ds_1ds_2 \nonumber \\
&e^{4\pi\tau_2\left[1/2 -\frac{k}{4}\left(s_1 - \frac{w}{N}+\frac{1}{k}\right)^2 -\frac{w}{2N}\right] -\frac{k\pi}{\tau_2}\left((s_1 - \frac{w}{N})\tau_1 +(s_2 - \frac{m}{N})\right)^2}.
\end{align}
For $N=w=1$, one obtains a saddle if
\begin{equation}
0 < 1-1/k < 1 \quad \Longleftrightarrow \quad 1/2 < k/2 < \frac{k+1}{2},
\end{equation}
which identifies $j=k/2$. The state is precisely marginal for any $k$.

\subsubsection*{w=0}
The most dominant states here come from terms with opposite signs in $\vartheta_3$ and $\vartheta_4$. One finds four possibilities from the expansion of
\begin{equation}
\left(q^{s_1+1/2}e^{2\pi i s_2} + q^{-s_1 + 1/2}e^{-2\pi i s_2}\right)\left(\bar{q}^{s_1+1/2}e^{-2\pi i s_2} + \bar{q}^{-s_1 + 1/2}e^{2\pi i s_2}\right),
\end{equation}
 whose interpretation will be quite diverse. \\

\noindent \emph{Option 1}: $q^{-s_1+1/2}\bar{q}^{-s_1+1/2}$. \\
This leads to
\begin{align}
Z_{w,m} &\approx \frac{1}{N}k\int_{\mathcal{F}}\frac{d\tau d\bar{\tau}}{4\tau_2} \int_{0}^{1}ds_1ds_2 \nonumber \\
&e^{4\pi\tau_2\left[ -\frac{k}{4}\left(s_1 - \frac{w}{N}-\frac{1}{k}\right)^2 +\frac{w}{2N}\right] -\frac{k\pi}{\tau_2}\left((s_1 - \frac{w}{N})\tau_1 +(s_2 - \frac{m}{N})\right)^2}.
\end{align}
For $N=1$ and $w=0$, one obtains a saddle if
\begin{equation}
0 < 1/k < 1 \quad \Longleftrightarrow \quad 1/2 < 1 < \frac{k+1}{2},
\end{equation}
which identifies $j=1$. The state is again precisely marginal for any $k$. \\
It can be identified with the discrete dilaton mode. \\

\noindent \emph{Option 2}: $q^{+s_1+1/2}\bar{q}^{+s_1+1/2}$. \\
\begin{align}
Z_{w,m} &\approx \frac{1}{N}k\int_{\mathcal{F}}\frac{d\tau d\bar{\tau}}{4\tau_2} \int_{0}^{1}ds_1ds_2 \nonumber \\
&e^{4\pi\tau_2\left[\frac{2}{k} -\frac{k}{4}\left(s_1 - \frac{w}{N}+\frac{3}{k}\right)^2 -\frac{3w}{2N}\right] -\frac{k\pi}{\tau_2}\left((s_1 - \frac{w}{N})\tau_1 +(s_2 - \frac{m}{N})\right)^2}.
\end{align}
For $N=1$ and $w=0$, one cannot obtain a saddle since
\begin{equation}
0 < -3/k < 1
\end{equation}
is not true. However, in the large $\tau_2$, one can approximate the integral by setting $s_1=0$. One uses
\begin{equation}
\int_0 ^{1}dxe^{-A(x+B)^2} \approx \frac{e^{-AB^2}}{2AB} + \hdots,
\end{equation}
for large $A$ and positive $B$. Note that here $AB$ is $k$-independent, and hence after the dust settles, a factor of $\sqrt{k}$ remains in the end. The exponential then simplifies into
\begin{equation}
e^{4\pi\tau_2\left[-\frac{1}{4k}\right]}.
\end{equation}
So we found that there is a state with this weight and which has a prefactor that does not cancel. We interpret this as the lowest contribution of a continuous state. \\

\noindent \emph{Options 3 \& 4} \\
The two final options, yield the same answer and can hence be treated simultaneously. We find
\begin{align}
Z_{w,m} &\approx \frac{1}{N}k\int_{\mathcal{F}}\frac{d\tau d\bar{\tau}}{4\tau_2} \int_{0}^{1}ds_1ds_2 \nonumber \\
&e^{4\pi\tau_2\left[-\frac{1}{4k} -\frac{k}{4}\left(s_1 - \frac{w}{N}\right)^2 +\frac{w}{2N}\right] -\frac{k\pi}{\tau_2}\left((s_1 - \frac{w}{N})\tau_1 +(s_2 - \frac{m}{N})\right)^2}.
\end{align}
To obtain this equation, it must be realized that the additional $e^{4\pi i s_2}$ factors cause the $s_2$-integral to vanish in the end. To remedy this (and hence to obtain a state that is present in the spectrum), one should expand the $\vartheta_1$-function to the next order:
\begin{equation}
\vartheta_1(s_1\tau+s_2,\tau) \approx iq^{1/8}q^{-s_1/2}e^{-i\pi s_2}\left(1-q^{s_1+1}e^{2\pi i s_2}\ - q^{-s_1+1}e^{-2\pi i s_2}+ \hdots \right),
\end{equation}
where the final term is the relevant one here. Simultaneously, after Poisson resummation of the variable $m$, one should choose $n=\pm1$. These two modifications can cause the $s_2$-integral not to vanish. In the end, the above equation can be obtained in this way. \\
Setting $n=\pm1$ yields an additional exponential: $e^{-\pi N^2\tau_2/k}$. So
\begin{equation}
0 < 0 < 1 \quad \Longleftrightarrow \quad 1/2 < 1/2 < \frac{k+1}{2}.
\end{equation}
Hence $j=1/2$ and these states contribute with half weight only, due to the saddle point integral over $s_1$. In the end, we get:
\begin{equation}
e^{4\pi\tau_2\left[-\frac{1}{4k}-\frac{1}{4k}\right]},
\end{equation}
which is indeed the expected contribution from the two half-weight states present in the spectrum.

\subsection*{R-NS sector}
From the $\vartheta_2$ function, one finds the contribution
\begin{equation}
-8(-)^{m}q^{1/2}q^{-s_1/2}e^{-i\pi s_2}.
\end{equation}
Extremely important here is the overall minus sign appearing. This causes the spacetime fermions on the thermal manifold to contribute negatively to the partition function and, after differentiation w.r.t. $N$, to the entropy. \\
For $w=1$, one obtains
\begin{align}
Z_{w,m} &\approx -(-)^m \frac{4}{N}k\int_{\mathcal{F}}\frac{d\tau d\bar{\tau}}{4\tau_2} \int_{0}^{1}ds_1ds_2 \nonumber \\
&e^{-i\pi s_2} e^{4\pi\tau_2\left[-1/4+1/2 -\frac{k}{4}\left(s_1 - \frac{w}{N}+\frac{1}{k} - \frac{1}{2k}\right)^2-\frac{w}{4N}-\frac{3}{16k}\right] -\frac{k\pi}{\tau_2}\left((s_1 - \frac{w}{N})\tau_1 +(s_2 - \frac{m}{N})\right)^2},
\end{align}
where hence $j=k/2+1/4$. After again a Poisson resummation in $m$ (which now yields half-integer values for $n$), one can perform the $s_2$-integral which enforces $n=-1/2$. This then, in its turn, yields an extra factor of $e^{-\pi N^2\tau_2/(4k)}$. Combining this with the previous expression, one readily finds again the same total contribution:
\begin{equation}
e^{4\pi\tau_2\left[-\frac{1}{4k}\right]},
\end{equation}
as expected. \\

\noindent For $w=0$, one again obtains 2 cases. One of these switches the sign of $n$, and yields
\begin{align}
Z_{w,m} &\approx -(-)^m \frac{4}{N}k\int_{\mathcal{F}}\frac{d\tau d\bar{\tau}}{4\tau_2} \int_{0}^{1}ds_1ds_2 \nonumber \\
&e^{i\pi s_2} e^{4\pi\tau_2\left[-\frac{k}{4}\left(s_1 - \frac{w}{N}- \frac{1}{2k}\right)^2 + \frac{w}{4N} - \frac{3}{16k}\right] -\frac{k\pi}{\tau_2}\left((s_1 - \frac{w}{N})\tau_1 +(s_2 - \frac{m}{N})\right)^2}.
\end{align}
The discrete momentum is forced to $n=+1/2$, giving an additional correction of $e^{-\pi N^2\tau_2/(4k)}$ in the end again. The same weight as above is again obtained, where we have now: $j=3/4$. \\

\noindent The other case, yields $n=-3/2$ and
\begin{align}
Z_{w,m} &\approx -(-)^m  \frac{4}{N}k\int_{\mathcal{F}}\frac{d\tau d\bar{\tau}}{4\tau_2} \int_{0}^{1}ds_1ds_2 \nonumber \\
&e^{-3i\pi s_2} e^{4\pi\tau_2\left[-\frac{1}{4k}-\frac{k}{4}\left(s_1 - \frac{w}{N}+\frac{2}{k}- \frac{1}{2k}\right)^2 -\frac{w}{N}+\frac{1}{k} - \frac{1}{2k}+ \frac{w}{4N} + \frac{1}{16k}\right] -\frac{k\pi}{\tau_2}\left((s_1 - \frac{w}{N})\tau_1 +(s_2 - \frac{m}{N})\right)^2}.
\end{align}
The momentum gives a contribution of $e^{-9\pi N^2\tau_2/(4k)}$ to be added again in the end. \\
The $s_1$-integral has no saddle point within the required interval and this sector can hence again be interpreted as a continuous state. 
As always, the NS-R sector is completely analogous and will not be discussed explicitly.

\subsection*{R-R sector}
Here one obtains for the partition function
\begin{align}
Z_{w,m} &\approx  \frac{16}{N}k\int_{\mathcal{F}}\frac{d\tau d\bar{\tau}}{4\tau_2} \int_{0}^{1}ds_1ds_2 \nonumber \\
& e^{4\pi\tau_2\left[-\frac{k}{4}\left(s_1 - \frac{w}{N}\right)^2 -\frac{1}{4k}\right] -\frac{k\pi}{\tau_2}\left((s_1 - \frac{w}{N})\tau_1 +(s_2 - \frac{m}{N})\right)^2},
\end{align}
always yielding a half weight state with $j=\frac{1}{2}$ for $w=0$ and $j=\frac{k+1}{2}$ for $w=1$. No separate study needs to be made for the $w=0$ case or the $w=1$ case. Again we see that the resulting weight is independent of $N$.\footnote{Half-weight states are somewhat special since they are just on the border of becoming continuous. Strictly speaking, they are no longer normalizable. 
Note that if the saddle is at $s_1=0$, the above expansion has a problem, since one cannot say $s_1\tau_2$ is large anymore. Related to this, the sine denominator of the $\vartheta_1$-function actually blows up, which is invisible in the limit that we were studying. Hence such saddles might actually have a portion of a continuous sector attached to them.} \\

\noindent To sum up, we have found all of the discrete states that are becoming marginal in the large $k$ limit. We also obtained hints of the appearance of the marginal continuous modes, although a more elaborate analysis of these will not be conducted here.

\section{Modes that appear or disappear from the spectrum: some toy series}
\label{toy}
Unlike in flat space, it is possible for modes to appear or disappear as we change the conical deficit. It is unclear then whether it makes sense to think about the entropy associated to a single mode on the thermal manifold as it might disappear upon changing $N$ (as one is instructed to do when computing the entropy $S$).

\noindent To investigate this, we want to draw attention to the situation at hand: we have a series of contributions, with each term an exponential in $\tau_2$, in which we look at the most dominant terms ($\tau_2\to+\infty$) in the entropy $S= \partial_N(NZ)$, evaluated in the end at $N=1$. \\
Let us look at some mathematical sums that exhibit similar features. \\

\noindent Suppose we try to evaluate the following sum:
\begin{equation}
\partial_N\sum_{n=0}^{N}\exp(-nf(N)t)
\end{equation}
in the large $t$ limit for an arbitrary positive function $f(N)$. Explicit evaluation shows that this is $-tf'(N)e^{-f(N)t}$. This is indeed expected as the dominant contribution to the entropy comes from the $n=1$ term (the $n=0$ term vanishes for the entropy). This mode is present for any choice of $N$ and is hence naturally continued in $N$, just as the thermal scalar state. \\
Note that this is precisely the situation of the thermal scalar, as this mode also formally disappears from the spectrum if $N$ would become \emph{smaller} than 1. Nonetheless, this mode provides the dominant contribution to the entropy indeed. \\

\noindent It is at least a bit reassuring that this class of examples can be generalized into
\begin{equation}
\partial_N\sum_{n=0}^{N}g(n,N)\exp(-nf(N)t).
\end{equation}
Assuming $g(n,N)$ can be Taylor-expanded around $n=0$ and $\partial_N g(0,N)=0$, the dominant contribution comes indeed from the $n=1$ mode as\footnote{If the condition $\partial_N g(0,N)=0$ is not satisfied, the $n=0$ term makes the dominant contribution $\partial_N g(0,N)$ as it should.}
\begin{equation}
-tf'(N)g(1,N)e^{-f(N)t} + \partial_N g(1,N)e^{-f(N)t}.
\end{equation}

\noindent The above examples have in common that the most dominant contribution is not at the boundary of the summation interval, in the sense that changing $N$ slightly will not cause it to be absent suddenly. In contrast, consider the sum
\begin{equation}
\partial_N\sum_{n=0}^{N}\exp(-(N-n)t)
\end{equation}
in the large $t$ limit. We expect the $n=N$ term to give the dominant contribution. Its entropy equals $-t$. This is similar to the situation we encountered here, as this does \emph{not} define a good continuation of this mode as one changes $N$.\footnote{One expects to continue this mode with a predefined, $N$-independent choice of $n$. This is analogous to the situation of the R-NS and NS-R spacetime fermion sectors.} The most dominant contribution is on the border of being excluded in the summation range. However, the above expression actually has for its dominant contribution $te^{-(N+1)t}$. This shows that the first few most dominant contributions all get cancelled! \\

\begin{figure}[h]
\centering
\includegraphics[width=0.5\textwidth]{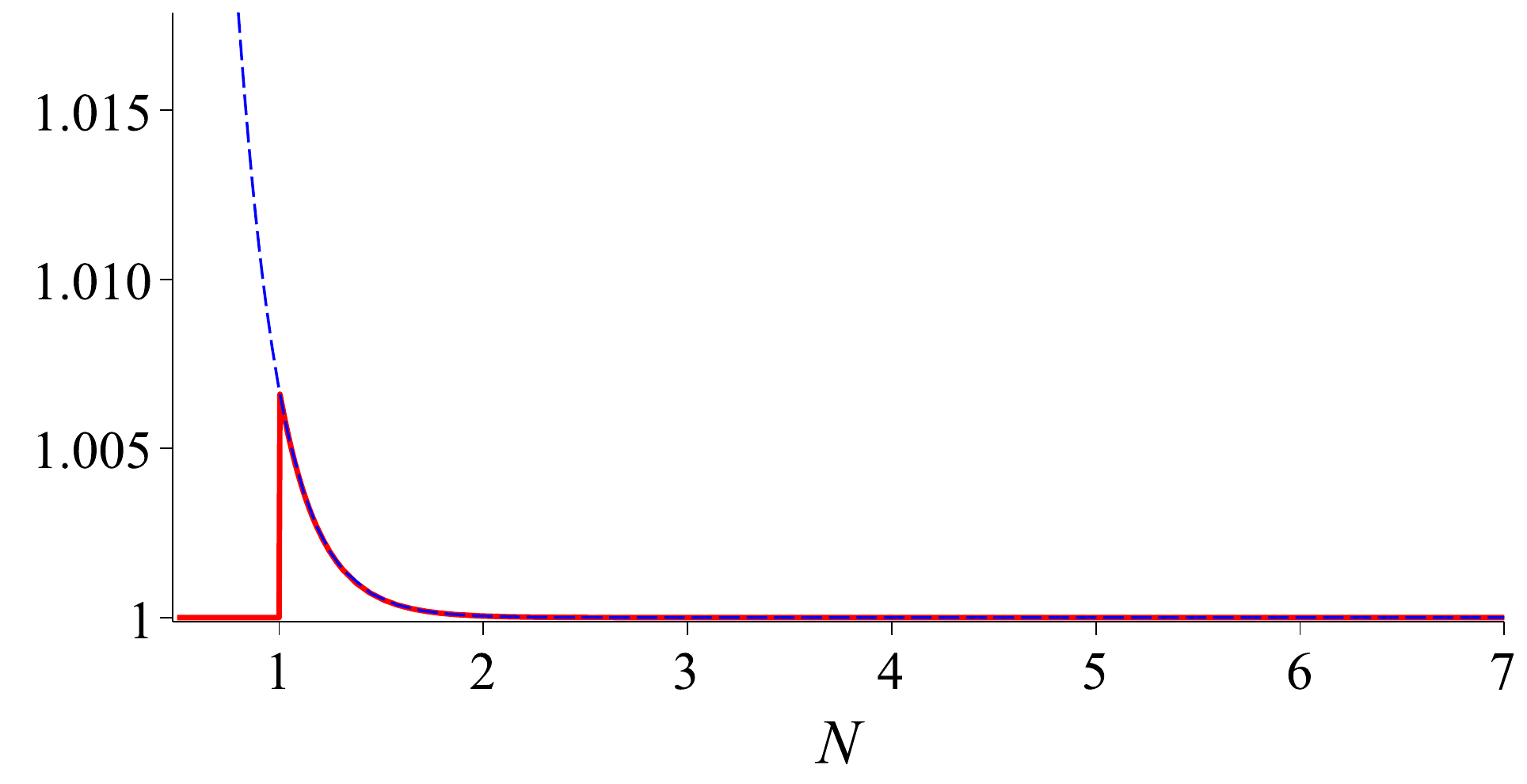}
\caption{Sum (\ref{summm}) as a function of $N$ for large $t$ ($t=5$ in this figure). The red curve represents the sharply truncated version. The blue dashed curve represents the continuation of the sum from its closed form expression. It is clearly seen that the entropy as defined through the closed form expression agrees with that from the truncated version if one computes it in the latter case with $N\geq1$.}
\label{sums}
\end{figure}

\noindent The above summations have in common that one can perform the summation analytically. In more interesting cases, such as when the exponential depends on $n^2$ for instance, this is not possible. Nonetheless, numerical experimentation does teach us the following lesson. \\
If one considers the sharp-cutoff continuation of the series, such as for instance ($p\in\mathbb{N}$)
\begin{equation}
\partial_N\sum_{n=0}^{N}\exp(-n^pNt) \quad \to \quad \partial_N\sum_{n=0}^{\left\lfloor N\right\rfloor}\exp(-n^pNt), 
\end{equation}
then one can demonstrate that the dominant large $t$ behavior corresponds to indeed taking the $n=1$ mode and computing its entropy.\footnote{We note that this cut-off continuation of the series is inherently discontinuous. However, in the large $t$ limit, the resulting jumps are suppressed.} This is interesting since this is precisely the way in which the cigar regularization of flat cones works: as explained in section \ref{meth3}, this continuation amounts to a sharp truncation in the large $k$ limit summations indeed \cite{Mertens:2014saa}. \\
As a further lesson on how the continuation in $N$ should be looked at, we plot in figure \ref{sums} the case of the sum
\begin{equation}
\label{summm}
\sum_{n=0}^{N}\exp(-nNt),
\end{equation}
continued either using its closed form expression or its sharply truncated definition (i.e. by putting a floor-function in the summation range). It is clear that both expressions agree in the large $t$ limit, as long as $N\geq1$: both are given by the dominant $n=1$ contribution. The entropy should be computed by a right-derivative. \\

\noindent It is intuitively obvious that if we differentiate with respect to $N$ for a dominant mode that is deeply engraved into the spectrum, then one can just differentiate this mode itself without worrying about subtleties happening for modes that appear or disappear, since this is expected to be subdominant anyway. However, if the dominant mode is on the brink of disappearing, then more care must be taken. The above mathematical exposition is an illustration of this fact.

\section{Some useful theta function formulas}
\label{appTheta}
We collect some useful identities for the Jacobi theta functions:
\begin{align}
\vartheta_3(\nu+1,\tau) &= \vartheta_3(\nu,\tau), \\
\vartheta_4(\nu + 1,\tau) &= \vartheta_4(\nu,\tau), \\
\vartheta_2(\nu+1,\tau) &= -\vartheta_2(\nu,\tau), \\
\vartheta_1(\nu+1,\tau) &= -\vartheta_1(\nu,\tau),
\end{align}
and
\begin{align}
\vartheta_3(\nu+\tau,\tau) &= e^{-\pi i \tau - 2\pi i \nu}\vartheta_3(\nu,\tau), \\
\vartheta_4(\nu + \tau,\tau) &= - e^{-\pi i \tau - 2\pi i \nu}\vartheta_4(\nu,\tau), \\
\vartheta_2(\nu+\tau,\tau) &= e^{-\pi i \tau - 2\pi i \nu}\vartheta_2(\nu,\tau), \\
\vartheta_1(\nu+\tau,\tau) &= - e^{-\pi i \tau - 2\pi i \nu}\vartheta_1(\nu,\tau).
\end{align}
We also require derivatives of theta-functions w.r.t. $\nu$:
\begin{align}
\partial_\nu \vartheta_3 (\nu,\tau) &= 2\pi i \sum_{m=1}^{+\infty}\left(\frac{q^{m-1/2}z}{1+zq^{m-1/2}} - \frac{q^{m-1/2}z^{-1}}{1+z^{-1}q^{m-1/2}}\right) \vartheta_3(\nu,\tau), \\
\partial_\nu \vartheta_4 (\nu,\tau) &= -2\pi i \sum_{m=1}^{+\infty}\left(\frac{q^{m-1/2}z}{1-zq^{m-1/2}} - \frac{q^{m-1/2}z^{-1}}{1-z^{-1}q^{m-1/2}}\right) \vartheta_4(\nu,\tau), \\
\partial_\nu \vartheta_2 (\nu,\tau) &= 2\pi i \left[-\frac{1}{2i} \tan(\pi \nu) + \sum_{m=1}^{+\infty}\left(\frac{q^{m-1/2 }z}{1+zq^{m-1/2}} - \frac{q^{m-1/2}z^{-1}}{1+z^{-1}q^{m-1/2}}\right)\right] \vartheta_2(\nu,\tau).
\end{align}
These derivatives have similar behavior under shifts by $\tau$ of $\nu$ as:
\begin{align}
\left(\partial_\nu \vartheta_3\right)(\nu+\tau,\tau) &= e^{-\pi i \tau - 2\pi i \nu}\left(\partial_{\nu}\vartheta_3\right)(\nu,\tau) - 2\pi i e^{-\pi i \tau - 2\pi i \nu} \vartheta_3(\nu,\tau), \\
\left(\partial_\nu \vartheta_4\right)(\nu+\tau,\tau) &= -e^{-\pi i \tau - 2\pi i \nu}\left(\partial_{\nu}\vartheta_4\right)(\nu,\tau) + 2\pi i e^{-\pi i \tau - 2\pi i \nu} \vartheta_4(\nu,\tau), \\
\left(\partial_\nu \vartheta_2\right)(\nu+\tau,\tau) &= e^{-\pi i \tau - 2\pi i \nu}\left(\partial_{\nu}\vartheta_2\right)(\nu,\tau) - 2\pi i e^{-\pi i \tau - 2\pi i \nu} \vartheta_2(\nu,\tau).
\end{align}
The last one of these formulas can be obtained by realizing that
\begin{align}
&\left(\partial_\nu \vartheta_2\right)(\nu+\tau,\tau) \nonumber \\ 
&= 2\pi i \left[-\frac{1}{2i} \tan(\pi (\nu+\tau)) + \sum_{m=2}^{+\infty}\frac{q^{m-1/2 }z}{1+zq^{m-1/2}} - \sum_{m=0}^{+\infty}\frac{q^{m-1/2}z^{-1}}{1+z^{-1}q^{m-1/2}}\right] e^{-\pi i \tau - 2\pi i \nu}\vartheta_2(\nu,\tau) \nonumber \\ 
&= 2\pi i \left[\frac{1}{2} \frac{zq-1}{zq+1} + \sum_{m=1}^{+\infty}\left(\frac{q^{m-1/2 }z}{1+zq^{m-1/2}} - \frac{q^{m-1/2}z^{-1}}{1+z^{-1}q^{m-1/2}}\right) - \frac{zq}{1+zq} - \frac{1}{z+1}\right] \nonumber \\ 
&\quad\quad\quad \times e^{-\pi i \tau - 2\pi i \nu}\vartheta_2(\nu,\tau). 
\end{align}
One rewrites
\begin{equation}
\frac{1}{2} \frac{zq-1}{zq+1} - \frac{zq}{1+zq} - \frac{1}{z+1} = \frac{1}{2}\frac{z-1}{z+1} - 1 = -\frac{1}{2i} \tan(\pi \nu) -1.
\end{equation}
Combining these formulas, one finds
\begin{align}
\left(\partial_\nu \vartheta_3\right)(u,\tau) \vartheta_3^3 + (-)^{w+1}\left(\partial_\nu \vartheta_4\right)(u,\tau) \vartheta_4^3 + (-)^{m+1}\left(\partial_\nu \vartheta_2\right)(u,\tau) \vartheta_2^3 = 0
\end{align}
where $u=w\tau + m$ and for any integer $w$ and $m$. In the main text, we need this formula for $w=0,1$ and $m=0,1$, but it holds more generally.\footnote{We checked this formula numerically as well. One finds also that these values of $u$ are the only ones where this vanishing occurs. Interestingly, it continues to hold for up to two more additional derivatives w.r.t. $\nu$; the first non-vanishing expression is hence ($u=w\tau + m$)
\begin{equation}
\left(\partial^4_\nu \vartheta_3\right)(u,\tau) \vartheta_3^3 + (-)^{w+1}\left(\partial^4_\nu \vartheta_4\right)(u,\tau) \vartheta_4^3 + (-)^{m+1}\left(\partial^4_\nu \vartheta_2\right)(u,\tau) \vartheta_2^3 \neq 0.
\end{equation}}

\end{document}